\documentclass{elsart}

\usepackage{amsmath}
\usepackage{amssymb}
\usepackage{graphicx}
\usepackage{color}

\usepackage[small]{caption}
\usepackage{eucal}
\usepackage{mathrsfs}
\usepackage{tabulary}
\usepackage{multirow}

\begin{document}

\begin{frontmatter}

\title{Extended Smoothed Boundary Method for Solving Partial Differential Equations with General Boundary Conditions on Complex Boundaries}
\author{Hui-Chia Yu},
\author{Hsun-Yi Chen}, and
\author{K. Thornton\corauthref{cor}}
\corauth[cor]{K. Thornton} 
\ead{kthorn@umich.edu}

\address{Department of Materials Science and Engineering, the University of Michigan, Ann Arbor, USA}


\begin{abstract} In this article, we describe an approach for solving partial differential equations with general boundary conditions imposed on arbitrarily shaped boundaries. A continuous function, the domain parameter, is used to modify the original differential equations such that the equations are solved in the region where a domain parameter takes a specified value while boundary conditions are imposed on the region where the value of the domain parameter varies smoothly across a short distance. The mathematical derivations are straightforward and generically applicable to a wide variety of partial differential equations. To demonstrate the general applicability of the approach, we provide four examples herein: (1) the diffusion equation with both Neumann and Dirichlet boundary conditions; (2) the diffusion equation with both surface diffusion and reaction; (3) the mechanical equilibrium equation; and (4) the equation for phase transformation with the presence of additional boundaries. The solutions for several of these cases are validated against corresponding analytical and semi-analytical solutions. The potential of the approach is demonstrated with five applications: surface-reaction-diffusion kinetics with a complex geometry, Kirkendall-effect-induced deformation, thermal stress in a complex geometry, phase transformations affected by substrate surfaces, and a self-propelled droplet.
\end{abstract}

\begin{keyword}
smoothed boundary method; diffuse interface method; complex microstructure; image-based simulation
\end{keyword}

\end {frontmatter}

\section{Introduction} 

The smoothed boundary method \cite{Bueno-Orovio:2006a,Bueno-Orovio:2006b,Bueno-Orovio:2006c} has recently been demonstrated as a powerful tool for solving diffusion equations with no-flux boundary conditions imposed at irregular boundaries within the computational domain. The method's origin can be traced to the embedded boundary method and the immersed interface (boundary) method, both of which embed a more complicated domain in a computational box with simpler geometry. These methods are advantageous because they eliminate the need for a structural mesh when solving partial differential equations within the embedded geometries because the grid system is obtained by a discretization of the regular computational box. (For an overview, see Refs. \cite{Badea:2001,Peskin:2002,Boyd:2005,Ito:2006,Lui:2009,Sabetghadam:2009}.) To impose boundary conditions at the immersed interfaces, a discretized Dirac delta function is employed to distribute a singular source over nearby grid points. Various studies have examined optimal discretization of the Dirac delta function \cite{Ito:2006,Peskin:1995,Tornberg:2003}. Similarly, the level set method can also be considered an immersed-interface-type method because the boundary defined by the contour of the zero level set is embedded within a regular computational box. Although the level set method was developed mainly for tracking moving boundaries \cite{Sethian:1999}, it is also applicable for solving partial differential equations with boundary conditions imposed at the zero-level-set contour using a technique similar to the immersed interface method \cite{Adalsteinsson:2003,Sethian:2008}. In addition to the methods above, the phase field approach possesses certain similarities to embedding interfaces within the computational box and also has the significant advantage of avoiding the need to explicitly track the interfaces. However, phase field methods are not widely employed in simulations that involve {\it explicit} boundary conditions along interfaces. While the Gibbs-Thomson boundary condition is automatically imposed in the standard phase field model, there have been only a few studies in which phase field models were used with explicit boundary conditions at interfaces. For example, in solidification problems, equilibrium conditions, such as equilibrium temperature or concentration \cite{Karma:1998a,Karma:1999a}, are imposed at solid-liquid interfaces in which the order parameter field and the temperature field are coupled via a latent heat term. Except for this type of phase field model, the direct application of boundary conditions at interfaces is rarely used because the construction of boundary conditions requires the tedious process of formally including an additional energy term in the energy functional, as suggested by Cahn \cite{Cahn:1977a}. Examples of imposed boundary conditions in the phase field model using modified energy functionals can be found in the recent works of Warren et al. \cite{Granasy:2007,Warren:2009}.

In contrast to the techniques for distributing a singular source of boundary conditions to grid points near the interfaces in the immersed interface method, the smoothed boundary method spreads the zero-thickness boundary into a finite-thickness diffuse interface using a phase-field-like, continuously transitioning domain indicator function (hereinafter termed ``the domain parameter''). Mathematically, this method approximates a Heaviside step function as a hyperbolic tangent function having one specified uniform value in a domain and continuously changing its value across the interface to another value specifying the other domain. Therefore, boundary conditions are straightforwardly distributed among the grid points residing within the interfacial regions in which the domain parameter varies smoothly across a short distance. This method has been successfully employed in simulating diffusion processes \cite{Kockelkoren:2003,Levine:2005} and wave propagation \cite{Bueno-Orovio:2006a,Bueno-Orovio:2006b,Bueno-Orovio:2006c,Fenton:2005,Buzzard:2007} constrained within geometries described by a domain parameter with a no-flux boundary condition imposed on the diffuse interfaces. Similar approaches have also been proposed to solve differential equations constrained in domains defined by order parameters in the phase field model \cite{Gal:2006,Wu:2007,Gal:2008}. These works demonstrated the potential for this type of numerical method that circumvents the difficulties associated with constructing a finite element mesh (e.g., meshing the surface and then building a volumetric mesh based on the surface mesh or combining regular subdomains that can be easily meshed). Such an approach is particularly useful when complex structures are involved. However, the method was only applicable to no-flux boundary conditions, and no further extensions to other types of equations or boundary conditions have been reported. Recently, Lowengrub and coworkers \cite{Li:2009,Lowengrub:2009,Ratz:2006,Teigen:2009a,Sohn:2010,Aland:2010,Teigen:2011} developed an alternative formulation for solving partial differential equations with various boundary conditions, based on asymptotic analyses commonly conducted in phase field modeling, which is different from the general derivation of the smoothed boundary method presented in this paper. Although such an implementation for imposing boundary conditions differs from the `formal' practice suggested by Cahn \cite{Cahn:1977a}, it dramatically simplifies the formulation, provides a justification of the method, and increases the applicability of the approach.  

In this study, we provide a mathematically consistent smoothed boundary method and a precise derivation for the equations, such that the method is generalized from its limited original application to a wide range of differential equations and boundary conditions. We consider the following specific equations: (1) the diffusion equation with Neumann and/or Dirichlet boundary conditions; (2) the bulk diffusion equation coupled with surface diffusion and reaction; (3) the mechanical equilibrium equation for linear elasticity; and (4) the Allen-Cahn or Cahn-Hilliard equations with contact angles as boundary conditions. The method is especially useful for three-dimensional image-based simulations because of its efficiency and flexibility in handling complex geometries without structural-mesh techniques.

\section{Background} 

The method is based on a diffuse interface description of different phases, similar to the continuously transitioning order parameters in the phase field method \cite{Cahn:1958,Cahn:1959,Allen:1979,Ginzburg:1950,Chen:2002,Emmerich:2003} often employed in simulating phase transformations and microstructural evolutions in materials. Phase field models are based on thermodynamics and kinetics of multiphase system, in which phases (e.g., liquid, solid, vapor, or two solids or liquids with different compositions) are described by one or more order parameters with prescribed bulk values for each phase. At the interface, the order parameter changes in a controlled manner. Asymptotic analyses \cite{Emmerich:2003} can be used to show that the phase-field governing equations approach the corresponding sharp interface equations in the sharp interface limit.

\begin{figure}[htb] 
\begin{center}
\includegraphics[width=1\textwidth]{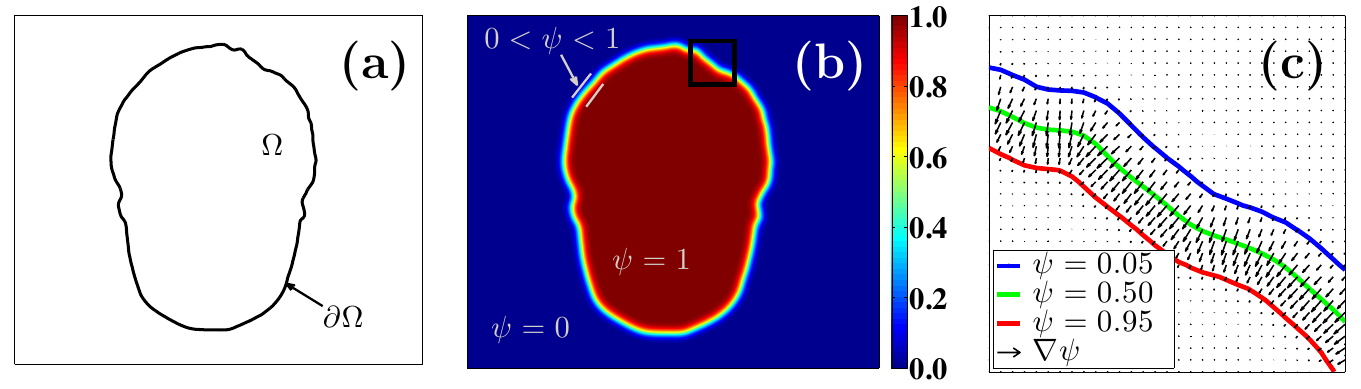}
\end{center}
\caption{(a) Conventional sharp interface description of a domain bound by a zero-thickness boundary. (b) Diffuse interface domain and boundary defined by a continuous domain parameter, $\psi$. (c) Inward normal vectors defined by $\nabla \psi$ plotted for the square region in (b).}
\label{Domain}
\end{figure}

Despite the advantages of phase-field-type diffuse interface methods for front tracking problems, we focus on another important advantage for efficiently solving differential equations within diffuse-interface-defined domains. Here, we adopt the concept to describe internal domain boundaries with an order-parameter-like domain parameter, which may or may not be stationary and takes a value of 1 inside the domain of interest and 0 outside. The equations are solved where the domain parameter is 1, with boundary conditions imposed where the domain parameter is at an intermediate value (approximately 0.5). Figure~\ref{Domain} schematically illustrates the sharp and diffuse interfaces. In the conventional sharp interface description, the domain of interest is $\Omega$ and is bounded by a zero-thickness boundary, denoted $\partial \Omega$; see Fig.~\ref{Domain}(a). Within $\Omega$, the partial differential equations are solved according to the boundary conditions imposed at $\partial \Omega$. Conversely, in the diffuse interface description, we employ a continuous domain parameter, which is uniformly 1 within the domain of interest and uniformly 0 outside. In this case, the originally sharp domain boundary is smoothed to yield a diffuse interface with a finite thickness given by $0<\psi<1$. The system thus determines the boundary by variation of the domain parameter. In addition, the gradient of the domain parameter $\nabla \psi$ automatically determines the inward normal vector of the contour level sets of $\psi$; see Fig.~\ref{Domain}(c). Our goal is to solve partial differential equations within the region where $\psi=1$ while imposing boundary conditions at the narrow transitioning interfacial region where $0<\psi<1$.  However, the convention can be reversed such that the domain is defined by $\psi=0$, in which case the following derivation could be modified by replacing $\psi$ with $1-\psi$ accordingly.  This could be used to solve a problem where multiple equations govern different regions within the computational domain.  Furthermore, these equations can be coupled through the shared boundary conditions, making the method highly versatile.

\section{Formulation}

\subsection{General Approach} 

The general approach is as follows. The domain parameter describes the domain of interest ($\psi=1$ inside the domain, and $\psi=0$ outside). The transition between the two values described is must be smooth so that the gradient is well defined. In this work, we have assumed the domain parameter to take the form of a hyperbolic tangent function for three reasons. First, it can be numerically implemented with ease. Second, it is consistent with the solution to the phase field equations, and thus this choice allows coupling of the two approaches; that is, one could for example simulate microstructural evolution using the phase field model, but impose interfacial flux boundary condition using the smoothed boundary method. Third, when given a non-smooth microstructural data, one could use the phase field equations to obtain the domain parameter from the discontinuous data. Other forms of domain parameters are possible, as long as the transition is monotonic and the gradient of the domain parameter has a narrow peak in the interfacial region. A better convergence could possibly be obtained using a function that has a more confined gradient; however, examining other forms of the domain parameter is beyond the scope of this paper.Ó As an example, we consider the Laplacian of the function, $H$. As the first step in deriving formulation for the Neumann boundary condition, $\nabla^2 H$ is multiplied by the domain parameter, $\psi$. Using identities of the product rule of differentiation such as:
\begin{equation} \label{ChainRule1}
\psi \nabla^2 H = \nabla \cdot (\psi \nabla H) - \nabla \psi \cdot \nabla H,
\end{equation}
we obtain terms proportional to $\nabla \psi$. Because the inward unit normal of the boundary  (pointing to the regions where $\psi=1$), $\vec{n}$, is given by $\nabla \psi/ |\nabla \psi |$, such terms can be written in terms of $\partial H/\partial n = \nabla H \cdot \vec{n}  = \nabla H \cdot \nabla \psi /|\nabla \psi|$, and thus reformulated to be the Neumann boundary condition imposed on the diffuse interface.  

Similarly, to derive the smoothed boundary formulation for the Dirichlet boundary condition, the differential equation is multiplied by the square of the domain parameter. Again using mathematical identities, $\psi^2 \nabla^2 H = \psi \nabla \cdot (\psi \nabla H) - \psi \nabla \psi \cdot \nabla H$, where $\psi \nabla \psi \cdot \nabla H = \nabla \psi \cdot \nabla \left( \psi H \right) - H \left| \nabla \psi \right|^{2}$, we obtain: 
\begin{equation} \label{ChainRule2}
\psi^2 \nabla^2 H = \psi \nabla \cdot (\psi \nabla H) - [\nabla \psi \cdot \nabla \left( \psi H \right)-H |\nabla \psi|^2].
\end{equation}
Note that $H$, associated with $|\nabla \psi|^2$ appearing in the last term, is the boundary value, $H |_{\partial \Omega}$, imposed on the diffuse interface. Specific details of the derivation depend on the equation to which the approach is applied, and we therefore provide four examples below.

\subsection{Diffusion Equation} \label{DiffEqn} 

The first example is the diffusion equation with Neumann and/or Dirichlet boundary conditions. The Neumann boundary condition is appropriate, for example, as the no-flux boundary condition, whereas the Dirichlet boundary condition is necessary when the diffusion equation is solved with a fixed concentration on the boundaries. For Fick's Second Law of diffusion, the original governing equation is expressed as: 
\begin{equation} \label{FSL1}
\frac{\partial C}{\partial t} = -\nabla \cdot \vec{j} + S = \nabla \cdot (D \nabla C) + S, 
\end{equation}
where $\vec{j}$ is the flux vector, $D$ is the diffusion coefficient, $C$ is the concentration, $S$ is the source term, and $t$ is time. Instead of directly solving the diffusion equation, we multiply Eq.~\eqref{FSL1} by $\psi$, the domain parameter that describes the domain in which diffusion occurs, and use the identity $\psi \nabla \cdot (D \nabla C) = \nabla \cdot (\psi D \nabla C) - \nabla \psi \cdot (D \nabla C)$ to obtain the smoothed boundary formulated diffusion equation:
\begin{equation} \label{SBM-FSL1}
\psi \frac{\partial C}{\partial t} = \nabla \cdot (\psi D \nabla C) - \nabla \psi \cdot (D \nabla C) + \psi S.
\end{equation}
Next, we consider the boundary condition in this formulation. The Neumann boundary condition is the inward flux across the domain boundary, mathematically the normal gradient of $C$ at the diffuse interface, and is treated as:
\begin{equation} \label{NBC1}
D B_N \equiv  D \frac{\partial C}{\partial n} \equiv \vec{n} \cdot \vec{j} = - \frac{\nabla \psi \cdot (D \nabla C) }{\left| \nabla \psi \right|}, 
\end{equation}
where $\vec{n} = \nabla \psi/|\nabla \psi|$ is the unit inward normal vector at the boundary defined in the diffuse interface description. Note that the flux at the interface is equal to $D B_{N}$.  Equation \eqref{NBC1} is rearranged to become $\nabla \psi \cdot (D \nabla C) = - \left| \nabla \psi \right| D B_{N}$ and substituted back into Eq.~\eqref{SBM-FSL1}; thus, we obtain:
\begin{equation} \label{SBM-FSL2}
\frac{\partial C}{\partial t} = \frac{1}{\psi}\nabla \cdot (\psi D \nabla C) + \frac{\left| \nabla \psi \right |}{\psi} D B_{N} + S,
\end{equation}
with the Neumann boundary condition appearing in the second term. When a no-flux boundary is imposed, the second term vanishes and the resulting equation is the same as that proposed in Refs. \cite{Bueno-Orovio:2006a,Bueno-Orovio:2006b,Bueno-Orovio:2006c,Kockelkoren:2003,Fenton:2005}. 

To impose the Dirichlet boundary condition, we manipulate the original governing equation in a procedure similar to the derivation of Eq.~\eqref{SBM-FSL2}. Multiplying Eq.~\eqref{SBM-FSL1} by $\psi$ and using the identity $\psi \nabla \psi \cdot( D \nabla C) = D [\nabla \psi \cdot \nabla \left( \psi C \right) - C \nabla \psi \cdot \nabla \psi] = D [\nabla \psi \cdot \nabla \left( \psi C \right) - C |\nabla \psi|^2]$ to replace the second term, we obtain:
\begin{equation} \label{SBM-FSL3}
\frac{\partial C}{\partial t}  =\frac{1}{\psi} \nabla \cdot (\psi D \nabla C) - \frac{1}{\psi^2} D[ \nabla \psi \cdot \nabla \left( \psi C \right) - B_D \left| \nabla \psi \right|^{2}] + S,
\end{equation}
where $B_D$ is the Dirichlet boundary condition imposed at the diffuse interface to replace $C$, associated with $\left| \nabla \psi \right|^{2}$ in the third term. The convergence to the imposed Neumann and Dirichlet boundary conditions is shown in Appendix \ref{SBM_proof}.

In this method, the boundary gradient, $B_N$, and the boundary value, $B_D$, are not specified to be constant values; they can vary spatially and/or temporally or be functions of $C$ or other parameters. In addition, it is convenient to use weighting factors to combine Eqs.~\eqref{SBM-FSL2} and \eqref{SBM-FSL3} to impose Neumann and Dirichlet boundary conditions simultaneously to yield mixed (or Robin) boundary conditions. The equation then becomes: 
\begin{equation}\label{Mixed-SBM-BC-01}
\frac{\partial C}{\partial t} = \frac{1}{\psi} \nabla \cdot (\psi D \nabla C) + \frac{| \nabla \psi |}{\psi} D B_N W_N- \frac{D}{\psi^2}  [ \nabla \psi \cdot \nabla (\psi C) - B_D |\nabla \psi |^2 ] W_D+ S,
\end{equation}
where $W_N$ and $W_D$ are the spatially dependent weighting factors for the Neumann and Dirichlet boundary conditions, respectively ($W_N + W_D = 1$). These factors can be a linear combination when imposing Robin boundary conditions or be employed to impose Neumann and Dirichlet boundary conditions at different regions of the interface. Moreover, a small nonzero value ($10^{-16}  < \upsilon < 10^{-6}$) should be added to the domain parameter appearing in the denominators to avoid singularities resulting from the terms $1/\psi$ and $1/\psi^2$ in regions where $\psi = 0$.

\subsection{Coupled Surface-Bulk Diffusion} \label{SurfDiffFormulation} 

The second example demonstrates that surface diffusion can be incorporated into the smoothed boundary equation derived above. For this case, we take the set of equations that includes the surface reaction, surface diffusion and bulk diffusion to describe an oxygen reduction model in a solid oxide fuel cell (SOFC) cathode \cite{Lu:2009}. The oxygen-vacancy concentration, $C$, on the cathode surface is governed by Fick's Second Law for surface diffusion:
\begin{equation} \label{FSL-S1}
- D_\text{b}\frac{\partial C}{\partial n} = \kappa C - D_\text{s} \nabla_\text{s}^2 C + L \frac{\partial C}{\partial t},
\end{equation}
where $n$ is the coordinate along the inward unit normal vector of the surface and $\nabla_\text{s}^2$ is the surface Laplacian. The parameters $D_\text{b}$, $\kappa$, $D_\text{s}$, and $L$ are the bulk diffusivity, reaction rate, surface diffusivity, and accumulation coefficient, respectively. Thus, the term on the left-hand side represents the flux from the bulk, and the terms on the right-hand side represent the surface reaction, surface diffusion and surface concentration accumulation, respectively \cite{Lu:2009}. Here, these parameters are all assumed to be constant for simplicity. In the bulk of the cathode material, the oxygen-vacancy diffusion is governed by Fick's Second Law for bulk diffusion:
\begin{equation} \label{FSL-B1}
\frac{\partial C}{\partial t} = D_\text{b} \nabla^2 C. 
\end{equation}
To simulate the oxygen-vacancy concentration evolution in the cathode, the two diffusion equations, Eqs.~\eqref{FSL-S1} and \eqref{FSL-B1}, are coupled and must be solved simultaneously, with the flux normal to the cathode surface as a common boundary condition. Recently, a similar set of equations was formulated using anothor diffuse interface approach combined with an asymptotic analysis \cite{Teigen:2009a}, which led to two differential equations coupled by a common boundary condition. We show below that we can eliminate the need for solving two separate equations by applying the smoothed boundary formulation described herein to obtain a single equation that governs both surface and bulk effects.

Similar to the derivation of Eq.~\eqref{SBM-FSL2}, we first multiply Eq.~\eqref{FSL-B1} by $\psi$ and apply the product rule of differentiation to obtain the bulk diffusion equation containing a boundary term, $D_\text{b} \nabla \psi \cdot \nabla C$, similar to Eq.~\eqref{SBM-FSL1}. As in Eq.~\eqref{NBC1}, the normal gradient at the diffuse interface is defined by $\partial C/\partial n = -\nabla C \cdot \nabla \psi /|\nabla \psi |$. Substituting this relationship back into Eq.~\eqref{FSL-S1} and rearranging terms gives:
\begin{equation} \label{SBM-FSL-BC-1}
\nabla \psi \cdot \nabla C = \frac{|\nabla \psi|}{D_\text{b}} \bigg[ \kappa C - D_\text{s} \nabla_\text{s}^2 C + L \frac{\partial C}{\partial t} \bigg].
\end{equation}
Using this relation, we replace the boundary term to obtain the smoothed boundary formulated equation:
\begin{equation} \label{SBM-FSL-BS1}
\frac{\partial C}{\partial t} =   \frac{1}{\psi} D_\text{b} \nabla \cdot (\psi \nabla C) - \frac{|\nabla \psi |}{\psi} \bigg[ \kappa C- D_\text{s}  \nabla_\text{s}^2 C + L\frac{\partial C}{\partial t} \bigg],
\end{equation}
which combines the bulk diffusion and surface diffusion terms into a single equation used in the examples presented in Sections \ref{bulkSurf_cylinder} and \ref{SOFC_Diff}. Again, a small nonzero value should be added to the domain parameter appearing in the denominators. In the bulk ($| \nabla \psi |=0$ and $\psi = 1$), Eq.~\eqref{SBM-FSL-BS1} reduces to Eq.~\eqref{FSL-B1}. When the interfacial thickness approaches zero, it reduces to Eq.~\eqref{FSL-S1} at the interface ($|\nabla\psi|\neq0$), as proven in Appendix \ref{SBM_proof}.

The surface Laplacian ($\nabla_\text{s}^2 =   \nabla_\text{s} \cdot  \nabla_\text{s}$) is calculated according to the surface gradient given by:
\begin{equation}
\nabla_\text{s} = ( \mathbf{I} - \mathbf{n} \otimes \mathbf{n} ) \nabla = \bigg( \mathbf{I} - \frac{\nabla \psi}{| \nabla \psi |} \otimes \frac{\nabla \psi}{| \nabla \psi |} \bigg) \nabla,
\end{equation}
where $\mathbf{I}$ is the unity tensor, `$\otimes$' is the Dyadic product, and $\mathbf{n}$ is the inward unit normal vector of the diffuse interface, as used in Ref. \cite{Teigen:2009a}. (In indicial notation, the surface gradient is expressed as: $(\delta_{ij}-n_i n_j) \partial /\partial x_j$, where $\delta_{ij}$ is the Kronecker delta ($\delta_{ij} = 1$ if $i = j$, and $\delta_{ij} = 0$ if $i \ne j$). The repeated indices indicate summation over the index. See Appendix \ref{SurLap_A2} for details.) To simulate only surface diffusion on a diffuse-interface-described geometry, one can simply eliminate the bulk-related and reaction terms in Eq.~\eqref{FSL-S1} to obtain $ L(\partial C/\partial t)= D_\text{s} \nabla_\text{s}^2 C$, such that the concentration evolves only over the interfacial region. 

\subsection{Mechanical Equilibrium} \label{MechEquim-Derivation} 

The smoothed boundary method can also be applied to the mechanical equilibrium equation. When a solid body is in mechanical equilibrium, all forces acting on the body are balanced in all directions, as represented by $\partial \sigma_{ij}/\partial x_j = 0$, where $\sigma_{ij}$ is a stress tensor component, which is the force per unit area along $j$th axis on the surface whose normal vector is along the $i$th axis. Repeated indices indicate summation over the index. For linear elasticity, the stress tensor is given by the generalized form of Hooke's Law: $\sigma_{ij} = C_{ijkl} (\varepsilon_{kl}- \rho \delta_{kl})$, where $C_{ijkl}$ is the elastic constant tensor and $\rho$ is a scalar body force, such as thermal expansion ($\alpha \Delta T$) or misfit eigenstrain ($ (a_p-a_m)/a_m$, where $a_p$ and $a_m$ are the lattice constants of the precipitate and matrix phases, respectively), depending on the governing physics. The total strain tensor is defined by the gradients of displacements as $ \varepsilon_{ij} = [(\partial u_i/\partial x_j)+(\partial u_j/\partial x_i)]/2$, where $u_i$ is the infinitesimal displacement in the $i$th direction. Substituting Hooke's Law and the total strain back into the mechanical equilibrium equation gives:
\begin{equation} \label{ME-2}
\frac{\partial }{\partial x_j} C_{ijkl}  \frac{1}{2} \left(\frac{\partial u_k}{\partial x_l}+\frac{\partial u_l}{\partial x_k} \right) = \frac{\partial }{\partial x_j} \bigg( \rho C_{ijkl}\delta_{kl} \bigg).
\end{equation}
Multiplying Eq.~\eqref{ME-2} by the domain parameter that distinguishes the elastic solid region ($\psi = 1$) from the environment ($\psi=0$) and using the product rule of differentiation yields the smoothed boundary formulation:
\begin{equation} \label{SBM-ME-1}
\begin{split}
\frac{\partial}{\partial x_j} \left[ \psi C_{ijkl}  \frac{1}{2} \left(\frac{\partial u_k}{\partial x_l}+\frac{\partial u_l}{\partial x_k} \right) \right]   & -  \\ \left(\frac{\partial \psi}{\partial x_j}\right) \bigg\{ C_{ijkl}  
\bigg[ \frac{1}{2} & \bigg(\frac{\partial u_k}{\partial x_l} +\frac{\partial u_l}{\partial x_k} \bigg)-\rho\delta_{kl} \bigg] \bigg\}   = \frac{\partial}{\partial x_j} \bigg( \psi \rho C_{ijkl}\delta_{kl} \bigg);
\end{split}
\end{equation}
see Appendix \ref{A3} for details of the derivation.

The traction exerted on the solid surface is defined by $N_{i} = -\sigma_{ij}n_{j}$, where $n_j = \nabla \psi /|\nabla \psi |$ is the inward unit normal of the solid surface. (In indicial notation, $ \partial \psi/\partial x_i = \nabla \psi$ and $\sqrt{(\partial \psi/\partial x_i)(\partial \psi/\partial x_i)} = |\nabla \psi|$.) Therefore, the surface traction force is given by:
\begin{equation} \label{Trac-1}
N_{i} = -\bigg\{ C_{ijkl} \left[ \frac{1}{2}\left(\frac{\partial u_k}{\partial x_l}+\frac{\partial u_l}{\partial x_k} \right)-\rho\delta_{kl} \right] \bigg\} \left(\frac{\nabla \psi}{|\nabla \psi|} \right).
\end{equation}
Substituting Eq.~\eqref{Trac-1} into Eq.~\eqref{SBM-ME-1} yields the smoothed boundary formulated mechanical equilibrium equation with a traction boundary condition on the solid surface:
\begin{equation} \label{SBM-ME-2}
\frac{\partial}{\partial x_j} \left[ \psi C_{ijkl} \frac{1}{2} \left(\frac{\partial u_k}{\partial x_l}+\frac{\partial u_l}{\partial x_k} \right) \right] + |\nabla \psi| N_{i} = \frac{\partial}{\partial x_j} \bigg( \psi \rho C_{ijkl}\delta_{kl} \bigg),
\end{equation}
where $\partial(\psi \rho C_{ijkl}\delta_{kl})/\partial x_j = \tilde{\rho}_i$ can be treated as an effective body force in the $i$th direction.

For linear elasticity problems with prescribed displacements at the solid surface, one can perform the smoothed boundary formulation, as in the derivation of the Dirichlet boundary condition in Section \ref{DiffEqn}, by multiplying Eq.~\eqref{ME-2} by $\psi^2$ and using the product rule to obtain:
\begin{equation} \label{SBM-ME-3}
\begin{split}
\psi\frac{\partial}{\partial x_j} \left[ \psi C_{ijkl} \frac{1}{2} \left(\frac{\partial u_k}{\partial x_l}+\frac{\partial u_l}{\partial x_k} \right) \right] & - \bigg\{ \bigg(\frac{\partial \psi}{\partial x_j}\bigg) \bigg[ C_{ijkl} \frac{1}{2}\bigg( \frac{\partial \psi u_k}{\partial x_l}+\frac{\partial \psi u_l}{\partial x_k} \bigg) \bigg] \\ - \bigg(\frac{\partial \psi}{\partial x_j} \bigg) C_{ijkl} \frac{1}{2} \bigg( u_k \frac{\partial \psi}{\partial x_l} & + u_l \frac{\partial \psi}{\partial x_k}\bigg) \bigg\} = \psi^2 \frac{\partial}{\partial x_j}\bigg(\rho C_{ijkl}\delta_{kl}\bigg),
\end{split}
\end{equation}
where the displacements $u_k$ and $u_l$ appearing in the third term on the left-hand side should be the boundary values of the displacements at the solid surface; see Appendix \ref{A3} for the derivation. A similar formulation for solving the mechanical equilibrium equation within a domain defined by a phase-field-like order parameter can also be obtained by the asymptotic approach by matching terms of different orders \cite{Voigt:2009}.

\subsection{Equations for Phase Transformations with Additional Boundaries} \label{ContactAngleFormulation} 

Phase transformations affected by a mobile or immobile surface or other boundaries are of importance in many materials processes, including heterogeneous nucleation that occurs at material interfaces \cite{Granasy:2007,Warren:2009}. Maintaining a proper contact angle at the three-phase boundary (where the interface between the two phases meets the surface) is necessary to capture the dynamics accurately because the contact angle represents the difference between the surface energies (tensions) of the different phase boundaries. Although researchers have previously developed methods for imposing contact-angle boundary conditions on sharp domain walls \cite{Granasy:2007,Warren:2009}, here we show that a similar model with diffuse domain walls can be obtained simply by applying the approach described above. Below, we assume that the boundary is immobile, but this assumption can be easily removed by describing the evolution of the domain parameter as dictated by the physics of the system.

In the Allen-Cahn and Cahn-Hilliard equations of the phase field model, the total free energy has the following form \cite{Cahn:1958,Cahn:1959}:
\begin{equation} \label{eqTeng}
F = \int_\Omega \bigg [  f(\phi) +\frac{\epsilon^{2}}{2} | \nabla \phi |^{2} \bigg]  d\Omega ,
\end{equation}
where $\phi$ is the phase field order parameter commonly used to define different phases, $f(\phi)$ is a double-well free energy functional (in terms of $\phi$), $\epsilon$ is the gradient energy coefficient, and $\Omega$ is the domain of interest. At the extremum of the functional $F$, the variational derivative of the total free energy vanishes: $\delta F/\delta \phi = 0$. This requirement provides the following conditions: $\partial f/\partial \phi - \epsilon^{2} \nabla^{2} \phi = 0 \in \Omega$, which can be reformulated as $\nabla f = \nabla ( \epsilon^2 | \nabla \phi |^{2})/2$, by multiplying both sides by $\nabla \phi$. We thus find a useful equality for deriving the contact angle boundary condition: $| \nabla \phi | = \sqrt{2 f}/\epsilon$; see Appendix \ref{A4} for details.

In the smoothed boundary method, we introduce a domain parameter $\psi$ to incorporate boundary conditions into the original governing equation. As mentioned previously, the level sets of this domain parameter $\psi$ describe the diffuse boundaries and should satisfy $\vec{n} = \nabla \psi/| \nabla \psi |$. On $\partial \Omega$, we impose a contact angle, $\theta$, such that $\vec n \cdot (\nabla \phi/ | \nabla \phi |) =  -\cos \theta$, where $\nabla \phi/ | \nabla \phi |$ is the unit normal vector of the phase boundary (pointing to regions where $\phi=1$). We can thus derive the following equation for the boundary condition:
\begin{equation}  \label{eqAngBCF}
\nabla \psi \cdot \nabla \phi =  - \left | \nabla \psi \right \vert \cos \theta \frac{\sqrt{2f}}{\epsilon}.
\end{equation}
This contact-angle boundary condition is similar to that suggested by Warren et al.~\cite{Warren:2009} for contacting a sharp interface, in which a Dirac delta function replaces $|\nabla \psi |$.

The chemical potential that drives the morphological evolution is defined by the variational derivative of the total free energy of the system: $\mu = \delta F/ \delta \phi = \partial f / \partial \phi - \epsilon^2 \nabla^2 \phi$. We can apply the smoothed boundary formulation to the chemical potential by multiplying it by the domain parameter $\psi$ and applying the product rule to obtain:
\begin{equation} \label{SBM-mu-1}
\mu  = \frac{\partial f}{\partial \phi} - \frac{\epsilon^2}{\psi} \nabla \cdot ( \psi \nabla \phi ) + \frac{\epsilon^2}{\psi} \nabla \psi \cdot \nabla \phi = \frac{\partial f}{\partial \phi} - \frac{\epsilon^2}{\psi} \nabla \cdot (\psi \nabla \phi) - \frac{\epsilon |\nabla \psi |}{\psi} \sqrt{2f} \cos{\theta},
\end{equation}
where Eq.~\eqref{eqAngBCF} was used in the third term.  

For a nonconserved order parameter in the phase field models, the evolution is governed by the Allen-Cahn equation \cite{Allen:1979} (also known as the time-dependent Ginzburg-Landau equation \cite{Ginzburg:1950}), in which the order parameter evolves according to the local chemical potential variation:
\begin{equation} \label{AC-1}
\frac{\partial \phi}{\partial t} = - M \mu = -M \bigg( \frac{\partial f}{\partial \phi} - \frac{\epsilon^2}{\psi} \nabla \cdot (\psi \nabla \phi) - \frac{\epsilon |\nabla \psi |}{\psi} \sqrt{2f} \cos{\theta} \bigg),
\end{equation}
where $M$ is the mobility coefficient.

For a conserved order parameter, the evolution of the order parameter is governed by the Cahn-Hilliard equation, where the rate of the order parameter change is equal to the divergence of the its flux, which is proportional to the gradient of the chemical potential \cite{Cahn:1958,Cahn:1959}, $\partial \phi/\partial t = \nabla \cdot (M \nabla \mu)$. The smoothed boundary formulation (derived in a similar manner as in the derivation in Section \ref{DiffEqn}) is given by $\psi (\partial \phi / \partial t) = \nabla \cdot (\psi M \nabla \mu) - \nabla \psi \cdot (M \nabla \mu)$.
Note that $-M \nabla \mu = \vec{j}$ is the flux of the conserved order parameter; thus, the second term represents the fluxes normal to the domain boundary (equivalent to Eq.~\eqref{NBC1}). The smoothed boundary formulation of the Cahn-Hilliard equation is thus written as: 
\begin{equation} \label{SBM-CH-3}
\frac{\partial \phi}{\partial t} = \frac{1}{\psi} \nabla \cdot \bigg[ \psi M \nabla \bigg( \frac{\partial f}{\partial \phi} - \frac{\epsilon^2}{\psi} \nabla \cdot (\psi \nabla \phi) - \frac{\epsilon| \nabla \psi |}{\psi} \sqrt{2f} \cos{\theta} \bigg) \bigg] + \frac{ |\nabla \psi |}{\psi} J_n,
\end{equation}
where $J_n = \vec{n}\cdot \vec{j}$. For a closed system, $J_n$ is zero. Note that a small nonzero value should to be added to the domain parameter in the denominators in Eqs.~\eqref{AC-1} and \eqref{SBM-CH-3} to avoid division by zero.

\section{Validation of the Presented Approach} \label{Validation} 

We herein demonstrate the validity and accuracy of the approach introduced in Sections \ref{DiffEqn} and \ref{SurfDiffFormulation} and the phase transformation with the presence of additional surface in Section \ref{ContactAngleFormulation}.

\subsection{1D Diffusion Equation} \label{1D_Diffusion_Test} 

First, we performed a 1D simulation to demonstrate that the Neumann and Dirichlet boundary conditions were satisfied on two different sides of the domain. Fick's second diffusion equation with given source and sink terms was solved within the domain defined by $\psi = 1$. The diffusion coefficient was set at 1, and the source and sink strengths were 0.02 and 0.01, respectively. On the right boundary of the diffusion domain, the gradient of $C$ was set at -0.1, and on the left boundary, the value of $C$ was set at 0.4. We selected the 1D computational box for $0 < x < 30$ and used a hyperbolic tangent function for the continuous domain parameter $\psi$:
\begin{equation}
\psi = \frac{1}{2}\bigg[\tanh{\bigg(\frac{x-5}{\zeta}\bigg)}-\tanh{\bigg(\frac{x-25}{\zeta}\bigg)}\bigg],
\end{equation}
where $\zeta$ is the coefficient for adjusting the interfacial thickness. The interfacial thickness is given approximately by $\xi_0 = 4.185\zeta$ where the interfacial region is defined by the range, $0.015< \psi < 0.985$. The left and right interfaces are located at $x=5$ and $x=25$, respectively. We applied the smoothed boundary formulation, as in the derivation of Eq.~\eqref{Mixed-SBM-BC-01}, to reformulate the original diffusion equation, $\partial C /\partial t = \partial^2 C/ \partial x^2 - C/0.01+0.02$, to:
\begin{equation} \label{SBM-Diff-A}
\begin{split}
\frac{\partial C}{\partial t} = \frac{1}{\psi}\frac{\partial}{\partial x} \bigg( \psi \frac{\partial C}{\partial x}\bigg) - \frac{0.1}{\psi} \bigg|\frac{\partial \psi}{\partial x} \bigg| H_V(l_m)  & - \\ \frac{1}{\psi^2} \bigg[ \frac{\partial \psi}{\partial x} \frac{\partial (\psi C)}{\partial x} - 0.4 \bigg|\frac{\partial \psi}{\partial x} \bigg|^2 \bigg] [1& - H_V(l_m)]  - \frac{C}{0.01} +0.02,
\end{split}
\end{equation}
where $H_{V}(l_m)$ is the Heaviside function used to specify the choice of the boundary condition and $l_m=15$ is the midpoint of the diffusion domain. Therefore, the second and third terms only apply to the right and left interfaces, respectively. The initial concentration was $C=0$ everywhere in the computational box. A standard central finite difference scheme in space and an Euler explicit time scheme were employed in the simulations.

\begin{figure}[htb] 
\begin{center}
\includegraphics[width=1\textwidth]{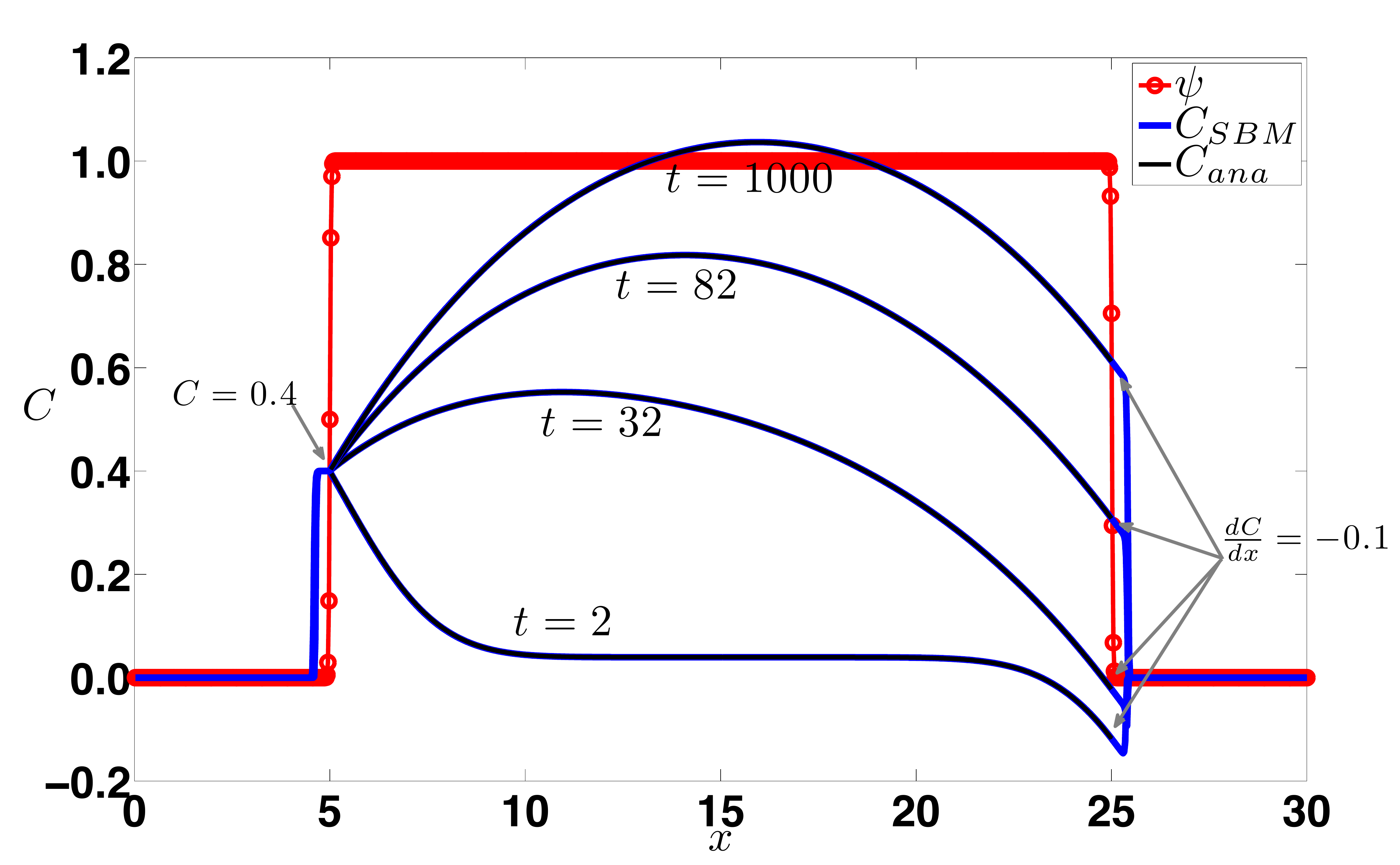}
\end{center}
\caption{Demonstration of the smoothed boundary method on the 1D diffusion equation. The red line with circular markers is the domain parameter, and the blue lines are the concentration profiles at different times. The Neumann and Dirichlet boundary conditions are imposed at the right and left boundaries, respectively. The black lines are the corresponding analytical solutions of the sharp interface version of the diffusion equation.}
\label{SBM_1D_BC_1}
\end{figure}

Figure \ref{SBM_1D_BC_1} shows the concentration profiles recorded at four different times (in solid blue lines). The domain parameter is plotted in the red line (the red circular markers indicate the position of grid points). The computational box was discretized to 1,200 grid points ($\Delta x = 2.5\times10^{-2}$), and $\zeta$ was taken to be $2.86\times10^{-2}$, such that the interfacial thickness is approximately $\xi_0 = 0.1197 = 4.79 \Delta x$. The parameters are given as Case 1b in Table \ref{Tbl-1}. On the right interface, it can be clearly observed that $dC/dx = -0.1$ at all times (except for a rapid change from $dC/dx=0$ to $dC/dx = -0.1$ in the very early transient period). In the early period, the concentration even took negative values to satisfy the gradient boundary condition imposed at the right interface. In contrast, the concentration remained at 0.4 at the left interface during the entire diffusion process (except in the very early transient period, during which $C$ changed from 0 to 0.4). The analytical solution for the original sharp interface equation is also plotted for comparison, showing excellent agreement between the two methods. This result clearly demonstrates that both Neumann and Dirichlet boundary conditions are satisfied on the diffuse interfaces, and the smoothed boundary formulated equation reproduces the same result to the corresponding sharp interface version.

\begin{figure}[htb] 
\begin{center}
\includegraphics[width=1\textwidth]{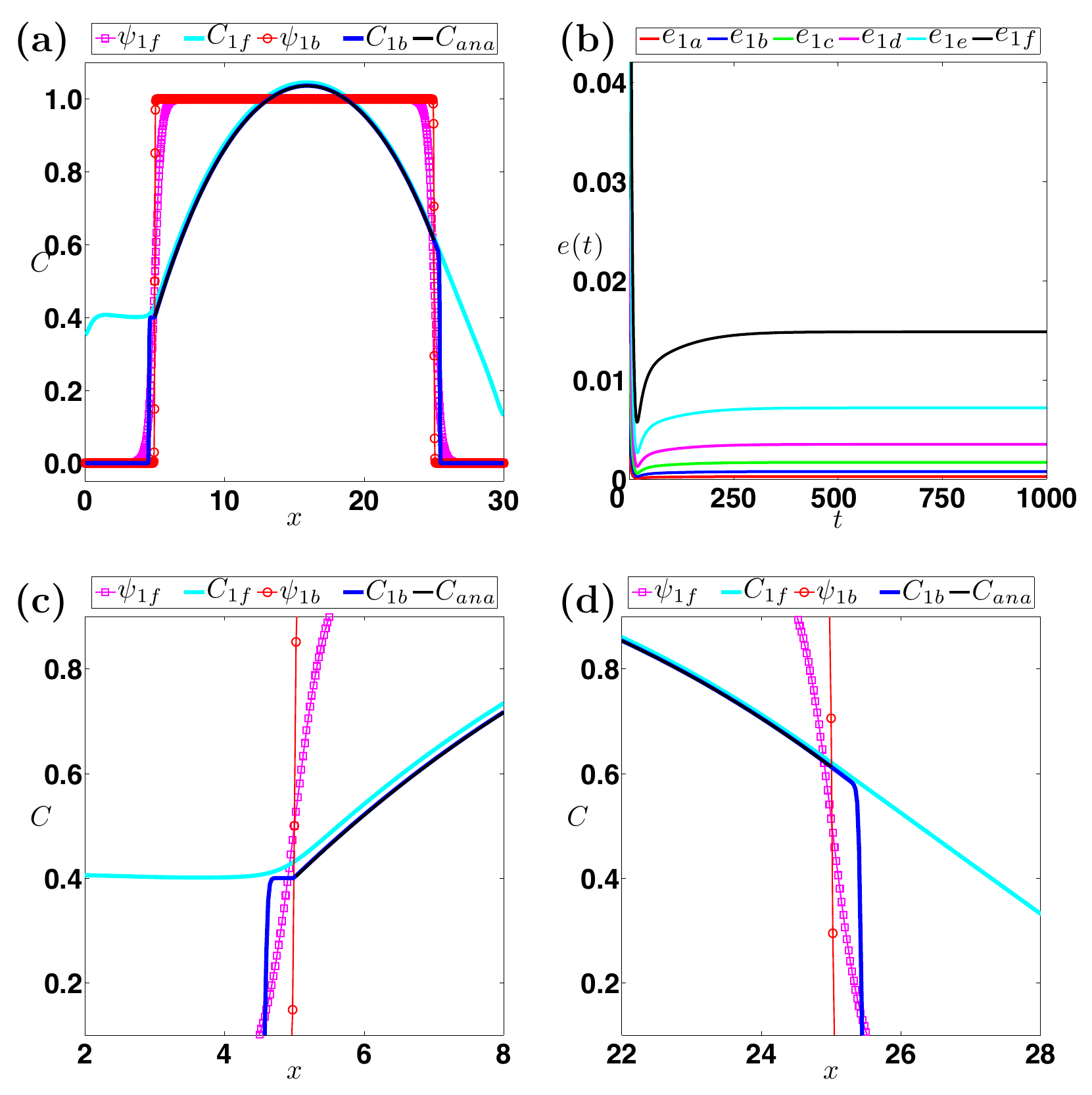}
\end{center}
\caption{(a) Equilibrium concentrations in domains of two different interfacial thicknesses, corresponding to Case 1f ($\zeta = 4.58\times10^{-1}$) and Case 1b ($\zeta=2.86\times10^{-2}$) in Table \ref{Tbl-1}. The black line is the analytical solution. (b) Relative errors for the smoothed boundary solutions during diffusion for different $\zeta$ values. Magnified views of (a) at the (c) left, and (d) right interfaces.}
\label{SBM_1D_BC_C1}
\end{figure}

To further analyze the effects of interfacial thickness and discretization resolution on the smoothed boundary method, various simulations were conducted. In the first case, various interfacial thicknesses were selected, as in Case 1 in Table \ref{Tbl-1}, while the grid size was kept at $\Delta x = 2.5\times10^{-2}$. Figure \ref{SBM_1D_BC_C1}(a) shows the concentration distributions at $t = 1,000$ (nearly equilibrium) for $\zeta=2.86\times10^{-2}$ (Case 1b) and $\zeta=4.58\times10^{-1}$ (Case 1f), for which the interfaces approximately span $4.79 \Delta x$ and $76.7 \Delta x$, respectively. It is clear that the calculated concentration deviates farther from the analytical solution when the interfacial thickness is greater (as shown in the derivation in Appendix \ref{SBM_proof}). Figure \ref{SBM_1D_BC_C1}(b) illustrates the relative errors during concentration evolution for various values of $\zeta$. Here, the relative error is defined by the root-mean-square deviation (between the smoothed boundary result and the analytical solution) divided by the average analytical concentration. The results clearly show that the error increases as the interfacial thickness increases. As the concentration evolution approaches equilibrium, the errors also converge to their equilibrium values, as listed in Table \ref{Tbl-1}. A scaling of the error to the interfacial thickness is observed as $\zeta$ is varied from $2.86\times10^{-2}$ to $4.58\times10^{-1}$. In addition, the deviation between the smoothed boundary results and the analytical solution is much larger near the left boundary than near the right boundary, indicating the error associated with a Dirichlet boundary condition is larger than that with a Neumann boundary condition; see Figs.~\ref{SBM_1D_BC_C1}(c)--(d).

\begin{figure}[htb] 
\begin{center}
\includegraphics[width=1\textwidth]{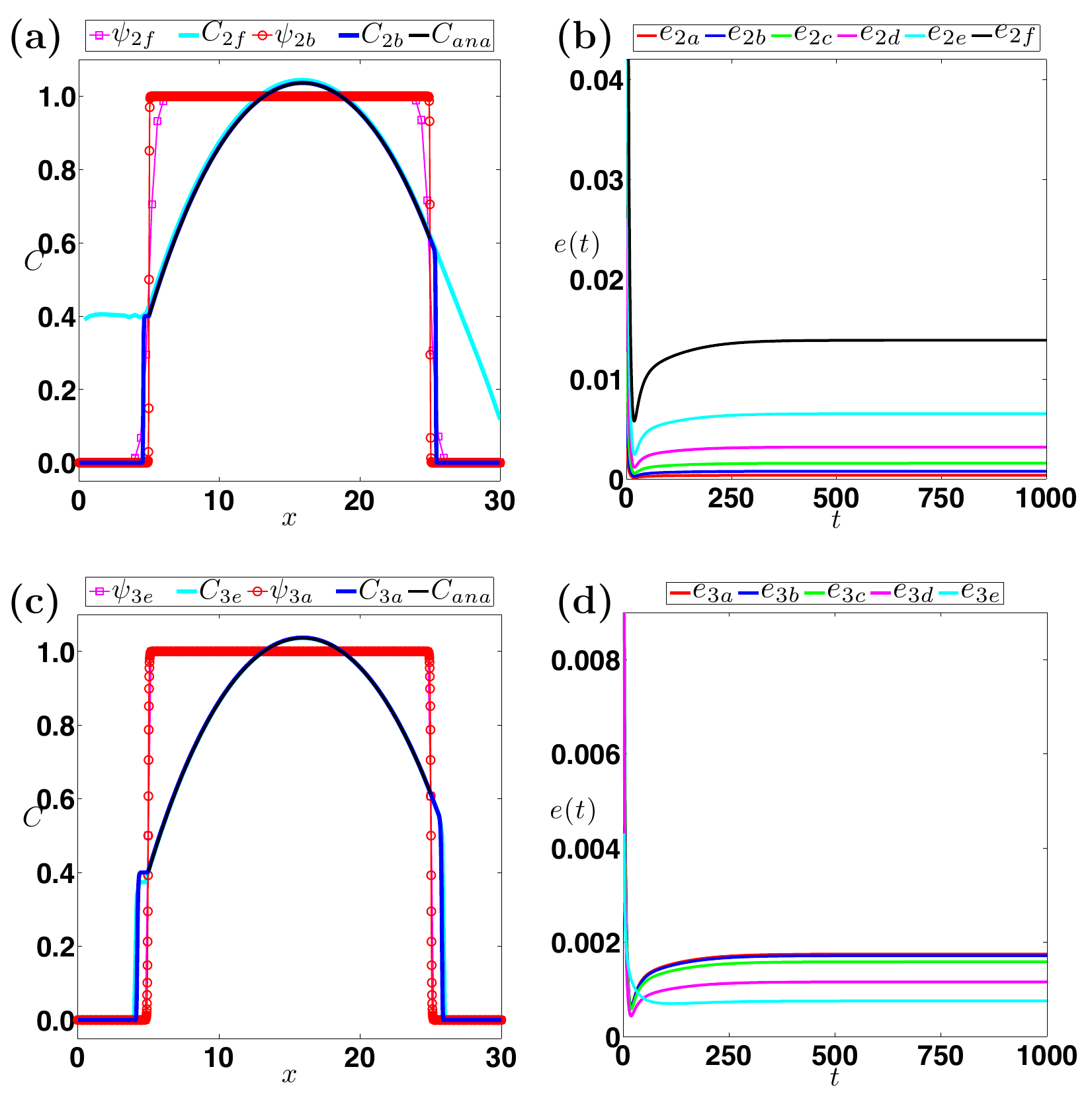}
\end{center}
\caption{(a) Equilibrium concentrations for Cases 2f and 2b in Table \ref{Tbl-1}. The black line is the analytical solution. (b) Relative errors for the smoothed boundary solutions during diffusion for the parameters in Case 2 in Table \ref{Tbl-1}. (c) Equilibrium concentrations for Cases 3e and 3a. (d) Relative errors for Case 3 in Table \ref{Tbl-1}.}
\label{SBM_1D_BC_C23}
\end{figure}

In the second case, we examined the effect of varying $\Delta x$ without changing the number of grid points across the interface. This was accomplished by selecting various grid sizes while maintaining the ratio of interfacial thickness to the grid size at $4.79$. Results similar to Case 1 were obtained (see Case 2 in Table \ref{Tbl-1} and Figs.~\ref{SBM_1D_BC_C23}(a)--(b)). Error increases with $\Delta x$.  Since the resolution of the interface is unchanged (i.e., the number of points across the interface is fixed), it implies that the increased interfacial thickness is the dominant source of error. However, the errors are in general smaller than those in Case 1, which may be due to the fact that the interfacial thickness is effectively reduced when the resolution is decreased (the parts of the interfacial regions where $\psi$ is near 0 and 1 are not resolved by large grid spacing). The same reason may explain the steep drop in the error for $\zeta = 1.43\times10^{-2}$ in Case 1a in Table \ref{Tbl-1}, for which the rapid transition of $\psi$ is not properly resolved by the discretization. 

In the third case, we selected various grid sizes to examine the effect of the resolution across the interface while maintaining the interfacial thickness (specifically, fixing $\zeta$ at $5.73\times10^{-2}$). The results show that error decreases when a larger grid size is selected; see Case 3 in Table \ref{Tbl-1} and Figs.~\ref{SBM_1D_BC_C23}(c)--(d).  This can be understood as follows.  As we observed earlier, the smoothed boundary formulation reduces to the bulk partial differential equation far from the interface, where the gradient of the domain parameter vanishes.  In the interfacial region, the bulk term and the boundary term together set the boundary condition, as shown in Appendix \ref{SBM_proof}.  In between, there is a region where the bulk equation is affected by the boundary term, which is small because the gradient is small, but not negligible.  When the resolution is sufficiently low, the domain parameter in these regions take the bulk values and vanishing the boundary term and thus increasing the accuracy.  As can be observed in Fig.~\ref{SBM_1D_BC_C23}(d), the error behavior in such case is very different from other cases.  In the specific example presented here, the discretized interface at the low resolution of $\Delta x = 0.2$ is nearly a Heaviside step function, which yields smaller error than the high-resolution cases.  (When the resolution is high enough, the error is not affected by the resolution.)  Therefore, in the 1D case, an interface does not need to be fully resolved, and in fact the accuracy can be increased by not doing so.  However, we found that numerical instability ensues when the resolution is further reduced.  It has been determined that at least one point with an intermediate value between the two bulk values is required in order to achieve numerical stability.

Above argument applies to only 1D case or when interfaces are very flat in multiple dimensions. In one dimension, the curvature of the interface is zero (i.e., the interface is flat), and therefore the effect of curvature can be neglected and a good resolution across the interface (which provides the smoothness of the curved interface) is not required. This is not the case when sufficiently large curvature is present, and thus smoothed interface with about three grid points are required to obtain accurate results for 2D or 3D calculations. 

In the fourth case, we varied the value of $\upsilon$ (the small value added to the denominators to avoid division by zero) from $1\times10^{-2}$ to $1\times10^{-11}$, while the grid size and interfacial thickness were maintained at $\Delta x = 2.5\times10^{-2}$ and $\xi_0 = 4.79 \Delta x$, respectively. In practice, a smaller $\upsilon$ would lead to a less stable numerical implementation because the values of $1/\psi$ or $1/\psi^2$ become much larger, which requires a much smaller
time step size. The results show that the error quickly converges to a small value when $\upsilon$ is smaller than $1\times10^{-5}$; see Case 4 in Table \ref{Tbl-1}. This suggests that once $\upsilon$ is small enough to yield converged results, further reduction is unnecessary and should be avoided so that a larger time step can be employed. 

In summarizing the above 1D test simulations, we found that the interfacial thickness is the dominant source of error. The errors are less sensitive to the resolution of the finite-differencing discretization (selection of $\Delta x$) and the parameter for singularity control (selection of $\upsilon$). When the diffuse interface is properly resolved, the error scales with the interfacial thickness. Moreover, in general, the error that results when a Dirichlet boundary condition is imposed is larger and more sensitive to the interfacial thickness than when a Neumann boundary condition is imposed.  This behavior can be understood from the results of analysis in Appendix \ref{SBM_proof}, where the scaling of the errors can be found in Eqs. \eqref{SBM-Nm-prf-03} and \eqref{Int-Dirich-03}.

\subsection{Surface Diffusion and Bulk Diffusion in a Cylinder} \label{bulkSurf_cylinder} 

To further demonstrate the validity of the smoothed boundary method, we applied the method to simulate oxygen-vacancy diffusion in a cylinder, for which a cylindrical coordinate grid system was used. We solved the coupled surface-bulk diffusion problem using both the smoothed boundary and the original sharp interface formulations in the same grid system for comparison. For the smoothed boundary method, we used a hyperbolic tangent function, $\psi(r,z) = \{ 1- \tanh{[(R-r)]/\zeta} \}/2$, of the continuous domain parameter to define a cylinder, where $r$ is the radial position, $z$ is the axial position, and $R$ is the cylinder radius. Therefore, the cylinder surface ($\psi=0.5$) where surface reaction and surface diffusion occur, is located at $r=R$, the solid region ($\psi = 1$) for bulk diffusion is defined at $r<R$, and the environment ($\psi=0$) is defined at $r>R$. We selected the cylinder radius $R$ and the cylinder axial length to be 1 and 12, respectively. The grid sizes were selected to be $\Delta r = 1.76\times10^{-2}$ and $\Delta z = 4\times10^{-2}$, such that the cylinder contains 57 and 300 grid points in the radial and axial directions, respectively. (The computational box is larger than the cylinder in the radial direction, and contains 75 and 300 grid points in the radial and axial directions, respectively.) The interfacial thickness was selected to be $4.26\Delta r$ by setting $\zeta = 1.0182 \Delta r$. Equation \eqref{SBM-FSL-BS1} was solved using a standard central finite difference scheme in cylindrical coordinate system and an Euler explicit time scheme; see Appendix \ref{SurLap_A2} for the discretization scheme. The parameters for the diffusion equation were selected to be $D_\text{b} = 1$, $D_\text{s} = 10$, and $L= 0$. The boundary conditions at the computation box boundary were set at $C=1$ at $z$=0 and $C=0$ at $z=12$, with no gradient on the remaining two sides. For comparison, the original sharp interface equations, Eqs.~\eqref{FSL-S1} and \eqref{FSL-B1}, were solved using the same discretization scheme with the same grid system and resolution. For this case, the surface-reaction-diffusion boundary condition, Eq.~\eqref{FSL-S1}, was explicitly imposed at the 57th grid points in the radial direction. The concentration evolution is implemented as follows. First, the surface concentration is updated by Eq.~\eqref{FSL-S1} according to the normal flux at the cylinder surface calculated from the normal gradient of the surface concentration obtained from Eq.~\eqref{FSL-B1}. Next, the normal surface flux is calculated using Eq.~\eqref{FSL-S1} with the updated surface concentration. The cylinder concentration is then evolved according to Eq.~\eqref{FSL-B1} with the normal flux boundary condition. This procedure is repeated within the Euler explicit time scheme for the concentration evolution.  For the smoothed boundary formulation, we simply solve a single equation that automatically includes coupled bulk and surface diffusion, Eq.~\eqref{SBM-FSL-BS1}.

\begin{figure}[htb]
\begin{center}
\includegraphics[width=1\textwidth]{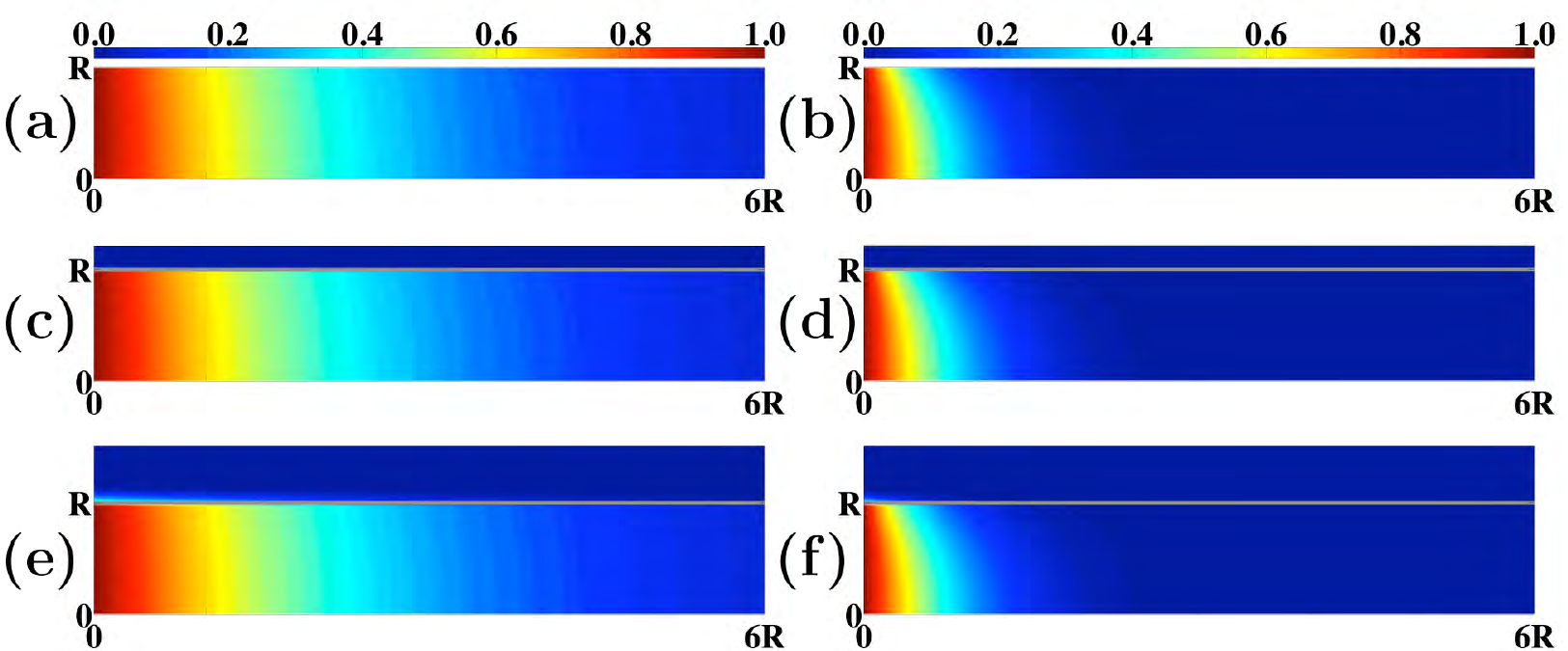}
\caption{Steady-state concentration profiles for $D_\text{b}=1$, $D_\text{s} = 10$.  The left column is for $\kappa=2.1$, and the right column is for $\kappa=50$.
The top row shows the sharp interface solution; the middle row is the thin-interface smoothed boundary method; and the bottom row is the thick-interface smoothed boundary method results with interfacial thickness four times larger than those in the middle row. The top regions of constant blue color in (c)--(f) represent the areas outside of the solid, whereas the solid white lines indicate the solid cylinder surface.}\label{Cyl-Con}
\end{center}
\end{figure}

Figures \ref{Cyl-Con}(a) and (b) show the steady-state concentration profiles of the sharp interface version for $\kappa = 2.1$ and $\kappa = 50$, respectively. The concentration decays along the axial direction according to boundary values prescribed at the box boundaries. The diffusion front bends because of the surface reaction, such that the concentration is lower near the cylinder surface. Shown in Figs.~\ref{Cyl-Con}(c) and (d) are the corresponding smoothed boundary results. For clarity, only the concentration in the region of $0 < z < 6R$ is presented. The results from the two methods are in excellent agreement, clearly demonstrating the utility and validity of the smoothed boundary method for incorporating two sharp-interface equations into one smoothed boundary equation.

To further examine the effect of interfacial thickness, we included two other radial grid sizes in the simulations, i.e., $\Delta r = 3.49\times10^{-2}$ and $\Delta r = 6.86\times10^{-2}$, such that the cylinders contain 29 and 15 radial grid points, respectively. By selecting the radial grid sizes in this way, each radial grid point in a lower-level resolution (thicker interface) overlaps with every other grid point in a higher-level resolution (thiner interface). The diffuse interface is maintained to span $4.26 \Delta r$, and the axial grid size is kept at $\Delta z = 4\times10^{-2}$. Hereafter, we refer the three interfacial thicknesses to as the ``thin'' ($\xi_0 = 0.075$), ``medium'' ($\xi_0 = 0.149$) and ``thick'' ($\xi_0 = 0.292$) interface cases. Shown in Figs.~\ref{Cyl-Con}(e) and (f) are the steady-state concentration profiles for the thick-interface results, corresponding to the cases in Figs.~\ref{Cyl-Con}(a) and (b). The results are still in reasonably good agreement with the original sharp interface results, even though the interfacial thickness is approximately 29.2\% of the cylinder radius.

\begin{figure}[htb]
\begin{center}
\includegraphics[width=1\textwidth]{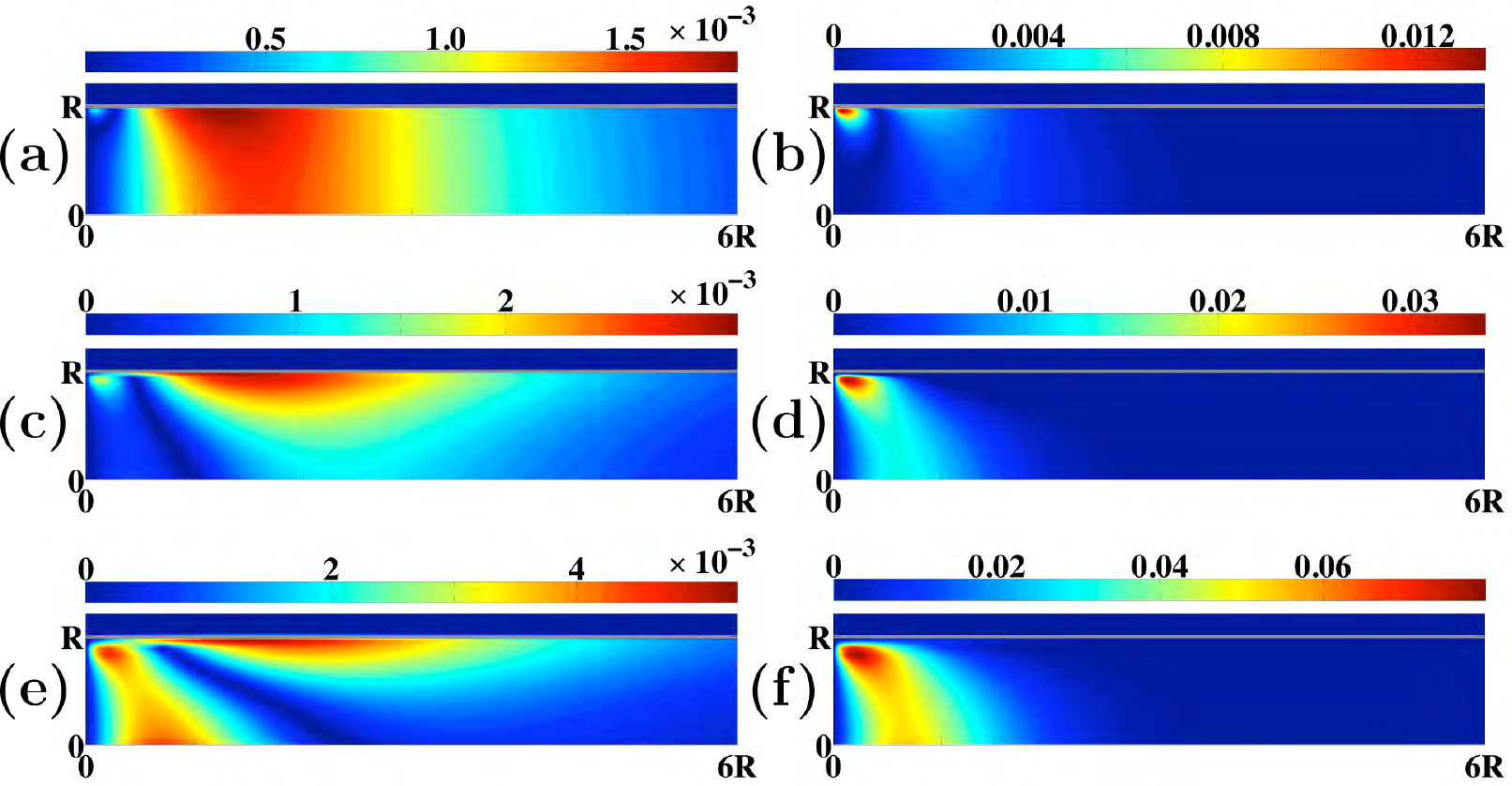}
\caption{Profiles of relative error between the sharp interface results and the smoothed boundary results having thin, medium and thick interfaces from the top to the bottom: (a), (c) and (e) have $D_\text{b}=1$, $D_\text{s} = 10$, and $\kappa=2.1$, and are compared to the sharp interface result shown in Fig.~\ref{Cyl-Con}(a); (b), (d) and (f) have $D_\text{b}=1$, $D_\text{s} = 10$, and $\kappa=50$, and are compared to the sharp interface result in Fig.~\ref{Cyl-Con}(b). }\label{Cyl-Con-Err} 
\end{center}
\end{figure}

The relative errors of the thin, medium, and thick interface smoothed boundary results are plotted in Fig.~\ref{Cyl-Con-Err}. The relative errors are calculated by dividing the differences between the smoothed boundary and sharp interface results by the average concentration of the sharp interface results. The average concentration is calculated for the active region between the plane at $z=0$ and the plane on which the maximum concentration is 0.01. Note that only the errors within in the cylinder defined by $\psi \ge 0.5$ are considered. For the thick-interface smoothed boundary result, the maximum local errors for $\kappa = 2.1$ and $\kappa = 50$ are approximately $5\times10^{-3}$ and $0.075$ (see Figs.~\ref{Cyl-Con-Err}(e) and (f)), whereas the average relative errors are $1.81\times10^{-3}$ and $2.90\times10^{-3}$, respectively; see Table \ref{Tbl-2}. The average relative errors, denoted by $e$ in Table \ref{Tbl-2}, are calculated by dividing the root-mean-square deviation between the smoothed boundary and the sharp interface results by the average sharp interface concentration in the cylinder. The root-mean-square deviation and average concentration are calculated in the cylindrical coordinate system. As expected, the error increases as the interface becomes thicker (i.e., as $\xi_0$ increases). However, in contrast to the 1D simple diffusion test in Section \ref{1D_Diffusion_Test}, the behavior of the error is inconsistent across the parameter sets; see Table \ref{Tbl-2}. This relatively complicated error behavior may originate from the coupling of the bulk and surface diffusion equations. In addition to the effect of the interfacial thickness, the error also increases with a larger reaction coefficient $\kappa$, which may be explained by the increase of the scaling coefficient for the error ($h_0$ defined above Eq.~\eqref{SBM-Nm-prf-02} in Appendix \ref{SBM_proof}) when the given boundary value is larger. In addition, the gradient of the concentration near the boundary increases in magnitude with increasing boundary condition value, which can lead to a larger error.  

One interesting phenomenon is observed in the high $\kappa$ results ($\kappa = 50$ and $100$). Although the errors in the bulk greatly increase with the interfacial thickness, the errors at the surface remain small; see Fig.~\ref{Cyl-Con-Err}(f) and Table \ref{Tbl-2}.  This indicates that the error originates from the boundary condition affecting the bulk solution, rather than from an increased error in the boundary condition value. The thicker interface thus leads to a larger bulk region that is affected by the boundary condition. Therefore, we compare the error associated with the bulk region and with the boundary condition. Here, the bulk errors, denoted by $e_\text{b}$ in Table \ref{Tbl-2}, are calculated by the same method as the average relative error but exclude the grid points on the nominal cylinder surface ($r = R$). The surface errors, denoted by $e_\text{s}$ in Table \ref{Tbl-2}, are calculated by the same method but with only the grid points at the nominal cylinder surface. In the case where $\kappa = 100$, the error at the surface even decreases with interfacial thickness. 

\subsection{Contact-Angle Boundary Condition} \label{ContAngBC-valid-1} 

We performed simple 2D simulations to validate the smoothed boundary formulation for the contact-angle boundary condition at the three-phase boundary. Equations \eqref{AC-1} and \eqref{SBM-CH-3} were tested for nonconserved and conserved order parameters, respectively. The equations were solved using the central finite difference scheme and the Euler explicit time scheme. The computational box sizes are $L_x= 100$ and $L_y=100$, and the parameters used are $\Delta x= 1$ and $M= 1$. A simple common double-well function was selected for the bulk free energy functional, $f(\phi) = w \phi^2 (1-\phi)^2$, such that the steady-state phase field order parameter profile is determined by $\phi = \{ 1-\tanh{[(\sqrt{w} \mathbf{x})/(\sqrt{2} \epsilon)}] \} /2$, where $\mathbf{x}$ is the coordinate variable indicating the distance to the phase boundary, and the characteristic thickness of the diffuse interface is determined by $\delta_{\phi} = \epsilon \sqrt{2/w}$. By setting $\epsilon = \sqrt{1/w}$, the characteristic thickness is controlled by $\delta_{\phi} = \sqrt{2}/w$, and the phase field interfacial energy is maintained at a constant value, $\gamma_{\phi} = \epsilon \sqrt{2w} /6 = \sqrt{2}/6$. A horizontal diffuse-interface flat substrate surface is defined by the hyperbolic tangent function $\psi = \{ 1 + \tanh{[(y-30)/\zeta]} \}$, such that $\psi=0.5$ is at $y=30$, and $\psi$ gradually transitions from 0 to 1 from below to above the substrate surface. Here, we have two diffuse interfaces: one for the phase field order parameter and the other for the smoothed boundary domain parameter. Both thicknesses can affect the accuracy when imposing contact-angle boundary conditions. To verify the contact angle boundary conditions, various combinations of substrate surface thickness and phase boundary thickness were selected by adjusting the values of $\zeta$ and $w$. The initial phase boundary was placed vertically in the middle of the domain ($x=50$), with Phase 1 ($\phi=1$) and Phase 0 ($\phi=0$) on the left and right halves, respectively. On the computational box boundaries, the normal gradients of the phase field order parameter were set at zero: $\partial \phi/\partial x = 0$ at $x = 0$ and $100$, and $\partial \phi/\partial y =0$ at $y=0$ and $100$, which can be interpreted as the no-flux boundary conditions.

\begin{figure}[htb]
\begin{center}
\includegraphics[width=1\textwidth]{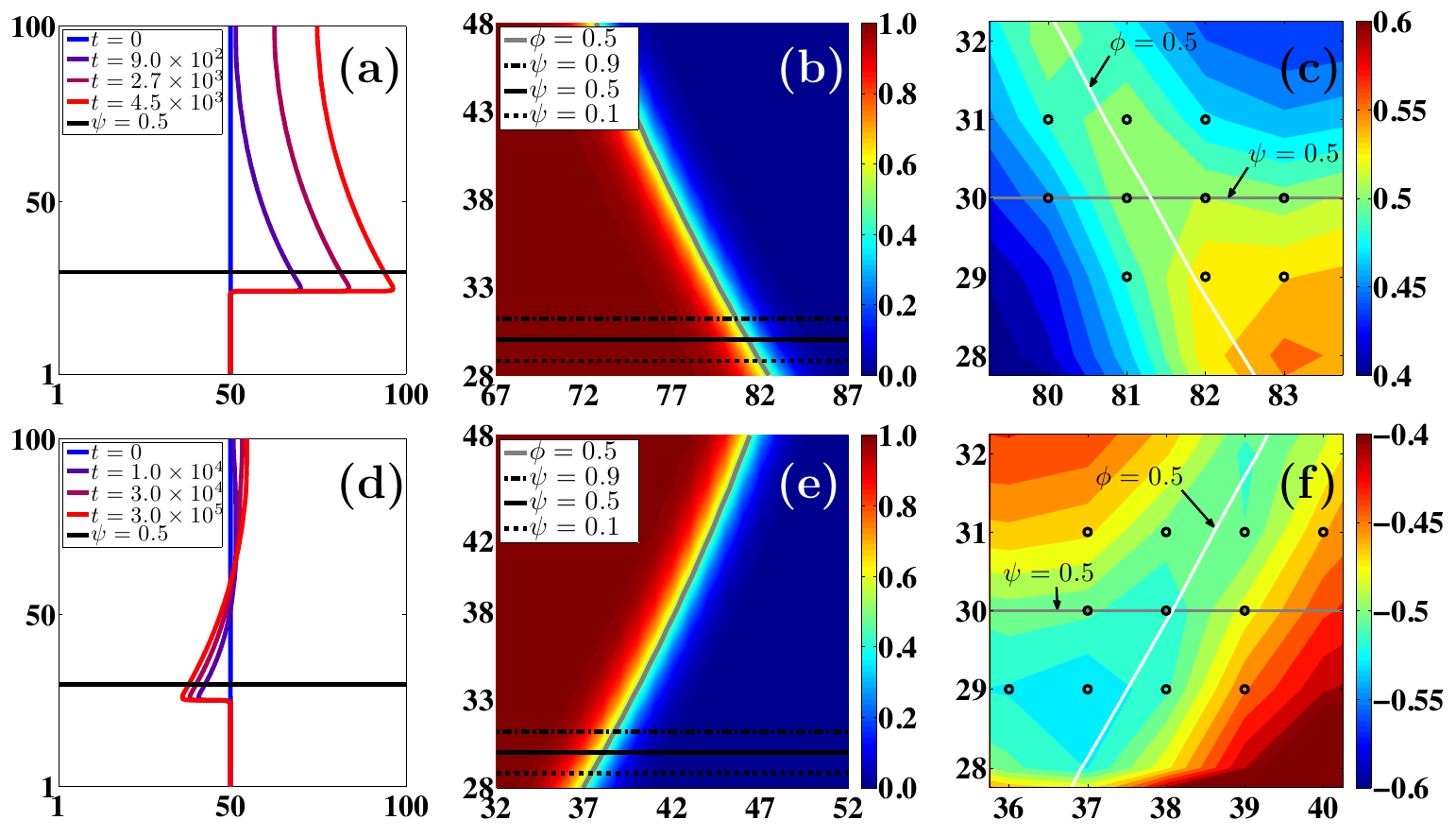}
\caption{(a) Contour lines of $\psi = 0.5$ at various times for Allen-Cahn-type evolution equation with a 60$^\circ$ contact-angle boundary condition. (b) Magnified view of the order parameter and (c) $\cos \theta$ profiles near the three-phase boundary (further magnified), corresponding to $t= 2.7\times10^{3}$ in (a). (d) Contour lines of $\psi = 0.5$ at various times for Cahn-Hilliard-type evolution equation with a 120$^\circ$ contact-angle boundary condition. (e) Magnified view of the order parameter and (e) $\cos \theta$ profiles at the three-phase boundary (further magnified), corresponding to $t= 3.0\times10^{5}$ in (d). The cosine values of the imposed contact angles are 0.5 and -0.5 for (a)-(c) and (d)-(f), respectively. The circular markers in (c) and (f) denote the grid points in the range of $0.1 < \psi < 0.9$ and $0.1< \phi < 0.9$. The order parameters in the region of $\psi < 0.5$ have no physical significance. For the Cahn-Hilliard case, the order parameter is conserved in the region of $\psi>0.5$.}\label{CA-Validate}
\end{center}
\end{figure}

In the first set of simulations, we evolved Eq.~\eqref{AC-1} for a nonconserved order parameter with a 60$^\circ$ contact angle. The result clearly shows a 60$^\circ$ contact angle at the three-phase boundary, as specified; see Fig.~\ref{CA-Validate}(a). The angle can be measured at the intersection between the two contours of $\psi = 0.5$ and $\phi=0.5$, as shown in Figs.~\ref{CA-Validate}(b) and (c). The 60$^\circ$ angle is maintained during the entire evolution, except for the very early transient period, when the contact angle changes from the initial 90$^\circ$ angle to the prescribed 60$^\circ$ angle. Because of the contact-angle boundary condition, the initially flat phase boundary bends and creates a negative curvature in Phase 1. As a result, the phase boundary moves toward Phase 0. Once the phase boundary evolves to a circular arc with a uniform curvature everywhere (other than regions in contact with the substrate), it moves at a uniform constant speed in a steady-state motion, and eventually only Phase 1 remains in the system.

In the second set of simulations, we evolved Eq.~\eqref{SBM-CH-3} for a conserved order parameter in a closed system with $J_n=0$ and a 120$^\circ$ contact angle. As expected, the phase boundary intersects the substrate surface at a 120$^\circ$ contact angle; see Fig.~\ref{CA-Validate}(d)--(f). In contrast to the Allen-Cahn-type dynamics, because of the conservation of the order parameter, the phase boundary near the substrate moves toward the left, whereas the phase boundary away from the substrate moves in the opposite direction. As a result, the phase boundary deforms into a curved shape. When the system reaches its equilibrium state, the phase boundary forms a circular arc with a uniform curvature everywhere (except where the phase is in contact with the substrate), such that the total surface energy is minimized; see Fig.~\ref{CA-Validate}(d) for $t=3.0\times10^5$.

Table \ref{Tbl-3} lists the average values of $\cos \theta$ calculated by $(\nabla \psi \cdot \nabla \phi) /(|\nabla \psi | |\nabla \phi|)$ at the grid points within the three-phase boundary region defined by $0.1 < \psi < 0.9$ and $0.1 < \phi < 0.9$ in the steady state (i.e., when the phase boundary becomes a circular arc). These results again clearly show that the error (for a given phase boundary thickness) increases as the interfacial thickness increases. This can be understood based on the analysis of the 1D test results in Section \ref{1D_Diffusion_Test} since the substrate surface is assumed to be flat in this test.  If the substrate interface is curved, the resolution of the interface will have more influence on the error.

In contrast to the effect of domain boundary thicknesses, the error is relatively insensitive to the phase boundary thickness once the phase boundaries are properly resolved; see cases with $\delta_{\phi} >= 1.4142$ in Table \ref{Tbl-3}. However, when the phase boundary is too thin, the error tends to increase because of the loss of resolution in the phase-field-order-parameter gradient; see cases with $\delta_{\phi} = 1.0607$ in Table \ref{Tbl-3}. In general, the results demonstrate that the contact-angle boundary condition is well imposed using the presented method. Even when the domain boundary thickness is as large as 16.74 grid spacings with $\zeta=4$, the contact angle only deviates less than 2$^\circ$ from the imposed values as long as the phase boundary is properly resolved.  Here, the error is sensitive to the resolution of the interface because the phase boundary is curved unlike the substrate surface.  

In addition to the contact angle, a no-flux boundary condition for a conserved order parameter is implicitly imposed at the substrate surface. The error associated with such a boundary condition was evaluated by examining the overall change in the value of the total order parameter. The conservation of the order parameter was met within a numerical error (well below 1\% in most cases) in these validation simulations; see Table \ref{Tbl-3}.

\section{Applications}

Although the details of the scientific calculations performed applying these methods to problems in materials science will be published elsewhere, it is worth presenting some of the results herein to demonstrate the potential of the method.  

\subsection{Oxygen-Vacancy Diffusion in SOFC Cathode} \label{SOFC_Diff} 

The first example is ionic transport through a complex microstructure. Here, ion diffusion is driven by a sinusoidal voltage perturbation. For the steady-state solution, the time dependence of the form $\exp(\mathrm{i} \omega t)$, where $\omega$ is the angular frequency and $\mathrm{i}=\sqrt{-1}$, can be removed as in the equation derived by Lu et al.~\cite{Lu:2009}. For this case, the smoothed boundary formulated equation is obtained from Eq.~\eqref{SBM-FSL-BS1} to: 
\begin{equation} \label{SBM-CxMs-1}
\nabla \cdot (\psi D_\text{b} \nabla \tilde{C}) - |\nabla \psi | ( \kappa \tilde{C} - D_\text{s} \nabla_\text{s}^2 \tilde{C} ) = \mathrm{i} \psi \omega \tilde{C},
\end{equation}
where $\tilde{C}$ is the complex concentration amplitude, which consists of real and imaginary parts and includes the amplitude of the concentration wave and the phase shift. Note that the surface accumulation term is ignored ($L=0$) here because its magnitude is usually very small in comparison with the bulk concentration \cite{Lu:2009}. This equation can be solved by an alternating direction line relaxation (ADLR) method in a second-order central-difference scheme in space; see Appendix \ref{SurLap_A2} for the numerical  implementation.

In this work, we adopted an experimentally reconstructed complex microstructure, the porous ceramic cathode and nonporous ceramic electrolyte of an SOFC, as the input geometry. The microstructure data is stored as a 3D array consisting of $321\times261\times297$ voxels that indicate the electrolyte (gadolinia-doped ceria: GDC), cathode (lanthanum strontium chromite: LSC) and pore phases by different values. To emphasize the convenience of image-based smoothed boundary simulations, we treat the center of each voxel as the location of the grid points in the calculation without further enhancement of the resolution from our initial reconstructed microstructure. For very high-accuracy scientific calculations, one can easily enhance the resolution by refining the grid sizes. To smooth the voxelated, discontinuous data, we first employed a level set distance function method \cite{Osher:2003} to determine the distances between grid points and the solid-pore interface, and then computed the hyperbolic tangent of the distance function to obtain the domain parameter profile; see Appendix \ref{LevelSetDist} for details.

\begin{figure}[htb]
\begin{center}
\includegraphics[width=1\textwidth]{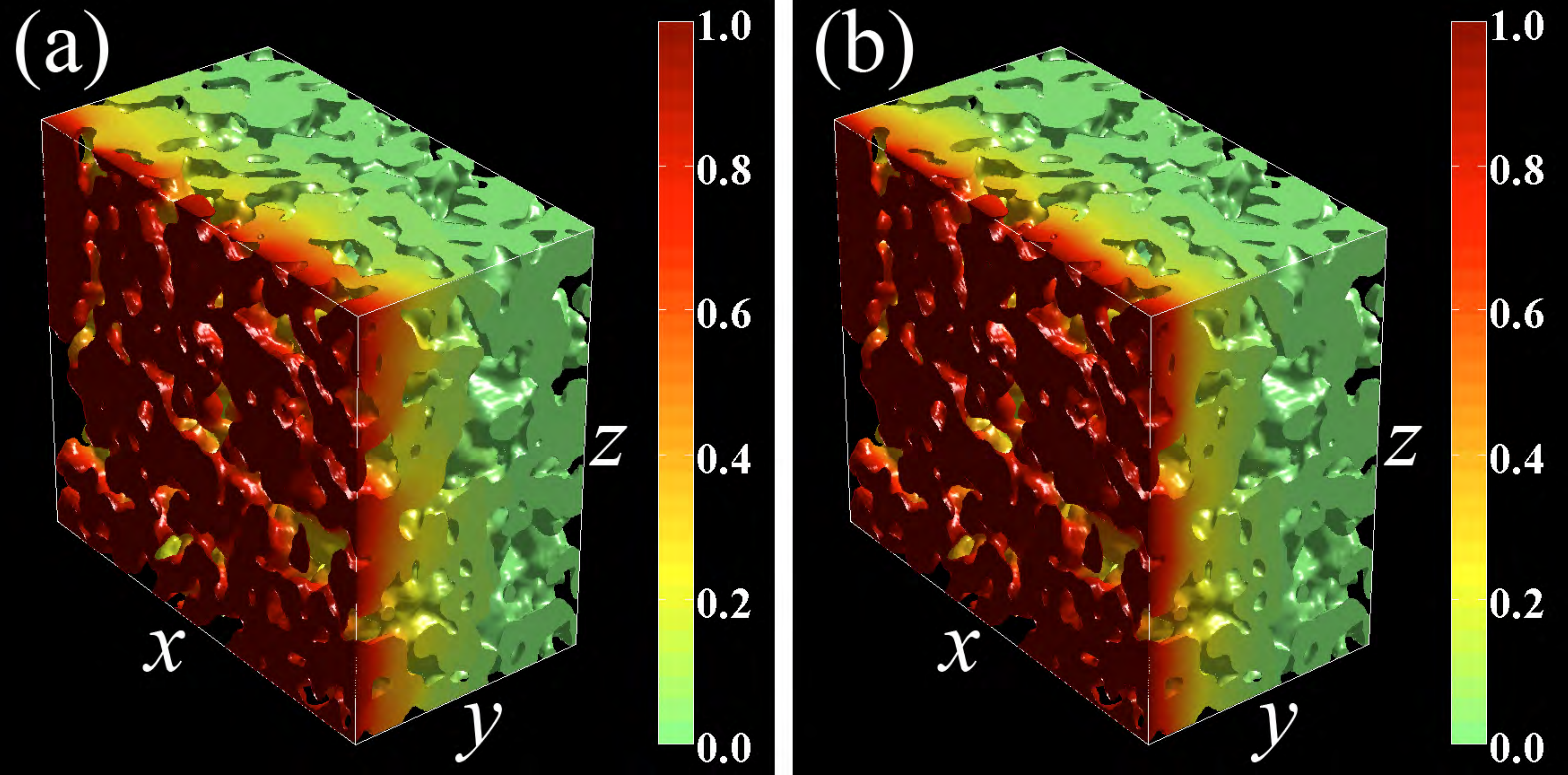}
\caption{Simulation results of dimensionless steady-state oxygen-vacancy concentrations under DC loading for $D_\text{b} = 1$ using an experimentally obtained SOFC cathode complex microstructure as the input geometry: (a) $\kappa = 0.1$ and $D_\text{s} = 0$; (b) $\kappa = 2.1$ and $D_\text{s} = 10$.}\label{CR-DC-1}
\end{center}
\end{figure}

For simulations of the concentration distribution in the porous cathode, the regions containing nonporous electrolyte are excluded, such that the computational box only consists of $321\times176\times297$ grid points. The grid spacing is set at $\Delta x= 6.285\times10^{-2}$. The boundary conditions along the main diffusion direction (the $y$-axis) on the computational box are $\text{Re}(\tilde{C}) = 1$ and $\text{Im}(\tilde{C}) = 0$ at $y=0$, and $\text{Re}(\tilde{C}) = 0$ and $\text{Im}( \partial \tilde{C}/\partial y) =0$ at $y=11.062$. The boundary conditions on the remaining four sides are zero-gradient for both the real and imaginary parts. As a demonstration of the method, the length scale and physical material properties are nondimensionalized. Figure \ref{CR-DC-1} shows the steady-state concentrations for the cases in which surface diffusion is excluded ($\kappa=0.1$ and $D_\text{s} = 0$) and included ($\kappa = 2.1$ and $D_\text{s}= 10$) with $D_\text{b} = 1$ and direct current (DC) loading ($\omega=0$). In these cases, the imaginary part vanishes, and the solution of the real part is equivalent to that of a homogeneous Helmholtz-like equation with the right-hand side of Eq.~\eqref{SBM-CxMs-1} equal to zero. As shown in Fig.~\ref{CR-DC-1}, the concentration decays from 1 to 0 along the $y$-axis over the complex cathode microstructure to satisfy the boundary conditions imposed on the box boundaries and at the cathode-pore interfaces. The utilization lengths (i.e., the length over which the cathode material is active) of the two cases are similar, as predicted by Lu et al.~\cite{Lu:2009} for a cathode with simplified cylindrical geometry, in which the effective diffusivities under DC loading with and without surface diffusion are found to be similar for the parameters given above. However, a slight difference in the concentration distributions of the two cases can be observed. Because of the faster transport path along the surface, the diffusion front with surface diffusion (Fig.~\ref{CR-DC-1}(b)) is more planar compared with that without surface diffusion (Fig.~\ref{CR-DC-1}(a)). 

\begin{figure}[htb]
\begin{center}
\includegraphics[width=1\textwidth]{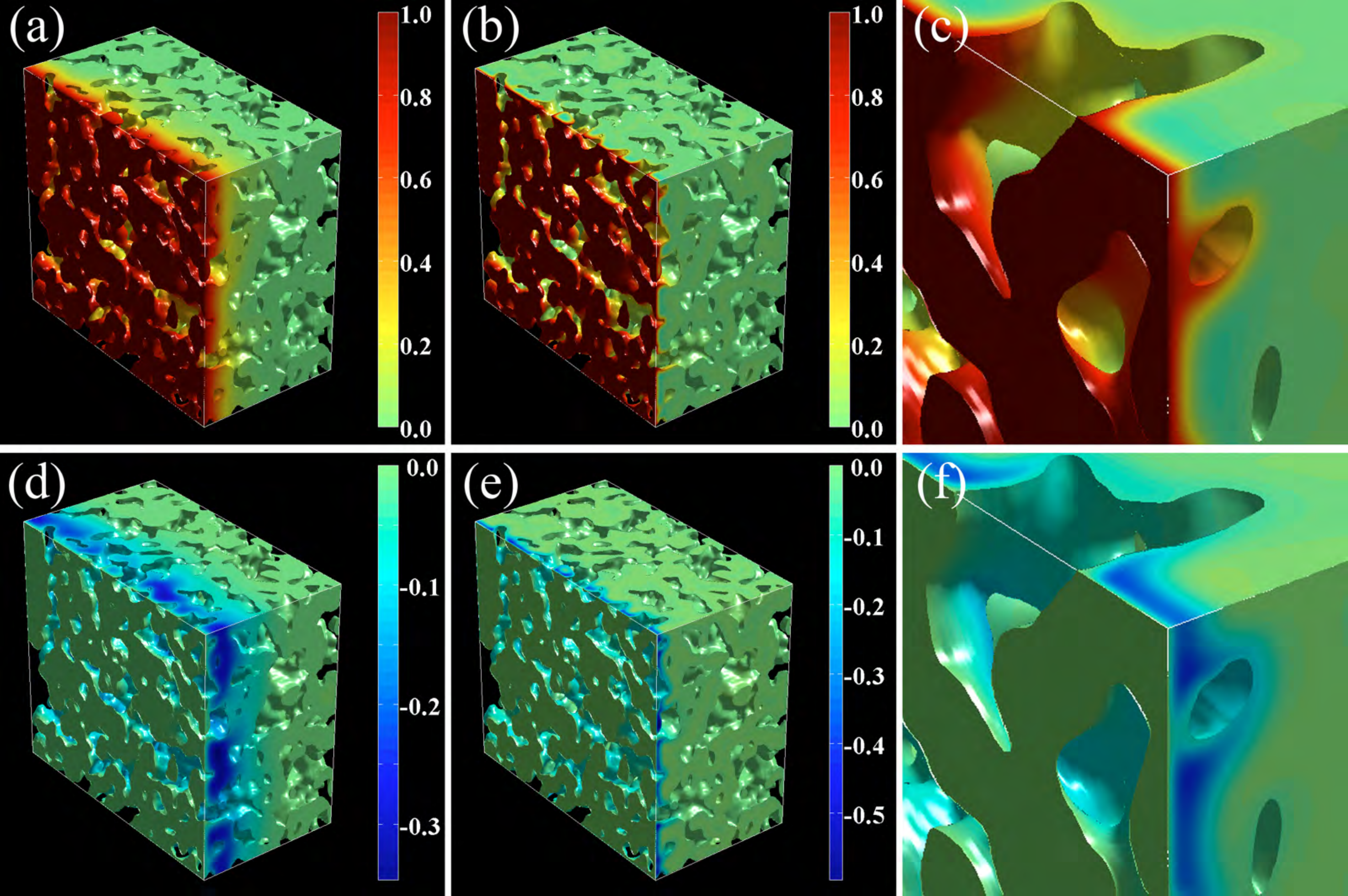}
\caption{Simulation results of dimensionless steady-state oxygen-vacancy concentrations under AC loading for $D_\text{b} = 1$, $\kappa = 2.1$, and $D_\text{s} = 10$ using an experimentally obtained SOFC cathode complex microstructure as the input geometry. In (a) and (d), the real and imaginary parts for $\omega = 1.5$ are shown, respectively. Similarly, (b) and (e) display the real and imaginary parts for $\omega = 51.5$, respectively. Shown in (c) and (f) are the magnified views of (b) and (e), respectively.}\label{SurfDiff_AC1}
\end{center}
\end{figure}

Figure \ref{SurfDiff_AC1} shows the real and imaginary parts of the steady-state concentration amplitude for the cases in which $D_\text{b} = 1$, $\kappa = 2.1$ and $D_\text{s} = 10$, with alternating current (AC) loading of the angular frequencies of $\omega=1.5$ and $51.5$. The boundary conditions on the computational box are the same as in the DC loading case above. In the low frequency case (Figs.~\ref{SurfDiff_AC1}(a) and (d)), the real part of the concentration, which represents the amplitude of the concentration wave decays and forms a planar diffusion front within the utilization length, where the material is active ($0 < y < 5$). Additionally, a negative value of the imaginary part occurs in the regions where the real part decays because of the phase shift resulting from the delayed response. The magnitude of the imaginary part then decays back to zero toward the inactive region. In the high frequency case, the enhancement of concentration along the surface is observed due to surface diffusion; this is evident in Figs.~\ref{SurfDiff_AC1}(b) and (c), which show larger values of the real part of the concentration amplitude within the utilization length ($0 < y < 2.5$). The real part of the concentration amplitude quickly decays from the surface into the bulk. In contrast to the enhanced real part at the cathode surface, the magnitude of the imaginary part is small near the surface because surface diffusion reduces the response time and the phase change is thus decreased. In an analogy to the low-frequency response, a negative imaginary part occurs in the region where the real part decays. The magnitude of the imaginary part decays toward the inactive region. This behavior can be more clearly discerned in the magnified views in Figs.~\ref{SurfDiff_AC1}(c) and (f).

The smoothed boundary method can also be used to impose Dirichlet boundary conditions on irregular surfaces. For example, if the ionic diffusivity in the electrolyte is assumed to be much larger than that in the cathode, the concentration in the electrolyte will be nearly uniform. To simulate this scenario, we impose a fixed concentration at the electrolyte-cathode contacting surface as the boundary condition. We used the experimentally reconstructed $321\times261\times297$ array that contains a porous cathode and a nonporous electrolyte as our input geometry. The voxelated data were smoothed to the hyperbolic tangent domain parameter profile by the level set distance function method mentioned in Appendix \ref{LevelSetDist}. Here, three domain parameters are employed to define the three regions: electrolyte ($\psi_1$), cathode ($\psi_2$), and pore ($\psi_3= 1-\psi_1-\psi_2$). The smoothed boundary formulated governing equation is obtained by modifying Eq.~\eqref{SBM-FSL-BS1} to: 
\begin{equation}
\begin{split}
\frac{\partial C}{\partial t} = \frac{\nabla \cdot (\psi_2 D_\text{b} \nabla C)}{\psi_2} - \frac{ | \nabla \psi_2 | }{\psi_2}  \bigg[ \kappa C  - D_\text{s}  & \nabla_\text{s}^2 C \bigg] W_N  - \\ \frac{D_\text{b}}{\psi_2^2} \bigg[ \nabla \psi_2 \cdot \nabla ( \psi_2 C) & - |\nabla \psi_2 |^2 B_D\bigg] W_D,
\end{split}
\end{equation}
where the weighting factors are given by $W_N = [|\nabla \psi_2 | |\nabla \psi_3|/ ( |\nabla \psi_1 | |\nabla \psi_2 |+ |\nabla \psi_2 | |\nabla \psi_3 | + |\nabla \psi_3 | |\nabla \psi_1 | )]^\beta$ and $W_D = [|\nabla \psi_1 | |\nabla \psi_2 |/ ( |\nabla \psi_1 | |\nabla \psi_2 |+ |\nabla \psi_2 | |\nabla \psi_3 | + |\nabla \psi_3 | |\nabla \psi_1 | )]^\beta$, such that the Neumann boundary condition (surface reaction and surface diffusion) is imposed only at the cathode-pore interface ($| \nabla \psi_2 | | \nabla \psi_3 | \neq 0$), and the Dirichlet boundary condition (a prescribed concentration value) is imposed only at the electrolyte-cathode interface ($| \nabla \psi_1 | | \nabla \psi_2 | \neq 0$). The exponent $\beta$ determines the transition profiles from the Neumann to the Dirichlet boundary conditions in the regions of three-phase boundaries. We selected $\beta=0.8$ for this numerical simulation. On the computational box boundaries, we set $C = 0$ at $y=16.404$ and the zero-gradient boundary condition for the remaining five sides. 

\begin{figure}[htb]
\begin{center}
\includegraphics[width=1\textwidth]{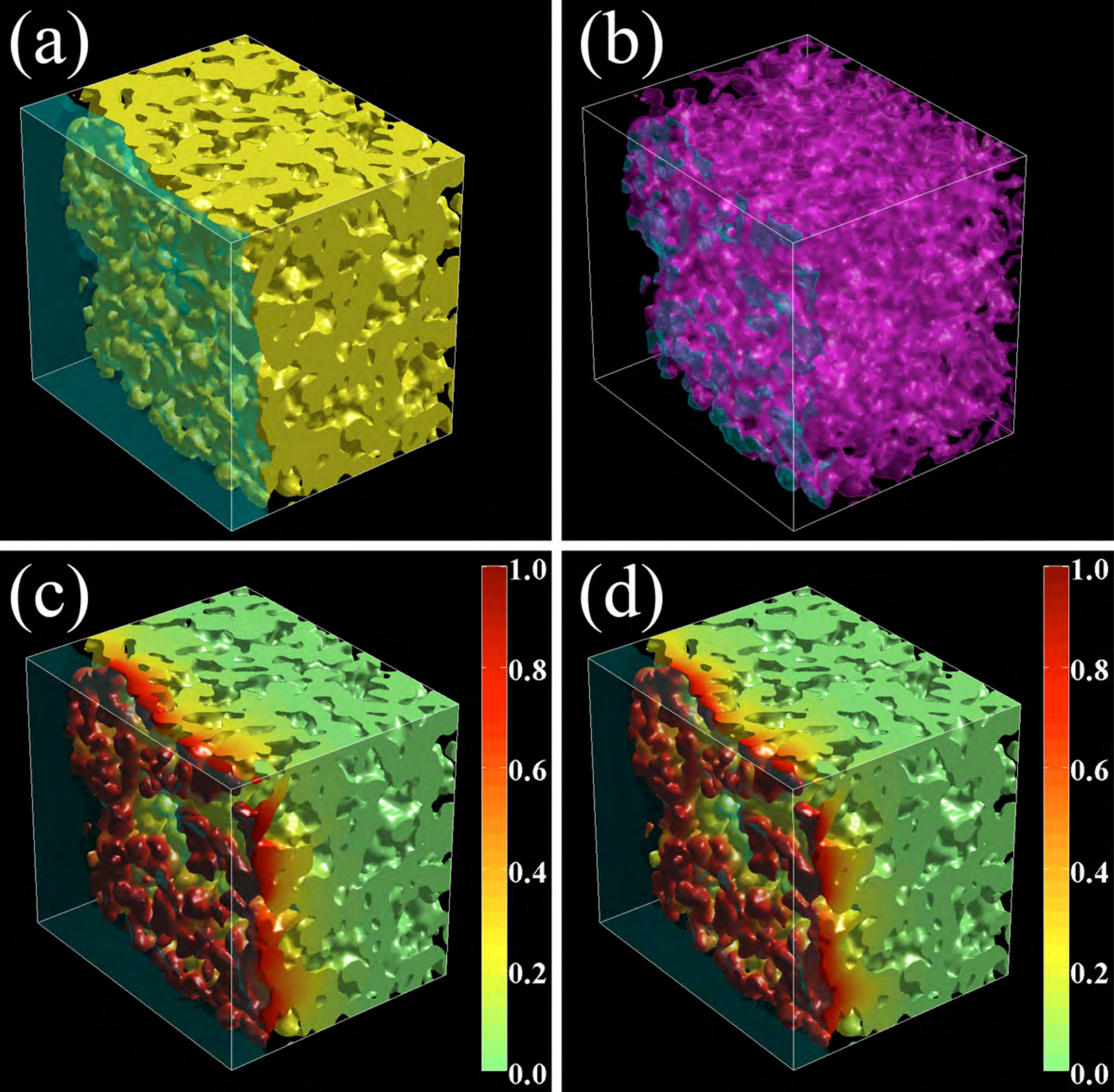}
\caption{(a) Solid phase containing cathode (yellow) and electrolyte (semitransparent cyan) in a solid oxide fuel cell material. (b) Interfaces defined by $\sqrt{|\nabla \psi_1||\nabla \psi_2|} > 0.2$ (semitransparent cyan) and $\sqrt{|\nabla \psi_2||\nabla \psi_3|} > 0.2$ (semitransparent purple). Dimensionless steady-state oxygen-vacancy concentration with the boundary condition $C=1$ at the electrolyte-cathode interface for $D_\text{b} = 1$: (c) $\kappa = 0.1$, and $D_\text{s} = 0$; (d) $\kappa = 2.1$, and $D_\text{s} = 10$.}\label{DiriBC}
\end{center}
\end{figure}

The same material parameters used in the cases of Fig.~\ref{CR-DC-1} were selected. Figures \ref{DiriBC} (a) and (b) illustrate the reconstructed SOFC complex microstructure and irregular surfaces defined by the values and gradients of the domain parameters, respectively. Figures \ref{DiriBC}(c) and (d) show the simulation results of the steady-state oxygen-vacancy concentration distributions with a fixed value of $C=1$ imposed at the cathode (LSC)-electrolyte (GDC) interfaces. The concentration distribution is very different from the ones shown in Fig.~\ref{CR-DC-1} because a larger portion of lateral diffusion occurs in the $x$ and $z$ directions, which results from the smaller contacting areas (compared to the cross-sectional area of LSC on the $x$-$z$ plane in Fig.~\ref{CR-DC-1}, where diffusion is mainly in the $y$ direction). As a result, the concentration drops rapidly within a short distance from the contacting areas, making the utilization length of the cathode material shorter and uneven. 

\subsection{Kirkendall-Effect Diffusion with a Moving Boundary} \label{Kirkendall-Effect-Diffusion} 

\subsubsection{ Kirkendall-Effect-Induced Deformation Modeled by Navier-Stokes-Cahn-Hilliard Equations}
The third application demonstrates the smoothed boundary method's broad applicability by applying it to the coupled Navier-Stokes-Cahn-Hilliard equations \cite{Gurtin:1996,Jacqmin:1999,Kim:2005,Zhou:2006,Villanueva:2008,Villanueva:2009}. This particular formulation aims to solve diffusion problems with the Kirkendall effect with efficient and abundant vacancy sources and sinks in the bulk of a solid \cite{Kirkendall:1939,Kirkendall:1942,Smigelskas:1947,Darken:1948,AtomMovements:Bardeen}. In this case, the solid experiences deformation because of vacancy generation and elimination. The Navier-Stokes-Cahn-Hilliard equations are coupled with the smoothed boundary formulation of the diffusion equation derived in Section \ref{DiffEqn} as a model of plastic deformation because of volume expansion and contraction resulting from vacancy flow.

When the diffusing species of a binary substitutional alloy have different mobilities, the diffusion fluxes of the two species are unbalanced, creating a net vacancy flux toward the side containing the fast diffuser. Here, we denote the quantities associated with the slow diffuser, fast diffuser and vacancy by the subscripts $A$, $B$, and $V$, respectively. Because of the accommodation/supply of excess/depleted vacancies, the solid locally expands/shrinks \cite{Strandlund:2004,Larsson:2006,Strandlund:2006,Yu:2007,Svoboda:2008} when maintaining the vacancy mole fraction at its thermal-equilibrium value. We treat the solid as a very viscous fluid \cite{Stephenson:1988,Boettinger:2005,Dantzig:2006,Boettinger:2007,Boettinger:2010} with a much larger viscosity than that of the surrounding environment. In this case, we solve the Navier-Stokes-Cahn-Hilliard equations to update the shape of the material as follows \cite{Yu:2009a}:
\begin{subequations}
\begin{equation} \label{NS-CH-1}
-\nabla P+\nabla \cdot \eta \bigg[ \nabla \mathbf{v} + (\nabla \mathbf{v})^T \bigg] - \nabla \bigg(\frac{2\eta}{d} g_V\bigg)+\frac{1}{C_\text{a}}\mu\nabla \psi=0,
\end{equation}
\begin{equation} \label{NS-CH-2}
\nabla \cdot \mathbf{v}= g_V,
\end{equation}
\begin{equation} \label{NS-CH-3}
\frac{\partial \psi}{\partial t} -\mathbf{v}\cdot \nabla \psi = M \nabla^2\bigg( \frac{\partial f}{\partial \psi} -\epsilon^2 \nabla^2\psi\bigg),
\end{equation}
\end{subequations}
where $P$ is the effective pressure, $\eta$ is the viscosity, $\mathbf{v}$ is the velocity vector, $d$ is the number of dimensions, the superscript $T$ denotes the transpose, $C_\text{a}$ is the Cahn number reflecting the capillary force compared to the pressure gradient, $g_V$ is the vacancy generation rate per unit volume, and $\psi$ is the domain parameter indicating the solid phase for diffusion. One great advantage in employing a phase-field type equation is that it automatically maintains the profile of the domain parameter, $\psi$, in the form of a hyperbolic tangent function because it is the equilibrium solution for the phase field equation (Eq.~\eqref{NS-CH-3}). Note that here we ignore the inertial force in the Navier-Stokes equation to obtain Eq.~\eqref{NS-CH-1} because the deformation is assumed to be a quasi-steady-state process. The vacancy generation rate that results in the local volume change (dilatational strain) is given by $g_V = -[\nabla \cdot (D_{VB} \nabla X_B)]/[\rho_l(1-X_V^{eq})]$, where $X_B$ is the mole fraction of the fast diffuser, $X_V^{eq}$ is the thermal-equilibrium vacancy mole fraction (which is assumed to be maintained throughout the solid in this model), $D_{VB}$ is the diffusivity for vacancy flux associated with $\nabla X_B$, and $\rho_l$ is the lattice site density of the solid. Here, $X_V^{eq}$ is taken to be $1.6\times10^{-6}$. The evolution of the fast diffuser mole fraction is governed by the advective Fick's diffusion equation, written as:
\begin{equation} \label{KE-Trad-1}
\frac{\partial X_B}{\partial t} -\mathbf{v} \cdot \nabla X_B= \nabla \cdot (D_{BB}^V\nabla X_B) -X_B g_V,
\end{equation}
where $D_{BB}^V$ is the diffusivity for the fast diffuser flux associated with $\nabla X_B$, and the advective term accounts for the lattice shift because of volume change. Because diffusing atoms cannot depart from the solid region, a no-flux boundary condition is imposed at the solid surface. Thus, the smoothed boundary formulation of Eq.~\eqref{KE-Trad-1} is written as:
\begin{equation} \label{KE-Trad-SBM-1}
\frac{\partial X_B}{\partial t} -\mathbf{v} \cdot \nabla X_B = \frac{\nabla \cdot (\psi D_{BB}^V\nabla X_B)}{\psi} - X_B g_V.
\end{equation}
As the concentration evolves, the shape of the solid is also updated by Eq.~\eqref{NS-CH-3} and by iteratively solving Eqs.~\eqref{NS-CH-1} and \eqref{NS-CH-2} through the application of a projection method \cite{Kim:2006a,Kim:2006b}; see Appendix \ref{ProjMethod} for the numerical implementation.

\begin{figure}[htb]
\begin{center}
\includegraphics[width=0.95\textwidth]{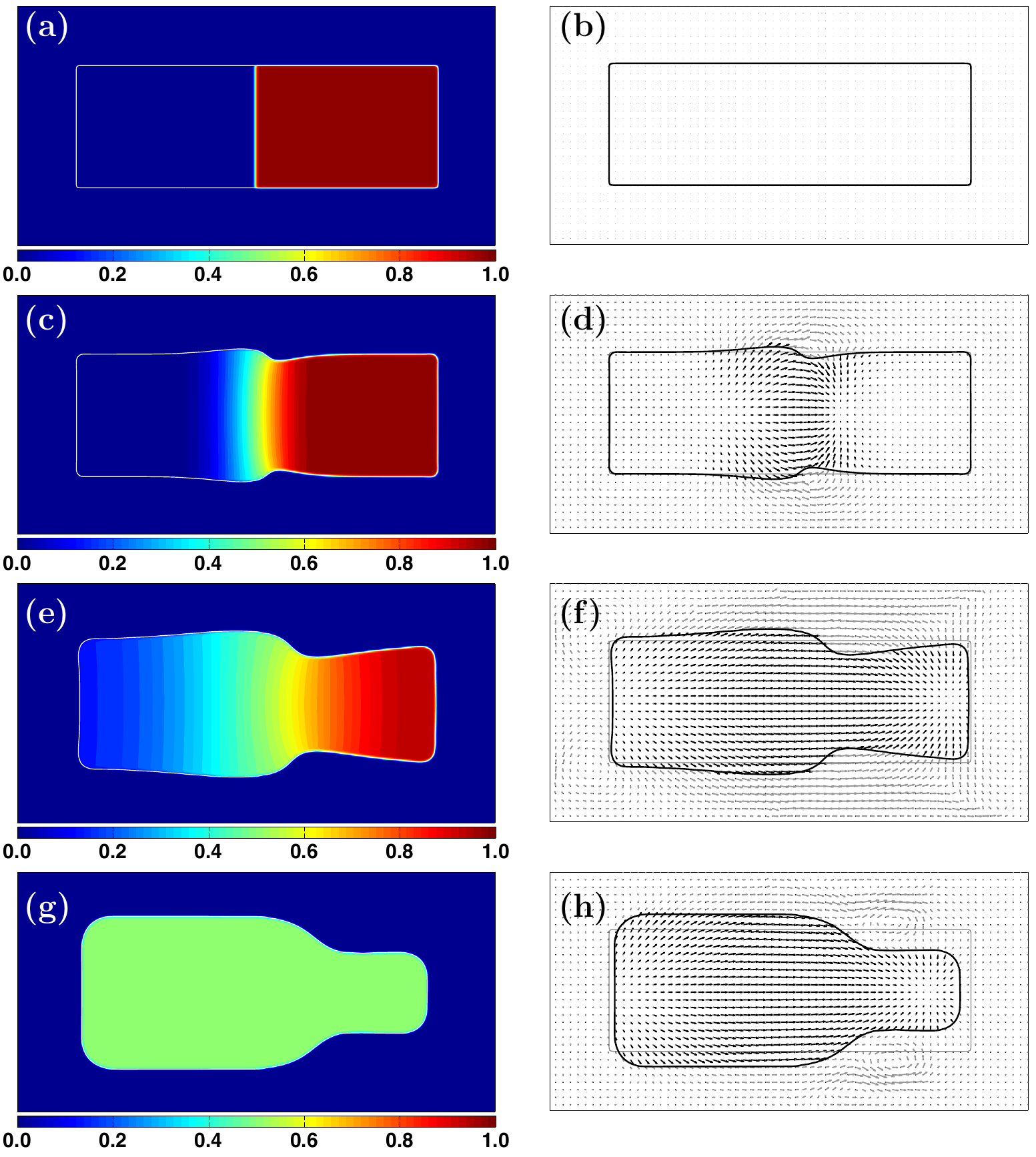}
\caption{ Left column: Fast diffuser mole fraction profiles (normalized to the lattice site density) recorded at dimensionless time of $t= 0$, $3.96\times10^7$, $5.09\times10^8$, and $4.32\times10^9$. Right column: Velocity fields corresponding to the mole fraction profile on the left. Black and gray arrows denote the flows inside and outside the material, respectively. The flow outside of the material has no physical significance for the shape change.  Simulation parameters were $\bar{\eta} = 1\times10^5$, $C_\text{a} = 1\times10^8$, $\epsilon = 1$ and $M = 1.25\times10^{-8}$. The fast diffuser hop frequency is four times larger than that of the slow diffuser.}\label{Def_Con}
\end{center}
\end{figure}

The slow and fast diffusers are initially placed in the left and right halves of the solid, respectively. We use their theoretically calculated diffusivities for this simulation \cite{Moleko:1989,Manning:1971,Belova:2000,VanDerVen:2010}. Here, we calculate the slow diffuser atomic hop frequency based on the material parameters of aluminum at 600 K, and set the fast diffuser atomic hop frequency four times larger than that of the slow diffuser. Figure \ref{Def_Con} shows snapshots of the mole fraction profiles (left column) and velocity fields (right column) from a 2D simulation. As the fast diffuser diffuses from the right to the left side, the vacancy elimination and generation cause contraction and expansion on the right and left sides, respectively. As a result, the initially rectangular slab deforms into a bottle-shaped object.

\subsubsection{Kirkendall Void Growth with Localized Vacancy Sources} \label{Kirkendall-Void} 

In another scenario in which the vacancy diffusion length is comparable to or smaller than the distance between vacancy sources and sinks, the explicit vacancy diffusion process must be considered \cite{Svoboda:2008,VanDerVen:2010,Yu:2008,Yu:2009}. In this case, vacancies diffuse in the same manner as the atomic species. In the bulk of a solid devoid of vacancy sources/sinks, the concentration evolutions are governed by:
\begin{subequations} \label{KE-Rig-1}
\begin{equation} 
\frac{\partial X_V}{\partial t} = \nabla \cdot (D_{VV} \nabla X_V + D_{VB} \nabla X_B),
\end{equation}
\begin{equation}
\frac{\partial X_B}{\partial t} = \nabla \cdot (D_{BV} \nabla X_V + D_{BB}^V \nabla X_B).
\end{equation}
\end{subequations}
Because the solid surfaces are very efficient vacancy sources/sinks \cite{Yu:2009a,Yu:2009}, we impose the thermal-equilibrium vacancy mole fraction at the solid surfaces as the Dirichlet boundary condition for solving Eq.~\eqref{KE-Rig-1}. In this case, the smoothed boundary formulation of Eq.~\eqref{KE-Rig-1} is given by:
\begin{subequations} \label{KE-Rig-2}
\begin{equation} \label{KE-Rig-2V}
\frac{\partial X_V}{\partial t} = \frac{1}{\psi} \nabla \cdot [\psi (D_{VV} \nabla X_V+D_{VB} \nabla X_B)]- \frac{K}{\psi^2},
\end{equation}
\begin{equation} \label{KE-Rig-2B}
\frac{\partial X_B}{\partial t} = \frac{1}{\psi} \nabla \cdot [\psi (D_{BV} \nabla X_V+D_{BB}^V \nabla X_B)]+\frac{X_B}{1-X_V^{eq}} \frac{K}{\psi^2},
\end{equation}
\end{subequations}
where $K = D_{VV} [\nabla \psi \cdot \nabla (\psi X_V)-|\nabla \psi|^2 X_V^{eq}]$. Because the vacancy generation and elimination in this scenario only occurs on the solid surfaces, internal volume change in the bulk is not considered. Therefore, instead of using a plastic deformation model as in the previous case, we adopt a typical Cahn-Hilliard type dynamics to model the shape change:
\begin{equation} \label{KE-Rig-Psi-1}
\frac{\partial \psi}{\partial t} = M\nabla^2 \bigg( \frac{\partial f}{\partial \psi} - \epsilon^2 \nabla^2 \psi \bigg)+\frac{\nabla \psi}{|\nabla \psi |}\cdot \frac{\vec{J}_V}{1-X_{V}^{eq}},
\end{equation}
where $\vec{J}_V = -( D_{VV}\nabla X_V + D_{VB}\nabla X_B)$ is the vacancy flux, and the last term represents the normal velocity of the solid surfaces because of vacancy injection into or ejection from the solid.

\begin{figure}[htb]
\begin{center}
\includegraphics[width=1\textwidth]{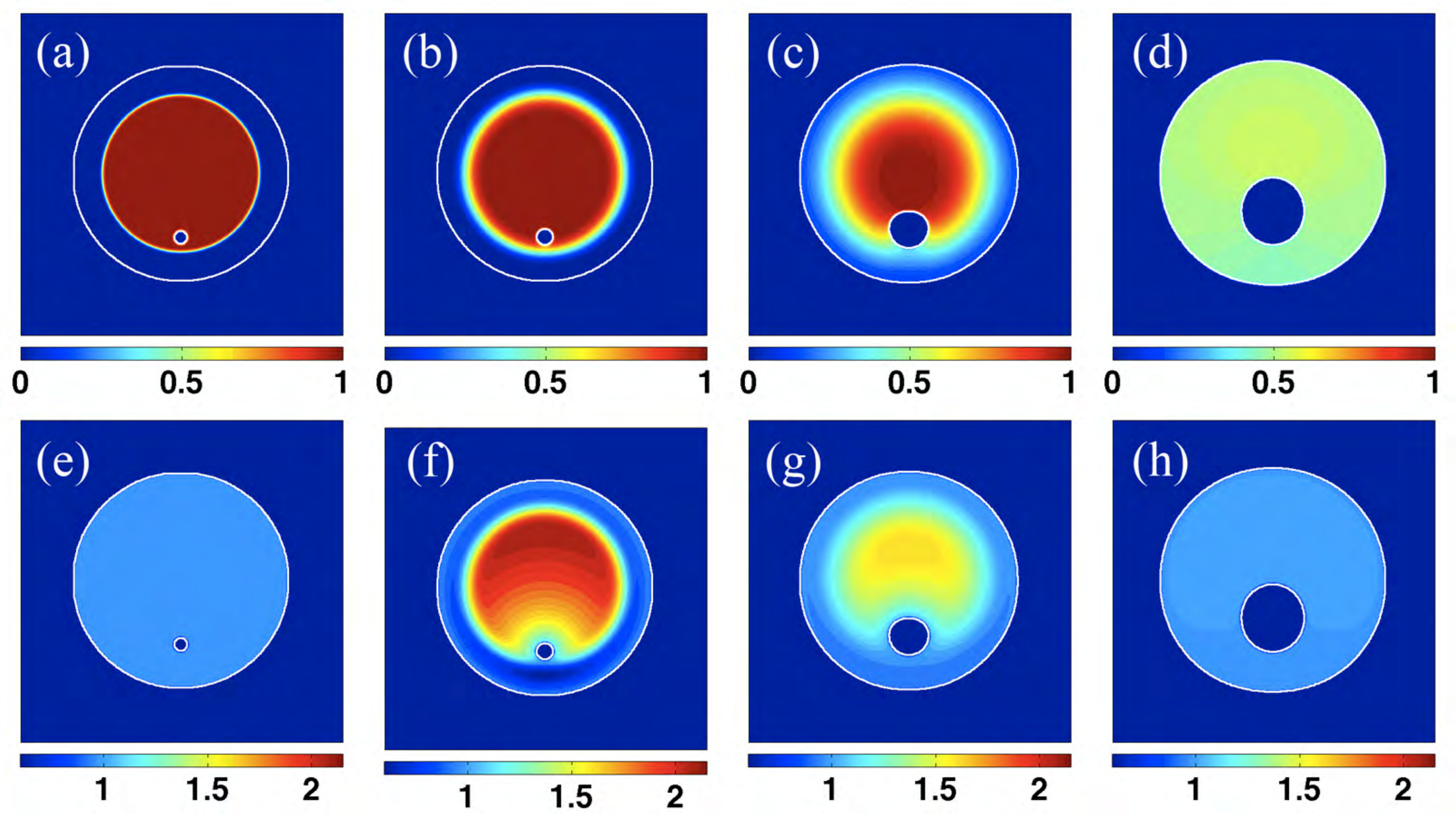}
\caption{Top row, (a)--(d): snapshots of the fast diffuser mole fraction recorded at four different dimensionless times, $t= 0$, $2.79\times10^2$, $2.62\times10^3$, and $3.74\times10^4$, respectively. Bottom row, (e)--(h): snapshots of vacancy mole fraction normalized to the equilibrium value, corresponding to (a)--(d). The solid white contour lines indicate the locations of the rod and void surfaces. The fast diffuser hop frequency is four times larger than that of the slow diffuser.}\label{Hallow-1}
\end{center}
\end{figure}

An example of the results obtained using this approach is the growth of a void in a rod \cite{Yu:2009,Fan:2006,Fan:2007,Glodan:2010}. The above equations were solved using a central difference scheme in space and an implicit time scheme (see Appendix \ref{ImplicitTime}). The fast diffuser was initially placed in the central region while the slow diffuser filled the outer region. A void was initially placed off-centered in the fast-diffuser region, where a 1D study found to be the likely nucleation site \cite{Yu:2009}. The vacancy mole fractions were fixed at the void and cylinder surfaces. Figure \ref{Hallow-1} shows snapshots of the fast diffuser mole fraction profile (normalized to the lattice density) and the vacancy mole fraction profile (normalized to its equilibrium value). As the fast diffuser diffuses outward, vacancies diffuse inward from the rod surface to the void surface, causing vacancy concentration enhancement and depletion in the central and outer regions, respectively. To maintain the equilibrium vacancy mole fraction at the rod and void surfaces, vacancies are injected and ejected at those surfaces. As a result, the rod radius increases, and the void grows. Similar dynamics were examined using a sharp interface approach \cite{Yu:2009}, but this new method provides the flexibility in geometry to examine cases where a void initially forms off-centered. 

\subsection{Thermal Stress} \label{T-Stress} 

Solid oxide fuel cells (SOFCs) usually operate at temperatures near $500^\circ - 1,000^\circ$C. Evaluating the thermal stress resulting from the differences in thermal expansion and elastic moduli is important for analyzing mechanical failure. We expand the generalized mechanical equilibrium equation, Eq.~\eqref{SBM-ME-2}, for a linear, elastic and isotropic solid. (Note that the derivation for the mechanical equilibrium equation is general and is not limited to isotropic solids. We selected an isotropic model because of the lack of available crystallographic information among the experimental data.) The equation is discretized in a central finite difference scheme and numerically solved by an ADLR solver; see Appendix \ref{ME_solverA} for details.

The thermal expansion rates of the ceramic electrolyte (GDC) and cathode (LSC) are taken to be $12.3\times10^{-6}~\text{K}^{-1}$ \cite{Wang:2003} and $10.6\times10^{-6}~\text{K}^{-1}$ \cite{Sorensen:2001}, such that the thermal expansions at operation temperature are $0.0123$ and $0.0106$, respectively.  (Here, we have assumed arbitrarily that the composite material is relaxed at a reference temperature, and assumed an operation temperature of $1000^\circ$ above the reference temperature.)  We chose the elastic constants of GDC to be isotropic ($\lambda_{11}-\lambda_{12}= 2\lambda_{44}$), and the values are $\lambda_{11}^{GDC}=375.94$ GPa, $\lambda_{12}^{GDC}=188.54$ GPa and $\lambda_{44}^{GDC} = 93.70$ GPa, calculated from a Young's modulus of 250 GPa and a Poisson's ratio of 0.334 \cite{Wang:2007,Serincan:2010}; see Appendix \ref{ME_solverA}. The LSC phase is softer than the GDC phase, and its elastic constant is also assumed to be isotropic. The values are selected to be $\lambda_{11}^{LSC}=269.23$ GPa, $\lambda_{12}^{LSC}=115.38$ GPa and $\lambda_{44}^{LSC} = 76.29$ GPa, based on a Young's modulus of 200 GPa and a Poisson's ratio of 0.3 \cite{Sorensen:2001}. As in Section \ref{SOFC_Diff}, we again use domain parameters to indicate the GDC phase ($\psi_{1} = 1$ inside the GDC and $\psi_{1} = 0$ outside the GDC) and the LSC phase ($\psi_{2}=1$ inside the LSC and $\psi_{2}=0$ outside the LSC). The entire solid phase is then represented by the sum of the two phases, $\psi = \psi_{1}+\psi_{2} = 1$. The body force term and elastic constant tensor are replaced by an interpolated, spatially dependent thermal expansion and elastic constant tensor according to the domain parameters; see Appendix \ref{ME_solverA}. The solid surface is assumed to be traction-free, $N_i=0$. 

In this simulation, we selected the same computational box as in the case of Fig.~\ref{DiriBC}(a), containing $321\times261\times297$ grid points in the $x$, $y$, and $z$ directions, respectively. Each grid point represents a voxel in the experimentally obtained microstructure. The yellow color indicates the LSC phase, and the semitransparent cyan color indicates the GDC phase; see Fig.~\ref{DiriBC}(a). The grid spacing is $\Delta x = 25$ nm, such that the computational box spans $8.025\times6.524\times7.425$ $\mu$m$^3$. We assumed a rigid computational box with frictionless boundaries on the six sides, which means that $u=\partial v/\partial x= \partial w/\partial x = 0$ on the two $y$-$z$ planes, $v= \partial u/\partial y =  \partial w/\partial y = 0$ on the two $x$-$z$ planes, and $w= \partial u/\partial z = \partial v/\partial z = 0$ on the two $x$-$y$ planes of the computational box boundaries, where $u$, $v$ and $w$ are the displacements along the $x$, $y$ and $z$ axes, respectively. While this set of boundary conditions is not realistic for SOFC material environment, we chose it for the demonstration purpose in order to avoid overlaps with a future publication of physically based SOFC simulations.

\begin{figure}[htb]
\begin{center}
\includegraphics[width=1\textwidth]{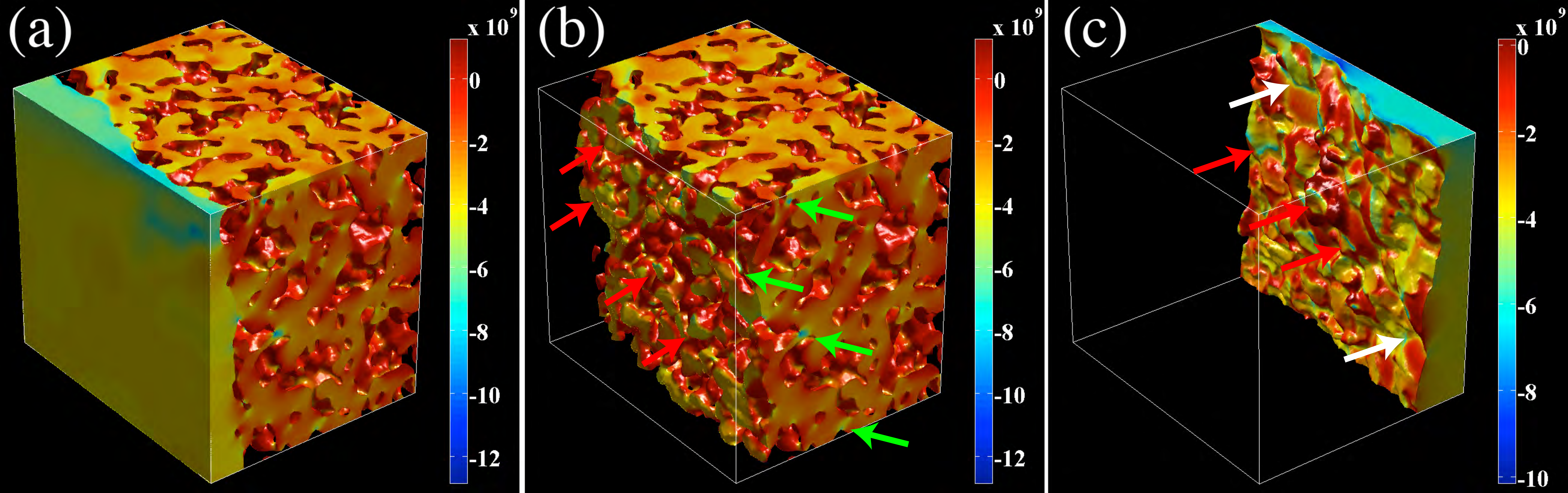}
\caption{The mean stresses resulting from thermal expansion in (a) the entire solid phase, (b) the cathode phase, and (c) the electrolyte phase after rotating the volume 180$^\circ$ around the $z$-axis. The unit of stress is Pa.} \label{TherStress}
\end{center}
\end{figure}

Shown in Fig.~\ref{TherStress}(a) are the calculated mean stress distributions resulting from thermal expansion in a confined sample. The mean stress is defined by: $\sigma_m = (\sigma_{xx}+\sigma_{yy}+\sigma_{zz})/3$, where the stress components are calculated according to the method provided in Appendix \ref{ME_solverA}. Here, we choose mean stress to illustrate the effective pressure in the solid. A negative mean stress indicates that the region is under compression. Despite a complicated stress distribution observed because of the complex geometry, the overall magnitude of the mean stress is roughly between 2 and 4 GPa, which can be roughly estimated by the product of Young's modulus and the thermal expansion with an enhancement resulting from the porosity of the solid. Additionally, an overall larger stress in the GDC phase is observed, reflecting the GDC phase is harder than the LSC phase. Figure \ref{TherStress}(b) shows the mean stress in the LSC phase and Fig.~\ref{TherStress}(c) shows the mean stress on the GDC surface after rotating the volume 180$^\circ$ around the $z$-axis.  Three types of stress enhancements can be observed in the simulation result. At the cathode-electrolyte contacting surfaces, stress is enhanced because of the mismatch of thermal expansion and elastic constants between the two materials; see the red arrows in Figs.~\ref{TherStress}(b) and (c). The second is the concentrated stress observed at the grooves on the electrolyte surface (not contacting the cathode), as shown by the white arrows in Fig.~\ref{TherStress}(c). The third type is the stress concentration effect at the bottlenecks in the cathode phase, where the stresses are roughly larger by a factor of three to four compared to the overall value, as shown by the green arrows in Fig.~\ref{TherStress}(b). The simulation results demonstrate that the smoothed boundary method can properly capture the linear elasticity and the geometric effects of the system based on a diffuse-interface defined geometry.

\subsection{Phase Transformations in the Presence of a Foreign Surface} \label{Contact angle AC} 

\begin{figure}[htbp]
\begin{center}
\includegraphics[width=1\textwidth]{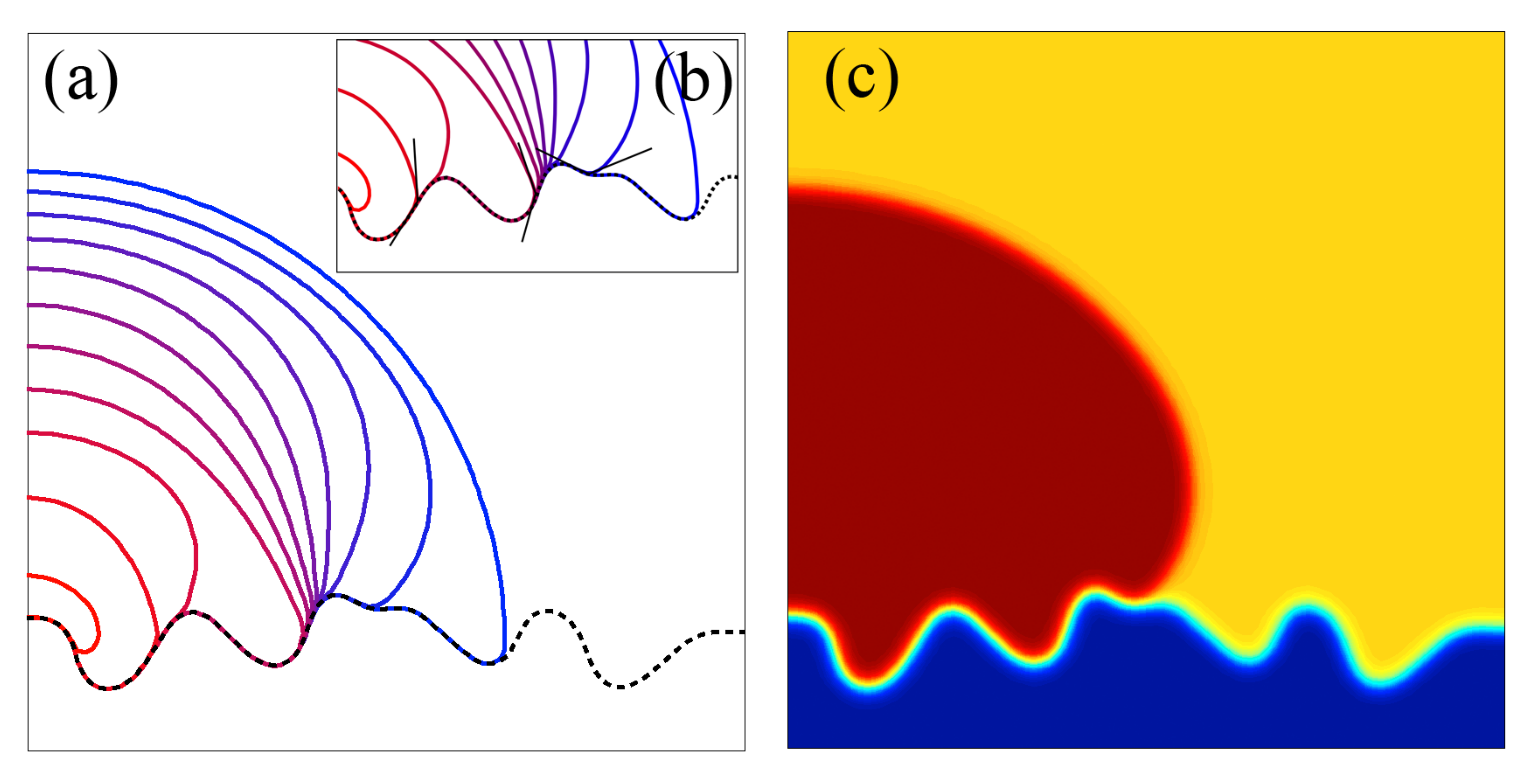}
\caption{(a) The evolution of an evaporating droplet on a rough surface (dashed line) governed by Allen-Cahn dynamics. The contact angle between the droplet and the surface is imposed at 135$^\circ$. Solid curves of various colors represent the profile of the droplet at different times. The outermost blue line represents the initial state, and the innermost red line represents the final state (recorded before complete evaporation in the simulation); the lines are plotted at time intervals of 270 dimensionless time units. The velocity of the three-phase boundary is greatly affected by the surface profile. (b) A magnified view of the three-phase boundary, showing that the contact angle is accurately set. The angle made by the thin black lines is 135$^\circ$. (c) The order and domain parameters are shown to illustrate the diffuse natures of the interface and boundary (recorded at $t = 270$).}\label{droplet}
\end{center}
\end{figure}

The Allen-Cahn equation describes the dynamics of a nonconserved order parameter, which can be taken as a model for the ordering of magnetic moments \cite{Chen:2002} and diffusionless phase transformations that involve only changes in crystalline order \cite{Chen:2002}. This equation can also be used as a model for evaporation-condensation dynamics \cite{Chen:2002,Emmerich:2003}. Here, we use the Allen-Cahn equation to examine the evaporation of a droplet on a rough surface. The domain parameter was given a ripple-like feature, as shown in Fig.~\ref{droplet}, having a hyperbolic-tangent-like profile continuously transitioning through the substrate surface ($\psi=1$ above the surface, and $\psi=0$ below the surface). The droplet phase was placed on top of the boundary, and its shape was evolved by the smoothed boundary formulation of the Allen-Cahn equation, Eq.~\eqref{AC-1}, using the standard central difference scheme in space and an Euler explicit scheme in time. The simulation was performed in two dimensions, using the parameters $\Delta x = 1$, $M=1$ and $\epsilon=1$, with a domain size of $L_x=100$ and $L_y=100$. The contact angle was set at 135$^\circ$, and a zero-gradient boundary condition of $\phi$ is set at the computational box boundaries.

The evolution of the droplet surface as it evaporates is illustrated in Fig.~\ref{droplet}(a) as a contour ($\phi=0.5$) plotted at equal intervals of 270 dimensionless time units. The color change from blue to red indicates various times from the initial to the final stages, respectively. As the surface evolves, it is clear that the contact angle is maintained, as shown in Fig.~\ref{droplet}(b). The dynamics of the motion of the three-phase boundary are interesting in that the velocity changes depending on the angle of the surface (with respect to the horizontal axis), which can be inferred from the change in the density of the contours. Because the interfacial energy is assumed to be constant, the droplet would prefer to have a circular cap shape. However, the contact angle imposes another constraint at the three-phase boundary. When the orientation of the surface is such that both of these conditions are nearly met, the motion of the three-phase boundary is slow as the droplet evaporates. When the orientation becomes such that the shape of the droplet near the three-phase boundary must be deformed (compared to the circular cap), the three-phase boundary moves very quickly, which leads to an unsteady motion of the three-phase boundary. In contrast, at the top of the droplet far from the substrate, the curvature is barely affected by the angle of the substrate surface; thus, the phase boundary there moves at a speed inversely proportional to the radius.

\subsection{Motion of a Droplet Due to Unbalanced Surface Tensions} \label{Contact angle CH}

In another example application, we modeled a self-propelled droplet. Here, two different contact-angle boundary conditions are imposed on the right and left sides of the droplet placed on a flat surface. The smoothed boundary formulation of the Cahn-Hilliard equation, Eq.~\eqref{SBM-CH-3}, is used with $J_{n} = 0$ in this simulation. The domain sizes are $L_x=240$ and $L_y=60$. The parameters and computational box boundary condition are the same as in Section \ref{Contact angle AC}. The contact angle on the right side of the droplet is set to 45$^\circ$ and that on the left side to 60$^\circ$ by imposing position-dependent boundary conditions. Note that this setup is equivalent to the situation in which the substrate-environment, droplet-substrate and droplet-environment surface energies satisfy the conditions of Young's equation:
\begin{subequations}
\begin{equation}
\gamma_\text{se}-\gamma_\text{sd} = \gamma_\text{de} \cos{60^{\circ}} ~~\text{for the left side},
\end{equation}
\begin{equation}
\gamma_\text{se}-\gamma_\text{sd} = \gamma_\text{de} \cos{45^{\circ}}~~\text{for the right side},
\end{equation}
\end{subequations}
where $\gamma_\text{se}$, $\gamma_\text{sd}$ and $\gamma_\text{de}$ are the interfacial energies of the substrate-environment, droplet-substrate and droplet-environment interfaces, respectively. Therefore, this model can be used to simulate a case where the surface energies are spatially and/or temporally dependent on other fields, such as surface temperature or surface composition. This specific case applies when the wetted substrate behind the droplet have a higher interfacial energy than the pristine substrate, as in Ref. \cite{Tersoff:2009}.

\begin{figure}[htbp]
\begin{center}
\includegraphics[width=1\textwidth]{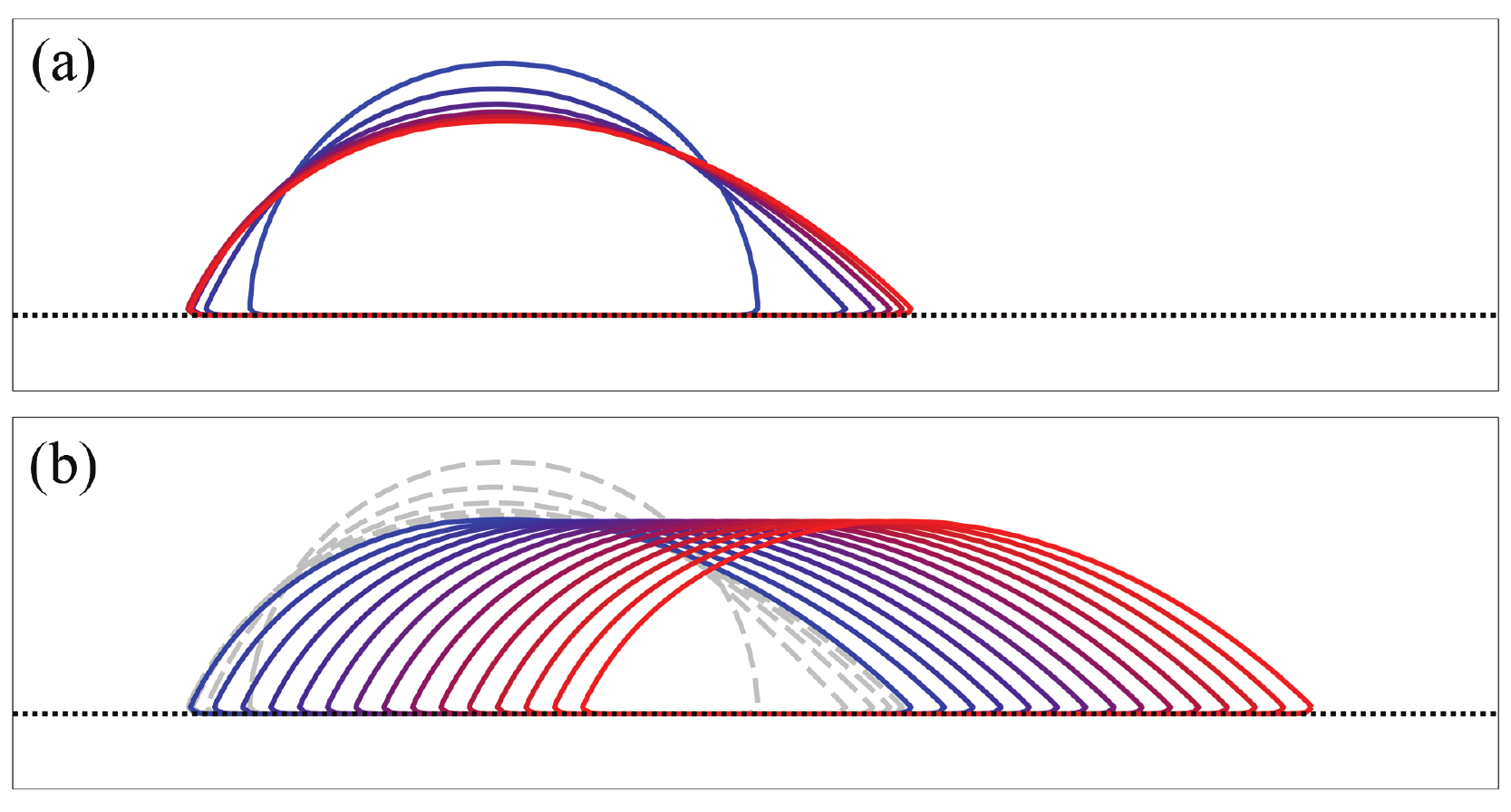}
\caption{A self-propelled droplet driven by unbalanced interfacial energies. The evolution is modeled by the Cahn-Hilliard equation with two different contact-angle boundary conditions on each side of the droplet. (a) The droplet shape changes during the relaxation period. The color contours are plotted at time intervals of $2\times10^{4}$ in dimensionless time units. (b) Droplet motion along the substrate surface. The color contours are plotted at time intervals of $1\times10^{5}$ in dimensionless time units. The droplet moves at constant speed at steady state.}\label{Self-propelled droplet}
\end{center}
\end{figure}

The evolution of the droplet surface is illustrated in Fig.~\ref{Self-propelled droplet}. The droplet initially has the shape of a hemisphere, with a 90$^\circ$ contact angle with the substrate surface. The early evolution is marked by the evolution of the droplet shape as it relaxes to satisfy the contact-angle boundary condition, as seen in Fig.~\ref{Self-propelled droplet}(a). The droplet then begins to accelerate. Once the contact angle reaches the prescribed value, it is maintained as the droplet moves toward the right; see Fig.~\ref{Self-propelled droplet}(b). In the steady state, the droplet moves at constant speed without other effects present. Such motions of droplets have been observed and explained as a result of an unbalanced surface tension between the head portion (with a nonwetting surface) and tail portion (with a wetting surface) because of the resulting spatially varying composition and composition-dependent surface energy \cite{Tersoff:2009}.  

\begin{figure}[htbp]
\begin{center}
\includegraphics[width=1\textwidth]{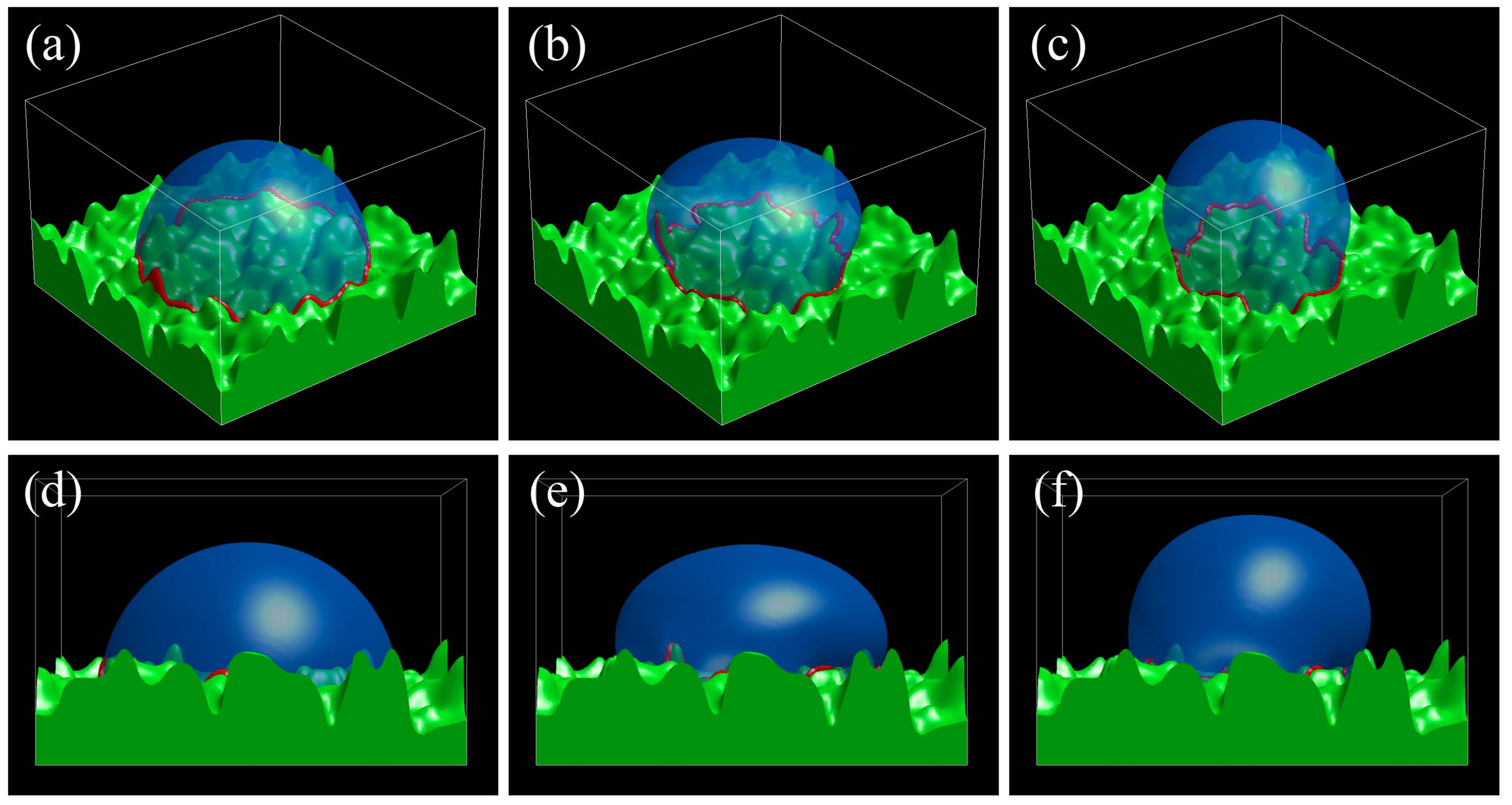}
\caption{A droplet relaxing toward its equilibrium shape. The evolution is modeled by the Cahn-Hilliard equation with a contact angle of 135$^\circ$ to the irregular substrate surface: (a) initial ($t=0$), (b) intermediate ($t=3\times10^{3}$), and (c) equilibrium ($t=2.35\times10^{4}$) states. The three-phase boundaries are delineated in red. Side views of the droplet are shown in (d), (e) and (f), corresponding to (a), (b) and (c), respectively. } \label{Relaxing_droplet}
\end{center}
\end{figure}

Figure \ref{Relaxing_droplet} shows the relaxation of an initially hemispherical droplet on an irregular substrate surface in a 3D simulation. The contact-angle boundary condition imposed at the three-phase boundary is 135$^\circ$. The computational box sizes are $L_x=L_y=120$ and $L_z= 80$. As shown here, the droplet changes its shape to satisfy the imposed contact angle, and the droplet evolves into a shape for which the total surface energy is minimized. The behavior favoring dewetting imposed by the contact angle ($\theta > 90^{\circ}$) is properly reflected in the lifting of the droplet, as shown in Figs.~\ref{Relaxing_droplet}(a)--(c) and (d)--(f). During this relaxation process, the three-phase boundary moves toward the center as the droplet-substrate contacting area decreases, as shown in Fig.~\ref{Relaxing_droplet}(a)--(c). This model and numerical method has been applied to simulate a nickel particle coarsening process in the complex channel within supporting porous ceramic microstructure (consisting of yttria-stabilized zirconia) in SOFC anodes, and to estimate the degradation of the anode material during SOFC operation \cite{Chen:2010}.

\section{Discussion and Conclusions} 

In this work, we demonstrated a generalized formulation of the smoothed boundary method. This method allows Neumann, Dirichlet, or mixed boundary conditions to be imposed on a diffuse interface to solve partial differential equations within the region where the domain parameter $\psi$ uniformly equals 1. The derivation of the method, as well as its implementation, is straightforward. The method can be used to solve differential equations numerically without complicated and time-consuming structural meshing of the domain of interest, as the domain boundary is specified by a spatially varying function. Instead, any grid system, including a regular Cartesian grid system, can be used with this method.

This smoothed boundary approach is flexible in coupling multiple differential equations. In Section \ref{SurfDiffFormulation}, we demonstrated how this method can be used to couple bulk diffusion with surface reaction-diffusion into a single equation while the two equations serve as complementary boundary conditions. In principle, this method can be used to couple multiple differential equations in different regions defined by different domain parameters. For example, if the physics within a domain defined by $\psi_i = 1$ are governed by a differential equation $H_i$, the overall phenomenon will be then represented by $H = \sum_{i} \psi_i H_i$, where the subscript `$i$' denotes the $i$th domain and $\sum_{i} \psi_i = 1$ represents the entire computational box. When sharing the diffuse interfaces between domains, the physical quantities can be interconnected as boundary conditions for each equation in each domain. Therefore, this method could be used to simulate coupled multiphysical and/or multiple-domain problems such as fluid-solid interaction phenomena or diffusion in multi-material polycrystalline solids.

We further demonstrated the capability of applying the smoothed boundary method to moving boundary problems in Section \ref{Kirkendall-Effect-Diffusion}. When the locations of domain boundaries are updated by a phase-field-type dynamics such that the domain parameters remain uniformly at 1 and 0 on either side of the interface, the smoothed boundary method can be conveniently employed to solve partial differential equations with moving boundaries. In addition to the phase-field-type dynamics, the smoothed boundary method is also applicable to moving boundary problems implementing the level set method \cite{Aland:2010,Teigen:2011,Yu:2011}, with the domain parameter obtained simply by taking the hyperbolic tangent of the distance function.

In addition to Neumann and Dirichlet boundary conditions, we also showed the capability of the smoothed boundary method for specifying contact angles between phase boundaries and domain boundaries (Sections \ref{ContAngBC-valid-1}, \ref{Contact angle AC} and \ref{Contact angle CH}). This type of boundary condition is difficult to impose using conventional sharp interface models.

Although the smoothed boundary method has many advantages, as shown in the results in Section \ref{Validation} and in the derivations in Appendix \ref{SBM_proof}, the nature of the diffuse interface inevitably introduces an error proportional to the interfacial thickness because we expand an originally zero-thickness boundary into a finite thickness interface. This spread of interface also leads to another error source depending on the resolution of the rapid transition of the domain parameter across the interfacial region. When numerically solving the smoothed boundary formulated equations, properly capturing the gradient of the domain parameter across the interface becomes very important. Based on our experience, 3 -- 5 grid points are necessary to properly resolve the diffuse interfaces while ensuring that the errors are well-controlled. Moreover, when solving time-dependent equations, one singularity occurs because of the terms $1/\psi$ and $1/\psi^2$ used to impose the Neumann and Dirichlet boundary conditions, respectively. In practice, a small value is necessarily added to $\psi$ to avoid singularity resulting from division by zero. In our simulations, the errors were quickly saturated when the value added to $\psi$ was selected to be smaller than $1\times10^{-5}$ when 3 -- 5 grid spacings were used for the interfacial regions, which suggests it is unnecessary to select a smaller value for the singularity-control term. However, when solving time-independent equations, such as the mechanical equilibrium equation and the steady-state diffusion equation, there are no singular terms in the equations. The small additional term is then merely used to condition the matrix solver. In this case, it can be on the order of numerical precision, such as $1\times10^{-16}$. 

Based on the general nature of the derivation, the smoothed boundary method is applicable to generalized boundary conditions, including time-dependent boundary values important for simulating the evolution of many physical systems. Because the domain boundaries are not specifically defined in the smoothed boundary method, this method can be applied to almost any geometry as long as it can be defined by the domain parameter. The developed method is thus a very powerful and convenient technique for solving differential equations in complex geometries that are often difficult and time-consuming to structurally mesh. As three-dimensional image-based calculations are increasingly prevalent in scientific and engineering research fields \cite{Langer:2001,Thornton:2008,Spanos:2008}, where voxelated data from serial scanning or sectioning are often utilized and are difficult to render as meshes, the smoothed boundary method is expected to be widely employed to simulate and study physics in complex geometries defined by 2D pixelated and 3D voxelated data with a simple process of smoothing the domain boundaries.

\textbf{Acknowledgements}: HCY and KT thank the National Science Foundation for financial support under Grant Nos. 0511232, 0502737, and 0854905. HYC and KT thank the National Science Foundation for financial support under Grant Nos. 0542619 and 0907030. KT also acknowledges the support of the National Science Foundation under Grant No. 0746424.  The authors thank John Lowengrub, Axel Voigt, Xiaofan Li, Anton Van der Ven and James Warren for valuable discussions and comments. The authors also thank Scott Barnett and Stuart Adler for providing the experimental 3D microstructures used in the demonstration.

\begin{appendix}

\section{Proof of Convergence for Neumann and Dirichlet Boundary Conditions for the Diffusion Equation} \label{SBM_proof} 

To demonstrate that the smoothed boundary formulated diffusion equation satisfies the assigned Neumann boundary condition (specifying the boundary flux or normal gradient), we use the one-dimensional version of Eq.~\eqref{SBM-FSL2} without loss of generality for cases with higher dimensions. By reorganizing terms and integrating over the interfacial region, we obtain:
\begin{equation} \label{SBM-Nm-prf-01}
\int_{a_i-\xi/2}^{a_i+\xi/2}\psi \left( \frac{\partial C}{\partial t} - S \right) dx = \left. \psi D \frac{\partial C}{\partial x} \right|_{a_i - \xi/2}^{a_i + \xi/2} - \int_{a_i - \xi/2}^{a_i + \xi/2} \left| \frac{\partial \psi}{\partial x} \right| DB_{N} dx,
\end{equation} 
where $a_i-\xi/2 < x < a_i+\xi/2$ is the region of the interface and $\xi$ is the thickness of the interface. Following Refs.~\cite{Bueno-Orovio:2006b,Bueno-Orovio:2006c,Kockelkoren:2003,Buzzard:2007}, we introduce the mean value theorem of integrals, which states that for a continuous function $g(x)$ there exists a constant value, $h_0$, such that $\min{g(x)} <  \int_{p}^{q} g(x) dx /(q-p) = h_0 < \max{g(x)}$, where $p<x<q$. By eliminating the second terms on the right-hand sides of Eqs.~\eqref{SBM-FSL2} and \eqref{SBM-Nm-prf-01}, the no-flux boundary condition can be imposed ($B_N = 0$); the resulting equation is similar to those proposed in Refs. \cite{Bueno-Orovio:2006a,Bueno-Orovio:2006b,Bueno-Orovio:2006c,Kockelkoren:2003,Buzzard:2007}. However, here we retain the term to maintain the generality of the method. Therefore, the analysis presented herein leads to an extension of the original method that greatly extends its applicability.

Because the function on the left-hand side of Eq.~\eqref{SBM-Nm-prf-01} is continuous and finite within the interfacial region, we can relate its value to the interfacial thickness by $h_0 \xi$, according to the mean value theorem of integrals. Using the conditions that $\psi = 1$ at $x=a_i + \xi/2$ and $\psi = 0$ at $x=a_i - \xi/2$, the first term on the right-hand side of Eq.~\eqref{SBM-Nm-prf-01} is written as $D(\partial C/\partial x)_{a_i+\xi/2}$. Because $| \partial \psi / \partial x | = 0$ for $x < a_i - \xi/2$ or $x > a_i+\xi/2$, the bounds of the integral can be extended to $-\infty$ and $\infty$. Therefore, we can rewrite Eq.~\eqref{SBM-Nm-prf-01} as 
\begin{equation} \label{SBM-Nm-prf-02}
h_0 \xi = D \frac{\partial C}{\partial x} \bigg|_{a_i+\xi/2} - \int_{-\infty}^{+ \infty} \bigg| \frac{\partial \psi}{ \partial x} \bigg| D B_{N} dx,
\end{equation}
and by taking the limit of this expression as $\xi \rightarrow 0$, we obtain:
\begin{equation} \label{SBM-Nm-prf-03}
\begin{split}
\left. D \frac{\partial C}{\partial x} \right|_{a_i} =  \int_{-\infty}^{+ \infty} \delta(x-a_i) DB_N dx =  DB_N \bigg|_{a_i},
\end{split}
\end{equation}
where $\partial C/\partial x|_{a_i+\xi/2} \cong \partial C/\partial x|_{a_i}$ and $\lim_{\xi \rightarrow 0} |\partial \psi / \partial x | = \delta(x-a_i)$ when $\psi$ takes the form of a hyperbolic tangent function and $\delta(x-a_i)$ is the Dirac delta function. The Dirac delta function has the property that $\int_{-\infty}^{+\infty} \delta(x-a_i) f(x) dx = f(a_i)$, providing the second equality in Eq.~\eqref{SBM-Nm-prf-03}. Therefore, Eq.~\eqref{SBM-Nm-prf-03} clearly shows that the smoothed boundary method recovers the Neumann boundary condition at the boundary when the thickness of the diffuse boundary approaches zero. This convergence has been observed for both stationary and moving boundaries \cite{Kockelkoren:2003,Aland:2010,Teigen:2011}. 

To demonstrate the convergence of the solution at the boundaries to the specified boundary value, we again use a one-dimensional version of the smoothed boundary formulated equation. Integrating Eq.~\eqref{SBM-FSL3} over the interfacial region and reorganizing terms, we obtain:
\begin{equation}\label{Int-Dirich-01}
\int_{a_i-\xi/2}^{a_i+\xi/2} \left[ \psi^2 \frac{\partial C}{\partial t} - \psi \frac{\partial}{\partial x} \left( \psi D \frac{\partial C}{\partial x} \right) - \psi^2 S \right] dx = -\int_{a_i-\xi/2}^{a_i+\xi/2} D \bigg(\frac{\partial \psi}{\partial x}\bigg) \left[ \frac{\partial \psi C}{\partial x}  - B_D \frac{\partial \psi}{\partial x} \right] dx.
\end{equation}
Similar to the derivation of Eq.~\eqref{SBM-Nm-prf-03}, the left-hand side of Eq.~\eqref{Int-Dirich-01} is proportional to the interfacial thickness and approaches zero in the limit of $\xi \rightarrow 0$. On the right-hand side of Eq.~\eqref{Int-Dirich-01}, the gradient of $\psi$ approaches the Dirac delta function, $\delta(x-a_i)$, as the interface thickness approaches zero. Therefore, we can reduce Eq.~\eqref{Int-Dirich-01} to $\lim_{\xi \rightarrow 0} h_0 \xi = -D [ \partial (\psi C)/\partial x - B_D \partial \psi /\partial x ]$ in the limit $\xi \rightarrow 0$. By integrating over the interfacial region again, we obtain:
\begin{equation} \label{Int-Dirich-03}
- \lim_{\xi \rightarrow 0} \frac{h_0 \xi^2}{D} = C \bigg|_{a_i+\xi/2} - \int_{a_i-\xi/2}^{a_i+\xi/2} B_D  \frac{\partial \psi}{\partial x} dx,
\end{equation}
which gives $C |_{a_i} = B_D |_{a_i}$ in the limit of $\xi \rightarrow 0$ because $C |_{a_i+\xi/2} \cong C |_{a_i}$ and $\lim_{\xi \rightarrow 0} (\partial \psi / \partial x ) = \delta(x-a_i)$. Therefore, the smoothed boundary formulation recovers the specified Dirichlet boundary condition: $C= B_D$ at $x=a_i$. 

\section{Surface Laplacian Operator and Alternating Direction Line Relaxation (ADLR) Method for Solving Coupled Surface Diffusion, Reaction and Bulk Diffusion Equation} \label{SurLap_A2} 

The surface gradient operator is defined by:
\begin{equation} \label{SG1}
\nabla_\text{s} = (\mathbf{I}-\mathbf{n} \otimes \mathbf{n}) \nabla = \left(
\begin{array}{ccc}
1-n_1 n_1&  -n_1n_2 & -n_1 n_3 \\ -n_2 n_1 & 1-n_2 n_2 & -n_2 n_3 \\ -n_3 n_1 & -n_3 n_2 & 1-n_3 n_3
\end{array}
\right) \left[
\begin{array}{c}
\partial /\partial x_1 \\ \partial /\partial x_2 \\ \partial /\partial x_3
\end{array}
\right], 
\end{equation}
where $n_i$ is the $i$th component of the inward unit normal vector (here, $i$ = 1, 2 and 3, corresponding to the $x$, $y$ and $z$ directions, respectively). In tensor notation, this operator is written as $\nabla_\text{s} = m_{ij} \partial/ \partial x_j$. The repeated indices indicate summation over the index. The coefficients $m_{ij}$ are related to the surface unit normal by $m_{11} = 1-n_1 n_1$, $m_{22} = 1-n_2 n_2$, $m_{33} = 1- n_3 n_3$, $m_{12} = m_{21} = -n_1 n_2$, $m_{13} = m_{31} = -n_1 n_3$, and $m_{23} = m_{32} = -n_2 n_3$. The surface Laplacian operator is defined by the surface divergence of the surface gradient: 
\begin{equation} \label{SL2}
\nabla_\text{s}^2 = \nabla_\text{s} \cdot \nabla_\text{s} = m_{ij} \frac{\partial}{\partial x_j} \bigg( m_{ik} \frac{\partial}{\partial x_k} \bigg).
\end{equation}
The scalar surface Laplacian is a sum of nine second-order-partial-differential-operator terms (where $j = k$) and 18 mixed-partial (cross) terms of the differential operator (where $j \ne k$). 

To solve Eq.~\eqref{SBM-CxMs-1}, which includes the surface Laplacian operator, we use an ADLR method in a second-order central difference scheme. We first separate the surface Laplacian operator in Eq.~\eqref{SL2} into a term involving pure second-order partial derivatives (``diagonal'') and another containing mixed partials (``cross''): $\nabla_\text{s}^2 = \nabla_{\text{diag}}^2 + \nabla_{\text{cross}}^2$, where:
\begin{equation}
\nabla_{\text{diag}}^2 = \sum_{i=1}^3 \sum_{j=1}^3 m_{ij}\frac{\partial}{\partial x_j} \bigg( m_{ij} \frac{\partial}{\partial x_j} \bigg)~~\text{and}~~\nabla_{\text{cross}}^2  = \sum_{i=1}^3 \sum_{j=1}^3 \sum_{k=1, k\ne j}^3 m_{ij} \frac{\partial}{\partial x_j} \bigg( m_{ik} \frac{\partial}{\partial x_k} \bigg).
\end{equation}
Thus, Eq.~\eqref{SBM-CxMs-1} is rearranged to $[ \nabla \cdot (\psi D_\text{b} \nabla) - |\nabla \psi | ( \kappa  - D_\text{s} \nabla_{\text{diag}}^2) ]  \tilde{C} = ( \mathrm{i} \omega \psi - | \nabla \psi | D_\text{s} \nabla_{\text{cross}}^2 ) \tilde{C}$, where the three axial components of the $\nabla \cdot ( \psi D_\text{b} \nabla) $ operator and the ``diagonal''  terms in the $\nabla_{\text{diag}}^2$ operator can be discretized by  central difference schemes similar to:
\begin{equation} \label{DiagDiffuOp-1}
\frac{\partial}{\partial x} \bigg( \psi \frac{\partial C}{\partial x} \bigg) = \frac{1}{\Delta x} \bigg( \psi_{i+1/2,j,k} \frac{C_{i+1,j,k}-C_{i,j,k}}{\Delta x} - \psi_{i-1/2,j,k} \frac{C_{i,j,k}-C_{i-1,j,k}}{\Delta x} \bigg),
\end{equation}
and $\psi_{i+1/2,j,k} = (\psi_{i+1,j,k}+\psi_{i,j,k})/2$. The discretization scheme along the $y$ and $z$ axes can be similarly obtained. Therefore, the operator $[ \nabla \cdot (\psi D_\text{b} \nabla) - |\nabla \psi | ( \kappa  - D_\text{s} \nabla_{\text{diag}}^2) ]$ can be lumped into an equivalent Helmholtz operator containing only second-order partial derivatives of the neighboring six points along the $x$, $y$ and $z$ axes and a coefficient term $| \nabla \psi | \kappa$ at the center point. Additionally, the 18 terms in the $ \nabla_{\text{cross}}^2$ operator can be calculated using a discretization scheme. For example, the $\partial ( m \partial C / \partial y)  \partial x$ can be discretized as:
\begin{equation} \label{CrossDiffuOp-1}
\frac{\partial}{\partial x} \bigg( m \frac{\partial C}{\partial y} \bigg) = \frac{1}{2 \Delta x} \bigg( m_{i+1,j,k} \frac{C_{i+1,j+1,k} - C_{i+1,j-1,k}}{2 \Delta y} - m_{i-1,j,k} \frac{C_{i-1,j+1,k} - C_{i-1,j-1,k}}{2 \Delta y} \bigg),
\end{equation}
and similarly for other components. As a result, the equation is represented by $\mathcal{L} \tilde{C} = \mathcal{S}$, where $\mathcal{L}$ is a linear Helmholtz-like operator and $\mathcal{S}$ is calculated as $( \mathrm{i} \omega \psi - | \nabla \psi | D_\text{s} \nabla_{\text{cross}}^2 ) \tilde{C}$. This equation can be solved using an ADLR solver \cite{VanDeVelde:1994,Hofhaus:1996} by decomposing the Helmholtz operator into the three axial directions:
\begin{subequations} \label{ADI-Solve-0}
\begin{equation} \label{ADI-Solver-1}
\mathcal{L}_{i+1,j,k} \tilde{C}_{i+1,j,k}^{(n+1/3)} - \mathcal{W}_{i,j,k}\tilde{C}_{i,j,k}^{(n+1/3)} + \mathcal{L}_{i-1,j,k} \tilde{C}_{i-1,j,k}^{(n+1/3)} = \mathcal{S}^{(n)} - \mathcal{L}_{yy} \tilde{C}_{yy}^{(n)} - \mathcal{L}_{zz} \tilde{C}_{zz}^{(n)}, 
\end{equation}
\begin{equation} \label{ADI-Solver-2}
\mathcal{L}_{i,j+1,k} \tilde{C}_{i,j+1,k}^{(n+2/3)}  - \mathcal{W}_{i,j,k}\tilde{C}_{i,j,k}^{(n+2/3)}  + \mathcal{L}_{i,j-1,k} \tilde{C}_{i,j-1,k}^{(n+2/3)}  = \mathcal{S}^{(n)} - \mathcal{L}_{xx} \tilde{C}_{xx}^{(n+1/3)}  - \mathcal{L}_{zz} \tilde{C}_{zz}^{(n+1/3)},  
\end{equation}
\begin{equation} \label{ADI-Solver-3}
\mathcal{L}_{i,j,k+1} \tilde{C}_{i,j,k+1}^{(n+1)}  - \mathcal{W}_{i,j,k}\tilde{C}_{i,j,k}^{(n+1)} + \mathcal{L}_{i,j,k-1} \tilde{C}_{i,j,k-1}^{(n+1)} = \mathcal{S}^{(n)} - \mathcal{L}_{xx} \tilde{C}_{xx}^{(n+2/3)}  - \mathcal{L}_{yy} \tilde{C}_{yy}^{(n+2/3)},  
\end{equation}
\end{subequations}
where $\mathcal{L}_{xx} \tilde{C}_{xx} = \mathcal{L}_{i+1,j,k} \tilde{C}_{i+1,j,k} + \mathcal{L}_{i-1,j,k} \tilde{C}_{i-1,j,k}$, $\mathcal{L}_{yy} \tilde{C}_{yy}= \mathcal{L}_{i,j+1,k} \tilde{C}_{i,j+1,k} + \mathcal{L}_{i,j-1,k} \tilde{C}_{i,j-1,k}$, $\mathcal{L}_{zz} \tilde{C}_{zz} = \mathcal{L}_{i,j,k+1} \tilde{C}_{i,j,k+1}+ \mathcal{L}_{i,j,k-1} \tilde{C}_{i,j,k-1}$, and the superscript $(n)$ denotes the $n$th iterative step. Within each iterative step, we first solve along the $x$ direction using Eq.~\eqref{ADI-Solver-1}, for which a simple tridiagonal matrix solver is employed for each column. Similarly, Eqs.~\eqref{ADI-Solver-2} and \eqref{ADI-Solver-3} are solved along the $y$ and $z$ directions, respectively, with the updated value on the right-hand sides. The above procedure is repeated until the solution converges to its equilibrium value.

For the simulations of oxygen-vacancy diffusion in a cylinder in Section \ref{bulkSurf_cylinder}, we consider a cylindrical symmetry for the differential operator, in which the dimensions reduce to effectively 2D, such that the ``bulk'' term in Eq.~\eqref{SBM-FSL-BS1} becomes:
\begin{equation} \label{LapOp-1}
\nabla \cdot (\psi \nabla C ) = \frac{1}{r} \frac{\partial }{\partial r} \bigg( r \psi \frac{\partial C}{\partial r} \bigg) + \frac{\partial}{\partial z} \bigg( \psi \frac{\partial C}{\partial z} \bigg), 
\end{equation}
with only components in the radial and axial directions. Here, we have set $D_\text{b}$ at 1 for clarify of the derivation. The first term on the right-hand side can be rewritten and discretized using the central difference scheme as: 
\begin{equation} \label{2nd-Dr-Op-1}
\begin{split}
\frac{2}{\partial (r^2)} & \partial \bigg( r \psi \frac{\partial C}{\partial r} \bigg) = \frac{2}{r_{i+1/2,j}^2 - r_{i-1/2,j}^2} \bigg( r_{i+1/2,j} \psi_{i+1/2,j}  \frac{C_{i+1,j}-C_{i,j}}{r_{i+1,j}-r_{i,j}} - r_{i-1/2,j} \psi_{i-1/2,j}  \frac{C_{i,j}-C_{i-1,j}}{r_{i,j}-r_{i-1,j}} \bigg),
\end{split}
\end{equation}
where the subscript $i$ and $j$ denote the $i$th and $j$th grid points in the radial and axial directions, respectively. If the radial grid spacing is selected to be uniform, $r_{i+1,j}-r_{i,j} = r_{i,j}-r_{i-1,j} = \Delta r$. The second term  on the right-hand side of Eq.~\eqref{LapOp-1} can be discretized using the scheme provided in Eq.~\eqref{DiagDiffuOp-1}.

The surface Laplacian with cylindrical symmetry is given as:
\begin{equation} \label{SurfLap-Cyl-1}
\begin{split}
\nabla_\text{s}^2 C = & m_{rr} \frac{1}{r} \frac{\partial}{\partial r} \bigg( r m_{rr} \frac{\partial C}{\partial r} \bigg) + m_{rr} \frac{\partial}{\partial r} \bigg( m_{rz} \frac{\partial C}{\partial z} \bigg) + m_{rz} \frac{\partial}{\partial z} \bigg( m_{rr} \frac{\partial C}{\partial r} \bigg) + m_{rz} \frac{\partial}{\partial z} \bigg( m_{rz} \frac{\partial C}{\partial z} \bigg) + \\ & m_{zr} \frac{1}{r} \frac{\partial}{\partial r} \bigg( r m_{zr} \frac{\partial C}{\partial r} \bigg) + m_{zr} \frac{\partial}{\partial r} \bigg( m_{zz} \frac{\partial C}{\partial z} \bigg) + m_{zz} \frac{\partial}{\partial z} \bigg( m_{zr} \frac{\partial C}{\partial r} \bigg) + m_{zz} \frac{\partial}{\partial z} \bigg( m_{zz} \frac{\partial C}{\partial z} \bigg),
\end{split}
\end{equation}
where $m_{rr} = 1-n_r n_r$, $m_{rz} = m_{zr} = -n_r n_z$, and $m_{zz} = 1-n_z n_z$. The ``diagonal'' terms along the radial and axial directions can be discretized in a manner similar to Eqs.~\eqref{2nd-Dr-Op-1} and \eqref{DiagDiffuOp-1}, respectively, and the ``cross'' terms can be discretized as in Eq.~\eqref{CrossDiffuOp-1} for solving Eq.~\eqref{SBM-FSL-BS1} with an additional factor of $m$.

\section{Derivation of the Mechanical Equilibrium Equation} \label{A3} 

To perform the smoothed boundary formulation on the tensorial mechanical equilibrium equation, we multiply Eq.~\eqref{ME-2} by $\psi$ and use the mathematical identity $\psi (\partial H_{ij}/\partial x_j) = \partial( \psi H_{ij} )/\partial x_j - ( \partial \psi / \partial x_j ) H_{ij} $ to obtain:
\begin{equation}
\frac{\partial}{\partial x_j} \bigg[ \psi C_{ijkl} \frac{1}{2} \bigg( \frac{\partial u_k}{\partial x_l} + \frac{\partial u_l}{\partial x_k} \bigg) \bigg] - \frac{\partial \psi}{\partial x_j} C_{ijkl} \frac{1}{2} \bigg( \frac{\partial u_k}{\partial x_l} + \frac{\partial u_l}{\partial x_k} \bigg) = \frac{\partial}{\partial x_j} \bigg( \psi \rho C_{ijkl} \delta_{kl} \bigg) - \frac{\partial \psi}{\partial x_j} \rho C_{ijkl} \delta_{kl}.
\end{equation}
By collecting the terms associated with $\partial \psi / \partial x_j$, we obtain Eq.~\eqref{SBM-ME-1} as given in Section \ref{MechEquim-Derivation}.

To impose a Dirichlet boundary condition (a specified displacement) on the mechanical equilibrium equation, we multiply the left-hand side of Eq.~\eqref{ME-2} by $\psi^2$ to obtain:
\begin{equation}
\begin{split}
\psi^2 \frac{\partial}{\partial x_j} C_{ijkl}\frac{1}{2} \bigg( \frac{\partial u_k}{\partial x_l} + \frac{\partial u_l}{\partial x_k} \bigg) =  \psi \frac{\partial}{\partial x_j} \bigg[ \psi C_{ijkl} \frac{1}{2} \bigg( \frac{\partial u_k}{\partial x_l} + \frac{\partial u_l}{\partial x_k} \bigg) \bigg] -  \psi \frac{\partial \psi}{\partial x_j} C_{ijkl} \frac{1}{2} \bigg( \frac{\partial u_k}{\partial x_l} + \frac{\partial u_l}{\partial x_k} \bigg),
\end{split}
\end{equation}
where the second term on the right-hand side can be replaced by:
\begin{equation}
\begin{split}
\psi \frac{\partial \psi}{\partial x_j} C_{ijkl} \frac{1}{2} \bigg(\frac{\partial u_k}{\partial x_l}+\frac{\partial u_l}{\partial x_k} \bigg) =  \frac{\partial \psi}{\partial x_j} C_{ijkl} \frac{1}{2} \bigg[ \frac{\partial (\psi u_k)}{\partial x_l} + \frac{\partial (\psi u_l )}{\partial x_k} \bigg] -   \frac{\partial \psi}{\partial x_j} C_{ijkl} \frac{1}{2} \bigg( u_k \frac{\partial \psi}{\partial x_l} + u_l \frac{\partial \psi}{\partial x_k} \bigg),
\end{split}
\end{equation} 
according to the product rule:
\begin{equation}
\begin{split}
\frac{\partial \psi}{\partial x_j} C_{ijkl} \bigg[ \frac{\partial (\psi u_k)}{\partial x_l} + \frac{\partial (\psi u_l )}{\partial x_k} \bigg] = & \frac{\partial \psi}{\partial x_j} C_{ijkl} \bigg[ \bigg( \psi \frac{\partial u_k}{\partial x_l} + u_k \frac{\partial \psi}{\partial x_l}\bigg) + \bigg( \psi \frac{\partial u_l}{\partial x_k} + u_l \frac{\partial \psi}{\partial x_k}\bigg) \bigg] \\ = & \psi \frac{\partial \psi}{\partial x_j}  C_{ijkl} \bigg(\frac{\partial u_k}{\partial x_l}+\frac{\partial u_l}{\partial x_k} \bigg) + \frac{\partial \psi}{\partial x_j} C_{ijkl} \bigg( u_k \frac{\partial \psi}{\partial x_l} + u_l \frac{\partial \psi}{\partial x_k} \bigg).
\end{split}
\end{equation}
Thereby, we obtain Eq.~\eqref{SBM-ME-3} in Section \ref{MechEquim-Derivation}.

\section{Relation Used in the Derivation of Contact Angle Boundary Condition} \label{A4} 

Here, we multiply the equilibrium criterion of the phase field model by $\nabla \phi$ to obtain:
\begin{equation} \label{PF-eqm-1}
\frac{\partial f}{\partial \phi} \nabla \phi - ( \epsilon^2 \nabla^2 \phi) \nabla \phi = \nabla f - \frac{\epsilon^2}{2} \nabla (\nabla \phi)^2 = 0.
\end{equation} 
Integrating the above, we obtain $f - \epsilon^2 | \nabla \phi |^2/2=c_1$,
where $c_1$ is a constant of integration. In the phase field model, the order parameter remains at a uniform value in the bulk away from the interface; thus giving $| \nabla \phi | = 0$ in the bulk.  Therefore, $c_1$ is equal to the bulk value of $f$.  For convenience, we have taken the free energy at the bulk values to be zero, and therefore $c_1 = 0$, leading to $\nabla \phi = \sqrt{2f}/\epsilon$. However, the choice of the free energy value at the bulk is arbitrary, and therefore does not affect the result of the calculation as long as it is taken into account by replacing $f$ appearing in Eq.~\eqref{eqAngBCF} by $f-c_1$.  

\section{Smoothing Voxelated Data Using a Distance Function} \label{LevelSetDist} 

The experimentally obtained microstructure is typically provided in a form of a 3D array containing voxels of different values indicating different phases. To incorporate the voxelated data into the smoothed boundary formulation, we must convert the discrete voxelated array into a domain parameter profile that continuously transitions from one phase to another. Here, we employ the distance function method commonly used for initialization in the level set method \cite{Osher:2003,Park:2010}. First, we construct the sign function by assigning positive and negative values to the voxels in the solid and pore phases, respectively: $Sgn(\mathbf{x}) = 1$ for the solid phase and $Sgn(\mathbf{x}) = -1$ for the pore phase, where $\mathbf{x}$ is the position of a voxel. The distance function indicating the distance between the center of a voxel and the solid-pore interface is calculated by evolving the time-dependent equation $\partial \varphi(\mathbf{x},t) / \partial t = Sgn(\mathbf{x})(1-| \nabla \varphi(\mathbf{x},t) |)$ to its equilibrium. This process is numerically implemented by:
\begin{equation} \label{DistFun1}
\varphi{(\mathbf{x},t+\Delta t)} = \bigg\{
\begin{array}{c}
\varphi{(\mathbf{x},t)} + \Delta t [ Sgn(\mathbf{x})(1-| \nabla \varphi(\mathbf{x},t) |)]~~\text{if}~~\varphi{(\mathbf{x},t+\Delta t)}\cdot \varphi{(\mathbf{x},t)} > 0 \\
\varphi{(\mathbf{x},t)} + \Delta t [Sgn(\mathbf{x}) \upsilon ]~~\text{if}~~\varphi{(\mathbf{x},t+\Delta t)} \cdot \varphi{(\mathbf{x},t)} \le 0
\end{array},
\end{equation}
where $\upsilon$ is a small nonzero value. The second case above prevents interfaces from moving more than one grid spacing by requiring that the sign of the function remain the same as the initial value. The absolute value of the gradient of the distance function is calculated using a Godunov upwind scheme \cite{Sethian:1999,Sussman:1994,Osher:2003a}:
\begin{equation}
\begin{split}
|\nabla \varphi_{i,j,k} | = [ & \max(\max(D_x^+ \varphi_{i,j,k},0)^2,\max(-D_x^- \varphi_{i,j,k},0)^2) + \\ & \max(\max(D_y^+ \varphi_{i,j,k},0)^2,\max(-D_y^- \varphi_{i,j,k},0)^2) + \\ & \max(\max(D_z^+ \varphi_{i,j,k},0)^2,\max(-D_z^- \varphi_{i,j,k},0)^2) ]^{1/2},
\end{split}
\end{equation}
where $i$, $j$ and $k$ are the indices of the grid points along the $x$, $y$ and $z$ axes, respectively, and: 
\begin{equation}
\begin{split}
D_x^+ \varphi_{i,j,k} = (\varphi_{i,j,k}-\varphi_{i-1,j,k})/\Delta x,~ D_x^- \varphi_{i,j,k} = (\varphi_{i+1,j,k}-\varphi_{i,j,k})/\Delta x, \\
D_y^+ \varphi_{i,j,k} = (\varphi_{i,j,k}-\varphi_{i,j-1,k})/\Delta y,~ D_y^- \varphi_{i,j,k} = (\varphi_{i,j+1,k}-\varphi_{i,j,k})/\Delta y, \\
D_z^+ \varphi_{i,j,k} = (\varphi_{i,j,k}-\varphi_{i,j,k-1})/\Delta z,~ D_z^- \varphi_{i,j,k} = (\varphi_{i,j,k+1}-\varphi_{i,j,k})/\Delta z.
\end{split}
\end{equation} 
In practice, for the smoothed boundary method, the function must take the form of the distance function only near the interfacial regions, and therefore the convergence condition can be placed in these regions only (and not in the bulk far from interfaces) as long as the values in the bulk are sufficiently large in magnitude.

From the distance function, we obtain a domain parameter based on the experimentally acquired voxelated data by taking the hyperbolic tangent of the distance function, $\psi(\mathbf{x}) = \{1+\tanh [\varphi(\mathbf{x})/\zeta ]\}/2$,
where $\psi = 0.5$ coincides the location of the zero level set ($\varphi = 0$), $\psi = 1$ in the solid, $\psi = 0$ in the pore, and the value of $\zeta$ controls the thickness of the interface.

\section{Projection Method} \label{ProjMethod} 

To simulate Kirkendall-effect-induced deformation, we model the solid diffusion couple as a very viscous fluid that deforms in a quasi-steady-state manner, namely, creep flow. In contrast, the environmental phase surrounding the solid is treated as a nearly inviscid fluid. A simple way to implement this model is to define the viscosity coefficient as $\eta(\psi) = \bar{\eta}\psi+\upsilon$, where $\bar{\eta}$ is a constant viscosity coefficient for the solid phase and $\upsilon \ll \bar{\eta}$ is a small value used to avoid numerical instability. To solve the velocity field with a variable viscosity coefficient, we adopt a projection method \cite {Kim:2006a,Kim:2006b}, in which the divergence of the viscous stress tensor is decomposed into a linear part and a residual part, giving:
\begin{equation} \label{Projection-01}
\nabla \cdot \eta [ \nabla \mathbf{v} + (\nabla \mathbf{v})^T ]  = \Lambda \nabla \cdot [ \nabla \mathbf{v} + (\nabla \mathbf{v})^T ] + \mathbf{r_v}, 
\end{equation}
where $\Lambda$ is a constant scalar numerical parameter for the scheme (normally set between $0.5 \bar{\eta}$ and $\bar{\eta}$) and $\mathbf{r_v}$ is a vector residual. Using the identity that $\nabla \cdot [\nabla \mathbf{v} + (\nabla \mathbf{v})^T ] = \nabla^2 \mathbf{v}+ \nabla (\nabla \cdot \mathbf{v})$, where $\nabla \cdot \mathbf{v} = g_V $, and $\nabla^2 \mathbf{v} = \partial^2 v_i / \partial x_j \partial x_j$ is a vector containing the Laplacian of each velocity component, we rewrite Eq.~\eqref{NS-CH-1} as:
\begin{equation} \label{NS-CreepFlow-1}
-\nabla P + \Lambda \nabla^2 \mathbf{v}+  \mathbf{r_v} +\nabla \bigg( \Lambda-\frac{2\eta}{d} \bigg) g_V + \frac{1}{C_\text{a}} \mu \nabla \psi =  0.
\end{equation}
By taking the divergence of Eq.~\eqref{NS-CreepFlow-1}, applying the relation $\nabla \cdot ( \nabla^2 \mathbf{v}) = \nabla^2 (\nabla \cdot \mathbf{v}) = \nabla^2 g_V$ and rearranging the terms, we obtain the Poisson equation of the scalar pressure field, which serves as one of the two equations for the iterative scheme:
\begin{equation} \label{Poisson-P-1}
\nabla^2 P^{(n)} = \nabla \cdot \mathbf{r_v}^{(n-1)} + \nabla^2 \bigg( 2 \Lambda - \frac{2 \eta}{d} \bigg) g_V + \frac{1}{C_\text{a}} \nabla \cdot (\mu \nabla \psi ) = 0,
\end{equation}
where the superscript $(n)$ denotes the $n$th iterative step. The second and third terms on the right-hand side are fixed during an evolution time step while the values of the velocity and pressure are updated during iteration. 

For the velocity field, we reorganize Eq.~\eqref{NS-CreepFlow-1} to obtain the Poisson equation for the velocity component in each coordinate direction:
\begin{equation} \label{Poisson-V-1}
\nabla^2 \mathbf{v}^{(n)} = \frac{1}{\Lambda} \bigg[ \nabla P^{(n)} - \mathbf{r_v}^{(n-1)} - \nabla \bigg( \Lambda -\frac{2 \eta}{d}  \bigg) g_V - \frac{1}{C_\text{a}} \mu \nabla \psi \bigg] = 0.
\end{equation}
The residual vector $\mathbf{r_v}$ is calculated using Eq.~\eqref{Projection-01} and is updated during the iteration: $\mathbf{r_v}^{(n)} = \nabla \cdot \eta [\nabla \mathbf{v}^{(n)} + (\nabla \mathbf{v}^{(n)})^T ] - \Lambda \nabla^2 \mathbf{v}^{(n)} - \Lambda \nabla g_V$. The Poisson equations can be solved using an ADLR method similar to that described in Appendix \ref{SurLap_A2}, except that the Helmholtz operator is replaced here by a Laplacian operator. Within each time step for the deformation (Eq.~\eqref{NS-CH-3}) and diffusion (Eq.~\eqref{KE-Trad-SBM-1}) of the diffusion couple, the pressure and velocity fields are solved iteratively until the values of pressure and the velocity components converges.  The convergence criteria is set to be $e \le 1\times10^{-5}$, where $e$ is the relative error taken by dividing the root-mean-square difference between the values of two consecutive iterative steps by the average magnitude of the values in the previous step. The velocity field is then substituted into the advective terms in the order parameter and concentration evolution equations.

\section{Implicit Time Scheme} \label{ImplicitTime} 

Equation \eqref{KE-Rig-2} contains the coupled diffusion equations for two species and is constrained by a very small time step for the Euler explicit time scheme because of the large diffusivity, $D_{VV}$ (nearly $10^4$ times larger than $D_{BB}^V$). Thus, here we use a semi-implicit time scheme, similar to that presented in Ref.~\cite{Yu:2009}, to solve the vacancy diffusion equation and to significantly enhance numerical efficiency. In the time-discretized form, the scheme is given by:
\begin{equation} \label{XV-Imp-1}
\begin{split}
\frac{X_V^{(n+1)}-X_V^{(n)}}{\Delta t} - & \frac{\chi \bar{D}_{VV}^{(n)}}{\psi^{(n)}} \nabla \cdot \psi^{(n)} \nabla X_V^{(n+1)} = \\
& \frac{1}{\psi^{(n)}} \nabla \cdot \psi^{(n)} [ (D_{VV}^{(n)} - \chi \bar{D}_{VV}^{(n)}) \nabla X_V^{(n)}+ D_{VB}^{(n)} \nabla X_B^{(n)}] - \frac{K^{(n)}}{\psi^{2(n)}}, 
\end{split}
\end{equation}
where the superscript $(n)$ denotes the $n$th time step and $\chi$ is a weighting factor that can be optimized to increase numerical stability. The diffusivities, $D_{VV}$ and $D_{VB}$, are mole-fraction dependent quantities. The average diffusivity, $\bar{D}_{VV}$, is calculated from $D_{VV}$ over the solid region in which diffusion occurs. The diffusion equation for $B$ atoms, Eq.~\eqref{KE-Rig-2B}, and the Cahn-Hilliard equation, Eq.~\eqref{KE-Rig-Psi-1}, are solved using the Euler explicit time scheme, as the diffusivities and mobilities for these equations are much smaller.

\section{Mechanical Equilibrium Equation Solver} \label{ME_solverA} 

Here, we expand the generalized mechanical equilibrium equation, Eq.~\eqref{SBM-ME-2}, for a linear elastic and isotropic solid. In this case, the components of the elastic constant tensor are expressed by:
\begin{subequations}
\begin{equation}
\lambda_{11} = C_{1111} = C_{2222} = C_{3333},
\end{equation}
\begin{equation}
\lambda_{12} = C_{1122} = C_{2211} = C_{2233} = C_{3322} = C_{3311} = C_{1133},
\end{equation}
\begin{equation}
\begin{split}
\lambda_{44} =&  C_{1212} = C_{1221} = C_{2112} = C_{2121} = C_{2323} = C_{2332} \\= &C_{3223} = C_{3232} = C_{1313} = C_{1331} = C_{3113} = C_{3131}.
\end{split}
\end{equation}
\end{subequations}
The remaining elastic constant components vanish. For an isotropic solid, the Young's moduli are related to Lame constants by $E = \lambda_{12}(1+\nu)(1-2\nu)/\nu$, where $\nu$ is the Poisson's ratio; the shear modulus is given by $\lambda_{44} = \lambda_{12}(1-2\nu)/2\nu$, and the elastic constant $\lambda_{11}$ is given by $\lambda_{11} = \lambda_{12}+2 \lambda_{44}$. The pair, $\lambda_{12}$ and $\nu$, forms the set of Lame constants.

We use coordinate notation to replace the indices $i = 1$, $2$ and $3$ with $x$, $y$ and $z$, respectively. With a traction-free boundary condition on the solid surface ($N_i = 0$), the smoothed boundary formulated mechanical equilibrium equation, Eq.~\eqref{SBM-ME-2}, can be written out for the $x$, $y$ and $z$ directions as: 
\begin{subequations} \label{ME-ISO-1}
\begin{equation} \label{ME-I1-1}
\begin{split}
&\frac{\partial}{\partial x}\bigg[ \psi \lambda_{11} \bigg(\frac{\partial u}{\partial x} \bigg) \bigg] +
\frac{\partial}{\partial y}\bigg[ \psi \lambda_{44} \bigg( \frac{\partial u}{\partial y} \bigg) \bigg] +
\frac{\partial}{\partial z}\bigg[ \psi \lambda_{44} \bigg( \frac{\partial u}{\partial z} \bigg) \bigg] =
\frac{\partial}{\partial x} [ \psi \rho (\lambda_{11}+2 \lambda_{12}) ] - \\ 
&\frac{\partial}{\partial x}\bigg[ \psi \lambda_{12} \bigg( \frac{\partial v}{\partial y}  + \frac{\partial w}{\partial z} \bigg) \bigg] - 
\frac{\partial}{\partial y}\bigg[ \psi \lambda_{44} \bigg( \frac{\partial v}{\partial x} \bigg) \bigg] - 
\frac{\partial}{\partial z}\bigg[ \psi \lambda_{44} \bigg(\frac{\partial w}{\partial x} \bigg) \bigg],
\end{split}
\end{equation}
\begin{equation} \label{ME-I2-1}
\begin{split}
&\frac{\partial}{\partial x}\bigg[ \psi \lambda_{44} \bigg( \frac{\partial v}{\partial x} \bigg) \bigg] + 
\frac{\partial}{\partial y}\bigg[ \psi \lambda_{11} \bigg( \frac{\partial v}{\partial y} \bigg) \bigg] +
\frac{\partial}{\partial z}\bigg[ \psi \lambda_{44} \bigg( \frac{\partial v}{\partial z} \bigg) \bigg] =
\frac{\partial}{\partial y} [  \psi \rho (\lambda_{11}+2 \lambda_{12}) ] - \\
&\frac{\partial}{\partial x}\bigg[ \psi \lambda_{44} \bigg( \frac{\partial u}{\partial y} \bigg) \bigg] -
\frac{\partial}{\partial y}\bigg[ \psi \lambda_{12} \bigg( \frac{\partial u}{\partial x} + \frac{\partial w}{\partial z} \bigg) \bigg] - 
\frac{\partial}{\partial z}\bigg[ \psi \lambda_{44} \bigg( \frac{\partial w}{\partial y} \bigg) \bigg],
\end{split}
\end{equation}
\begin{equation} \label{ME-I3-1}
\begin{split}
& \frac{\partial}{\partial x}\bigg[ \psi \lambda_{44} \bigg( \frac{\partial w}{\partial x} \bigg) \bigg] + 
\frac{\partial}{\partial y}\bigg[ \psi \lambda_{44} \bigg( \frac{\partial w}{\partial y} \bigg) \bigg] +
\frac{\partial}{\partial z}\bigg[ \psi \lambda_{11} \bigg( \frac{\partial w}{\partial z} \bigg) \bigg] = 
\frac{\partial}{\partial z} [ \psi \rho (\lambda_{11}+2 \lambda_{12}) ] - \\
& \frac{\partial}{\partial x}\bigg[ \psi \lambda_{44} \bigg( \frac{\partial u}{\partial z} \bigg) \bigg] -
\frac{\partial}{\partial y}\bigg[ \psi \lambda_{44} \bigg( \frac{\partial v}{\partial z}  \bigg) \bigg] -
\frac{\partial}{\partial z}\bigg[ \psi \lambda_{12} \bigg(\frac{\partial u}{\partial x} + \frac{\partial v}{\partial y} \bigg) \bigg],
\end{split}
\end{equation}
\end{subequations}
where $u$, $v$ and $w$ are the displacements along the $x$, $y$ and $z$ axes, respectively. Here, we provide the long form of the mechanical equilibrium equation for the sake of clarity for the readers. To solve Eq.~\eqref{ME-ISO-1}, we use an ADLR method similar to that provided in Appendix \ref{SurLap_A2}. We keep the ``diagonal'' terms, such as $\partial (\psi \lambda \partial /\partial x)/\partial x$, $\partial (\psi \lambda \partial /\partial y)/\partial y$, and $\partial (\psi \lambda \partial /\partial z)/\partial z$ on the left-hand sides, and move the ``cross'' terms of the differential operator to the right-hand sides. As a result, we obtain three equations with second-order-partial-differential operators for the three displacement components. The ``diagonal'' terms on the left-hand sides can be discretized using the scheme in Eq.~\eqref{DiagDiffuOp-1}. The ``cross'' terms moved to the right-hand sides can be calculated in a scheme similar to Eq.~\eqref{CrossDiffuOp-1}. The ADLR method shown in Eq.~\eqref{ADI-Solve-0} is employed to solve the displacements, and the solutions are updated and iterated until all the displacement components reach their equilibrium values.

For the case in Section \ref{T-Stress}, the solid phase ($\psi = \psi_1 + \psi_2$) includes two different materials: GDC ($\psi_1$) and LSC ($\psi_2$). Therefore, we smoothly interpolate material properties appearing in the differential operators:
$\psi \lambda_{ij} = \psi_1 \lambda_{ij}^{GDC} +  \psi_2 \lambda_{ij}^{LSC}$. Similarly, the body force term is interpolated by $\psi \rho (\lambda_{11} + 2 \lambda_{12}) = \psi_1 \rho^{GDC}(\lambda_{11}^{GDC} + 2 \lambda_{12}^{GDC}) + \psi_2 \rho^{LSC}(\lambda_{11}^{LSC} +2 \lambda_{12}^{LSC})$, where $\rho = \alpha \Delta T$ is the thermal expansion. The thermal stress is calculated according to Hooke's Law, which is written with the domain-parameter-interpolated elastic constants and thermal expansions as: 
\begin{equation} \label{ThermoStrs1}
\sigma_{ij} = (\psi_1 C_{ijkl}^{GDC} + \psi_2 C_{ijkl}^{LSC})\frac{1}{2}\bigg(\frac{\partial u_k}{\partial x_l} + \frac{\partial u_l}{\partial x_k} \bigg) - ( \psi_1 \rho^{GDC} C_{ijkl}^{GDC} + \psi_2 \rho^{LSC} C_{ijkl}^{LSC}) \delta_{kl}.
\end{equation}

\end{appendix}


\newpage

\begin{table}[h] 
\centering
{\scriptsize\tt
\begin{tabulary}{0.5cm}{| c || c | c || c | c || c | c ||  c | c |}
\hline
Case & \multicolumn{2}{c||}{1} & \multicolumn{2}{c||}{2} & \multicolumn{2}{c||}{3}  & \multicolumn{2}{c|}{4} \\ \hline
\multirow{3}{*}{~} & \multicolumn{2}{c||}{$\Delta x =2.5\times10^{-2}$} & \multicolumn{2}{c||}{$\zeta=1.145\Delta x$}  & \multicolumn{2}{c||}{$\zeta=5.73\times10^{-2}$} & \multicolumn{2}{c|}{$\Delta x = 2.5\times10^{-2}$}  \\ 
~ &  \multicolumn{2}{c||}{$\upsilon =1\times10^{-7}$} & \multicolumn{2}{c||}{$\upsilon =1\times10^{-7}$} & \multicolumn{2}{c||}{$\upsilon =1\times10^{-7}$} & \multicolumn{2}{c|}{$\zeta =2.86\times10^{-2}$} \\ \cline{2-9}
~ & $\zeta$ & $e$ & $\Delta x$ & $e$ & $\Delta x$ & $e$ & $\upsilon$ & $e$ \\ \hline
a & $1.43\times10^{-2}$& $2.74\times10^{-4}$ & $1.25\times10^{-2}$ & $3.93\times10^{-4}$ & $1.25\times10^{-2}$ & $1.75\times10^{-3}$  & $1.0\times10^{-2}$ & $7.75\times10^{-3}$ \\ \hline
b & $2.86\times10^{-2}$ & $^{\ast}7.88\times10^{-4}$ & $2.50\times10^{-2}$ & $^{\ast}7.88\times10^{-4}$ & $2.50\times10^{-2}$ & $^{\diamond}1.72\times10^{-3}$ & $1.0\times10^{-3}$ & $1.39\times10^{-3}$ \\ \hline
c & $5.73\times10^{-2}$ & $^{\diamond}1.72\times10^{-3}$ & $5.00\times10^{-2}$ & $^{\triangleleft}1.58\times10^{-3}$ &  $5.00\times10^{-2}$ & $^{\triangleleft}1.58\times10^{-3}$ & $1.0\times10^{-5}$& $7.93\times10^{-4}$ \\ \hline
d & $1.15\times10^{-1}$ & $3.53\times10^{-3}$ & $1.00\times10^{-1}$ & $3.20\times10^{-3}$ & $1.00\times10^{-1}$ & $1.16\times10^{-3}$ & $1.0\times10^{-7}$& $^{\ast}7.88\times10^{-4}$ \\ \hline
e & $2.29\times10^{-1}$ & $7.20\times10^{-3}$ & $2.00\times10^{-1}$ & $6.54\times10^{-3}$ & $2.00\times10^{-1}$ & $7.53\times10^{-4}$ & $1.0\times10^{-9}$ & $7.88\times10^{-4}$ \\ \hline
f & $4.58\times10^{-1}$ & $1.49\times10^{-2}$ & $4.00\times10^{-1}$ & $1.39\times10^{-2}$ & $4.00\times10^{-1}$ & unstable & $1.0\times10^{-11}$ & $7.88\times10^{-4}$ \\ \hline
\end{tabulary} }
\caption{Relative errors, $e$, for the 1D smoothed boundary diffusion equations to the analytical solutions with various parameters. Each of the markers, $\ast$, $\diamond$ and $\triangleleft$, denotes the identical result from a set of parameters.} \label{Tbl-1}
\end{table} 

\begin{table}[h] 
\centering
{\scriptsize\tt
\begin{tabulary}{0.5cm}{| c || c | c || c | c || c | c ||  c | c |}
\hline
~ & \multicolumn{2}{c||}{$\kappa = 2.1$} & \multicolumn{2}{c||}{$\kappa=20$} & \multicolumn{2}{c||}{$\kappa=50$}  & \multicolumn{2}{c|}{$\kappa=100$} \\ \hline
thin-interface  & $e$ & $7.99\times10^{-4}$ & $e$ & $2.26\times10^{-3}$ & $e$ & $2.46\times10^{-3}$ & $e$ & $7.26\times10^{-3}$  \\ \cline{2-9}
\multirow{2}{*}{$\xi_0=0.075$} & $e_\text{b}$ & $7.99\times10^{-4}$ & $e_\text{b}$ & $2.23\times10^{-3}$ & $e_\text{b}$ & $2.39\times10^{-3}$ & $e_\text{b}$ & $7.37\times10^{-3}$ \\ \cline{2-9}
~ &  $e_\text{s}$ & $8.03\times10^{-4}$ & $e_\text{s}$ & $3.02\times10^{-3}$ & $e_\text{s}$ & $4.02\times10^{-3}$ & $e_\text{s}$ & $2.63\times10^{-3}$ \\ \hline
medium-interface & $e$ & $1.08\times10^{-3}$ & $e$ & $3.06\times10^{-3}$ & $e$ & $8.32\times10^{-3}$ & $e$ & $2.74\times10^{-2}$  \\ \cline{2-9}
\multirow{2}{*}{$\xi_0=0.149$}  & $e_\text{b}$ & $1.04\times10^{-3}$ & $e_\text{b}$ & $2.89\times10^{-3}$ & $e_\text{b}$ & $8.51\times10^{-3}$ & $e_\text{b}$ & $2.84\times10^{-2}$ \\ \cline{2-9}
~ &  $e_\text{s}$ & $1.50\times10^{-3}$ & $e_\text{s}$ & $4.83\times10^{-3}$ & $e_\text{s}$ & $5.12\times10^{-3}$ & $e_\text{s}$ & $1.86\times10^{-3}$ \\ \hline
thick-interface & $e$ & $1.81\times10^{-3}$ & $e$ & $1.08\times10^{-2}$ & $e$ & $2.90\times10^{-2}$ & $e$ & $7.34\times10^{-2}$  \\ \cline{2-9}
\multirow{2}{*}{$\xi_0=0.292$} & $e_\text{b}$ & $1.63\times10^{-3}$ & $e_\text{b}$ & $1.13\times10^{-2}$ & $e_\text{b}$ & $3.11\times10^{-2}$ & $e_\text{b}$ & $7.89\times10^{-2}$ \\ \cline{2-9}
~ &  $e_\text{s}$ & $2.69\times10^{-3}$ & $e_\text{s}$ & $6.78\times10^{-3}$ & $e_\text{s}$ & $5.75\times10^{-3}$ & $e_\text{s}$ & $1.06\times10^{-3}$ \\ \hline
\end{tabulary} }
\caption{Relative errors for the coupled surface reaction-diffusion and bulk diffusion model in a cylinder, where $e$ denotes the overall error, $e_\text{b}$ denotes the bulk error excluding the surface points, and $e_\text{s}$ denotes the surface error calculated only with the surface points.} \label{Tbl-2}
\end{table} 

\begin{table}[h] 
\centering
{\scriptsize\tt
\begin{tabulary}{0.5cm}{| c || c | c | c | c | c |}
\hline
\multicolumn{6}{|c|}{Contact angles for Allen-Cahn equation} \\ \hline
$\zeta$ &$\delta_{\phi}=1.0607$ & $\delta_{\phi}=1.4142$ & $\delta_{\phi}=1.7678$  & $\delta_{\phi}=2.1213$ & $\delta_{\phi}=2.8284$ \\ \hline
$0.75$ & $0.5050~(59.67^\circ)$ & $0.5048~(59.68^\circ)$ & $0.4965~(60.23^\circ)$ & $0.5001~(59.99^\circ)$ & $0.5004~(59.97^\circ)$ \\ \hline
$1.00$ & $0.4900~(60.66^\circ)$ & $0.5039~(59.74^\circ)$ & $0.4966~(60.22^\circ)$ & $0.4982~(60.12^\circ)$ & $0.4956~(60.29^\circ)$ \\ \hline
$1.50$ & $0.4865~(60.89^\circ)$ & $0.4962~(60.25^\circ)$ & $0.4927~(60.48^\circ)$ & $0.4938~(60.41^\circ)$ & $0.4927~(60.48^\circ)$ \\ \hline
$2.00$ & $0.4886~(60.75^\circ)$ & $0.4918~(60.54^\circ)$ & $0.4883~(60.77^\circ)$ & $0.4901~(60.65^\circ)$ & $0.4901~(60.65^\circ)$ \\ \hline
$4.00$ & $0.4825~(61.15^\circ)$ & $0.4782~(61.43^\circ)$ & $0.4783~(61.43^\circ)$ & $0.4795~(61.35^\circ)$ & $0.4790~(61.38^\circ)$ \\ \hline \hline
\multicolumn{6}{|c|}{Contact angles for Cahn-Hilliard equation} \\ \hline
$\zeta$ &$\delta_{\phi}=1.0607$ & $\delta_{\phi}=1.4142$ & $\delta_{\phi}=1.7678$  & $\delta_{\phi}=2.1213$ & $\delta_{\phi}=2.8284$ \\ \hline
$0.75$ & $-0.4831~(118.89^\circ)$ & $-0.5003~(120.02^\circ)$ & $-0.4937~(119.59^\circ)$ & $-0.4979~(119.86^\circ)$ & $-0.4931~(119.54^\circ)$ \\ \hline
$1.00$ & $-0.4923~(119.49^\circ)$ & $-0.4926~(119.51^\circ)$ & $-0.4897~(119.32^\circ)$ & $-0.4929~(119.53^\circ)$ & $-0.4881~(119.21^\circ)$ \\ \hline
$1.50$ & $-0.4841~(118.95^\circ)$ & $-0.4871~(119.15^\circ)$ & $-0.4868~(119.13^\circ)$ & $-0.4890~(119.28^\circ)$ & $-0.4867~(119.13^\circ)$ \\ \hline
$2.00$ & $-0.4639~(117.64^\circ)$ & $-0.4857~(119.06^\circ)$ & $-0.4861~(119.09^\circ)$ & $-0.4862~(119.09^\circ)$ & $-0.4853~(119.13^\circ)$ \\ \hline
$4.00$ & $-0.4372~(115.93^\circ)$ & $-0.4713~(118.12^\circ)$ & $-0.4752~(118.37^\circ)$ & $-0.4745~(118.33^\circ)$ & $-0.4752~(118.37^\circ)$ \\ \hline
\multicolumn{6}{|c|}{Order parameter conservation for Cahn-Hilliard equation} \\ \hline
$\zeta$ &$\delta_{\phi}=1.0607$ & $\delta_{\phi}=1.4142$ & $\delta_{\phi}=1.7678$  & $\delta_{\phi}=2.1213$ & $\delta_{\phi}=2.8284$ \\ \hline
$0.75$ & $0.9929$ &  $0.9972$ & $0.9979$ & $0.9982$ & $0.9986$ \\ \hline
$1.00$ & $0.9930$ &  $0.9973$ & $0.9979$ & $0.9982$ & $0.9986$ \\ \hline
$1.50$ & $0.9933$&  $0.9974$ & $0.9980$ & $0.9983$ & $0.9987$ \\ \hline
$2.00$ & $0.9976$ & $0.9991$ & $0.9993$ & $0.9994$ & $0.9996$ \\ \hline
$4.00$ & $0.9982$&  $0.9993$ & $0.9995$ & $0.9996$ & $0.9997$ \\ \hline
\end{tabulary} }
\caption{Contact angle results from the validation simulations using the Allen-Cahn equation and the Cahn-Hilliard equation. In addition, a measure of the order parameter conservation, evaluated by $\int \psi \phi(t_\text{ss}) d\Omega / \int \psi \phi(t=0) d\Omega$, where $t_\text{ss}$ is the time required to reach steady-state conditions and $\Omega$ is the computational domain, are presented for simulations with the Cahn-Hilliard equation.} \label{Tbl-3}
\end{table} 

\newpage

\end{document}